\documentclass[superscriptaddress,aps,preprintnumbers,amsmath,amssymb,prd,nofootinbib,preprint]{revtex4-1}
\pdfoutput=1

\usepackage{graphicx}
\usepackage{epstopdf}
\usepackage{dcolumn}
\usepackage{bm}
\usepackage{bbm}
\usepackage{hyperref}
\usepackage[usenames, dvipsnames]{color}
\usepackage{amsmath,amssymb}
\usepackage[sort&compress]{natbib}
\usepackage{float}
\usepackage{multirow}
\usepackage{slashed}
\usepackage{cancel}
\usepackage[table]{xcolor}
\usepackage{multirow}
\usepackage{booktabs}
\usepackage{makecell}
\usepackage{mathtools}
\usepackage{pifont}
\usepackage{enumitem}

\numberwithin{equation}{section}

\makeatletter
\renewcommand{\p@subsection}{}
\renewcommand{\p@subsubsection}{}
\makeatother

\makeatletter
\def\l@subsubsection#1#2{}
\makeatother

\makeatletter
    \def\CT@@do@color{%
      \global\let\CT@do@color\relax
            \@tempdima\wd\z@
            \advance\@tempdima\@tempdimb
            \advance\@tempdima\@tempdimc
    \advance\@tempdimb\tabcolsep
    \advance\@tempdimc\tabcolsep
    \advance\@tempdima2\tabcolsep
            \kern-\@tempdimb
            \leaders\vrule
                    \hskip\@tempdima\@plus  1fill
            \kern-\@tempdimc
            \hskip-\wd\z@ \@plus -1fill }
    \makeatother
 
\makeatletter
\def\@ssect@ltx#1#2#3#4#5#6[#7]#8{%
  \def\H@svsec{\phantomsection}%
  \@tempskipa #5\relax
  \@ifdim{\@tempskipa>\z@}{%
    \begingroup
      \interlinepenalty \@M
      #6{%
       \@ifundefined{@hangfroms@#1}{\@hang@froms}{\csname @hangfroms@#1\endcsname}%
       {\hskip#3\relax\H@svsec}{#8}%
      }%
      \@@par
    \endgroup
    \@ifundefined{#1smark}{\@gobble}{\csname #1smark\endcsname}{#7}%
  }{%
    \def\@svsechd{%
      #6{%
       \@ifundefined{@runin@tos@#1}{\@runin@tos}{\csname @runin@tos@#1\endcsname}%
       {\hskip#3\relax\H@svsec}{#8}%
      }%
      \@ifundefined{#1smark}{\@gobble}{\csname #1smark\endcsname}{#7}%
      \addcontentsline{toc}{#1}{\protect\numberline{}#8}%
    }%
  }%
  \@xsect{#5}%
}%
\makeatother

\hypersetup{colorlinks,citecolor= blue,linkcolor= black}

\begin{document}


\def\a{\alpha}
\def\b{\beta}
\def\c{\varepsilon}
\def\d{\delta}
\def\e{\epsilon}
\def\f{\phi}
\def\g{\gamma}
\def\h{\theta}
\def\k{\kappa}
\def\l{\lambda}
\def\m{\mu}
\def\n{\nu}
\def\p{\psi}
\def\q{\partial}
\def\r{\rho}
\def\s{\sigma}
\def\t{\tau}
\def\u{\upsilon}
\def\v{\varphi}
\def\w{\omega}
\def\x{\xi}
\def\y{\eta}
\def\z{\zeta}
\def\D{\Delta}
\def\G{\Gamma}
\def\H{\Theta}
\def\L{\Lambda}
\def\F{\Phi}
\def\P{\Psi}
\def\S{\Sigma}

\def\o{\over}
\def\beq{\begin{align}}
\def\eeq{\end{align}}
\newcommand{\gsim}{ \mathop{}_{\textstyle \sim}^{\textstyle >} }
\newcommand{\lsim}{ \mathop{}_{\textstyle \sim}^{\textstyle <} }
\newcommand{\vev}[1]{ \left\langle {#1} \right\rangle }
\newcommand{\bra}[1]{ \langle {#1} | }
\newcommand{\ket}[1]{ | {#1} \rangle }
\newcommand{\1}{\mbox{1}\hspace{-0.25em}\mbox{l}}
\newcommand{\headline}[1]{\noindent{\bf #1}}
\def\diag{\mathop{\rm diag}\nolimits}
\def\Spin{\mathop{\rm Spin}}
\def\SO{\mathop{\rm SO}}
\def\O{\mathop{\rm O}}
\def\SU{\mathop{\rm SU}}
\def\U{\mathop{\rm U}}
\def\Sp{\mathop{\rm Sp}}
\def\SL{\mathop{\rm SL}}
\def\tr{\mathop{\rm tr}}
\def\mpl{M_{\rm Pl}}

\def\IJMP{Int.~J.~Mod.~Phys. }
\def\MPL{Mod.~Phys.~Lett. }
\def\NP{Nucl.~Phys. }
\def\PL{Phys.~Lett. }
\def\PR{Phys.~Rev. }
\def\PRL{Phys.~Rev.~Lett. }
\def\PTP{Prog.~Theor.~Phys. }
\def\ZP{Z.~Phys. }

\def\dd{\mathrm{d}}
\def\ff{\mathrm{f}}
\def\BH{{\rm BH}}
\def\inf{{\rm inf}}
\def\ev{{\rm evap}}
\def\eq{{\rm eq}}
\def\SM{{\rm sm}}
\def\Mpl{M_{\rm Pl}}
\newcommand{\Red}[1]{\textcolor{red}{#1}}
\newcommand{\MeV}{\textrm{ MeV}}
\newcommand{\GeV}{\textrm{ GeV}}
\newcommand{\TeV}{\textrm{ TeV}}
\newcommand{\eV}{\textrm{ eV}}

\newcommand{\dec}{{\rm dec}}
\newcommand{\nr}{{\rm nr}}

\newcommand{\nn}{\nonumber}
\newcommand{\bea}{\begin{eqnarray}}
\newcommand{\eea}{\end{eqnarray}}

\newcommand{\LC}{\ensuremath{\Lambda\mathrm{CDM}}}

\newcommand{\tot}{\mathrm{tot}}
\newcommand{\mpc}{\mathrm{Mpc}}

\newcommand{\dm}{\mathrm{dm}}
\newcommand{\cdm}{\mathrm{cdm}}
\newcommand{\wdm}{\mathrm{wdm}}
\newcommand{\DN}{\Delta N_\mathrm{eff}}

\newcommand{\ta}{{\tilde{a}}}
\newcommand{\grav}{{3/2}}

\newcommand{\mH}{\mathcal{H}}
\newcommand{\mO}{\ensuremath{\mathcal{O}}}
\newcommand{\mL}{\ensuremath{\mathcal{L}}}
\newcommand{\mV}{\ensuremath{\mathcal{V}}}
\newcommand{\mM}{\ensuremath{\mathcal{M}}}

\newcommand{\Sec}[1]{Sec.~\ref{#1}}
\newcommand{\Secs}[2]{Secs.~\ref{#1} and \ref{#2}}
\newcommand{\App}[1]{Appendix~\ref{#1}}
\newcommand{\Tab}[1]{Table~\ref{#1}}
\newcommand{\Fig}[1]{Fig.~\ref{#1}}
\newcommand{\Eq}[1]{Eq.~(\ref{#1})}
\newcommand{\Eqs}[2]{Eqs.~(\ref{#1}) and (\ref{#2})}
\newcommand{\Eqst}[2]{Eqs.~(\ref{#1})-(\ref{#2})}
\newcommand{\eg}{\textit{e.g.}\ }
\newcommand{\ie}{\textit{i.e.}\ }
\newcommand{\draftnote}[1]{\textbf{#1}}

\newcommand{\bl}{\left}
\newcommand{\br}{\right}

\title{Lepto-Axiogenesis}
\preprint{LCTP-20-12}

\author{Raymond T.~Co}
\affiliation{\small Leinweber Center for Theoretical Physics, Department of Physics, University of Michigan, Ann Arbor, MI 48109, USA}
\author{Nicolas~Fernandez}
\affiliation{\small Department of Physics, University of Illinois at Urbana-Champaign, Urbana, IL 61801, USA}
\author{Akshay~Ghalsasi}
\affiliation{\small Department of Physics, University of California Santa Cruz, Santa Cruz, CA 95064, USA}
\affiliation{\small Santa Cruz Institute for Particle Physics, Santa Cruz, CA 95064, USA}
\author{Lawrence J.~Hall}
\affiliation{\small Department of Physics, University of California, Berkeley, CA 94720, USA}
\affiliation{\small Theoretical Physics Group, Lawrence Berkeley National Laboratory, Berkeley, CA 94720, USA}
\author{Keisuke Harigaya}
\affiliation{\small School of Natural Sciences, Institute for Advanced Study, Princeton, NJ 08540, USA}

\begin{abstract}
We propose a baryogenenesis mechanism that uses a rotating condensate of a Peccei-Quinn (PQ) symmetry breaking field and the dimension-five operator that gives Majorana neutrino masses. The rotation induces charge asymmetries for the Higgs boson and for lepton chirality through sphaleron processes and Yukawa interactions. The dimension-five interaction transfers these asymmetries to the lepton asymmetry, which in turn is transferred into the baryon asymmetry through the electroweak sphaleron process. QCD axion dark matter can be simultaneously produced by dynamics of the same PQ field via kinetic misalignment or parametric resonance, favoring an axion decay constant $f_a \lesssim 10^{10}$~GeV, or by conventional misalignment and contributions from strings and domain walls with $f_a \sim 10^{11}$~GeV. The size of the baryon asymmetry is tied to the mass of the PQ field. In simple supersymmetric theories, it is independent of UV parameters and predicts the supersymmtry breaking mass scale to be $\mathcal{O}(10-10^4)$~TeV, depending on the masses of the neutrinos and whether the condensate is thermalized during a radiation or matter dominated era.
The high supersymmetry breaking mass scale may be free from cosmological and flavor/CP problems.
We also construct a theory where TeV scale supersymmetry is possible.  Parametric resonance may give warm axions, and the radial component of the PQ field may give signals in rare kaon decays from mixing with the Higgs and in dark radiation.
\end{abstract}

\maketitle

\begingroup
\hypersetup{linkcolor=black}
\renewcommand{\baselinestretch}{1.12}\normalsize
\tableofcontents
\renewcommand{\baselinestretch}{2}\normalsize
\endgroup

\newpage

\section{Introduction}
\label{sec:intro}

Over several decades, as the standard models of particle physics and the early universe have solidified, the outstanding problems left unaddressed by this theory have become ever more pressing.  These include the smallness of dimensionless parameters, such as the strong CP parameter and the light quark and lepton Yukawa couplings, as well as hierarchies between mass scales, e.g.~those associated with dark energy, neutrino masses, and the weak and gravitational scales.  Two cosmological issues are particularly pressing: the nature and abundance of dark matter (DM) and the origin of the small but crucial baryon asymmetry.

Indeed, with so much to explain, it is interesting to pursue simple ideas for new physics that make progress on several fronts.  Grand unification \cite{Georgi:1974sy} explains both the gauge quantum numbers of a fermion generation and the quantization of electric charge, as well as providing a framework for addressing gauge coupling unification~\cite{Georgi:1974yf}, the baryon asymmetry and neutrino masses. Supersymmetry provides a dark matter candidate~\cite{Witten:1981nf,Pagels:1981ke,Goldberg:1983nd} and precise gauge coupling unification~\cite{Dimopoulos:1981yj,Dimopoulos:1981zb,Sakai:1981gr,Ibanez:1981yh,Einhorn:1981sx,Marciano:1981un}, and partially explains the smallness of the electroweak scale~\cite{Maiani:1979cx,Veltman:1980mj,Witten:1981nf,Kaul:1981wp}.

In this paper, we introduce and study a framework that simultaneously addresses the strong CP problem, neutrino masses, the baryon asymmetry, and dark matter. We postulate a complex scalar field $P$ with the following features:
\begin{itemize}
    \item It spontaneously breaks a Peccei-Quinn (PQ) symmetry \cite{Peccei:1977hh,Peccei:1977ur} so that its phase $\theta(x)$ is the axion field \cite{Weinberg:1977ma,Wilczek:1977pj}, which solves the strong CP problem~\cite{Peccei:1977hh,Peccei:1977ur,Vafa:1984xg}.
    
    \item The radial component $S$ of the field has a flat potential and a large field value early in the cosmological evolution, for example during inflation. When the field starts to oscillate, higher-dimensional PQ-breaking operators in the potential lead to a velocity of the angular field component $\dot{\theta} \neq 0$. Subsequently, this rotating PQ condensate carries a conserved PQ charge density, relative to entropy density $s$, of $Y_{\rm PQ} \sim \dot{\theta}S^2/s$. 
    
    \item The QCD anomaly of the PQ symmetry leads to a strong sphaleron process in the early universe where $\dot{\theta}$ sources a particle-antiparticle asymmetry for quark chirality and, via Yukawa couplings and the electroweak sphaleron process, for the Higgs boson, $H$, and for lepton chirality. We assume that at these high temperatures neutrino masses are described by the dimension-five interaction $\ell \ell H^\dagger H^\dagger$~\cite{Weinberg:1979sa}, which acts to transfer the Higgs asymmetry to a lepton asymmetry via freeze-in with an amplitude proportional to the neutrino mass.
    
    \item The standard model violates $B+L$ via an anomaly from the electroweak gauge interaction \cite{Klinkhamer:1984di,Kuzmin:1985mm}, and at temperatures above the weak scale this acts to redistribute the above lepton asymmetry among both quarks and leptons, yielding a baryon asymmetry $Y_B \propto Y_{\rm PQ} m_\nu^2$. Note that both out-of-equilibrium and CP-violation requirements are satisfied by the PQ condensate.
    
    \item At temperatures near the GeV scale, $\dot{\theta}$ may be sufficiently large that the conventional misalignment mechanism~\cite{Preskill:1982cy,Abbott:1982af,Dine:1982ah} for axion dark matter is inoperative. Instead, axion dark matter is produced by the kinetic misalignment mechanism (KMM)~\cite{Co:2019jts} with a density $\rho_a \propto Y_{\rm PQ}/ f_a$.  The ratio of dark matter and baryon densities is independent of $Y_{\rm PQ}$. The QCD axion can explain the dark matter density of the universe for $f_a < 10^{11}$ GeV, while large $f_a$ requires entropy production after the QCD phase transition.
    
    \item After the field starts rotating, depending on the shape of the potential and the rotations, parametric resonance (PR)~\cite{Dolgov:1989us,Traschen:1990sw,Kofman:1994rk,Kofman:1997yn} may be effective at producing axions. If not thermalized, these axions contribute to the dark matter density~\cite{Co:2017mop,Harigaya:2019qnl} and may be warm enough to affect structure formation at an observable level.
\end{itemize}

Our work builds on several developments beginning in the 1980s. Affleck and Dine~\cite{Affleck:1984fy} introduced a rotating condensate that carried baryon charge, with a baryon asymmetry generated from the decay of the condensate.  With spontaneous baryogenesis \cite{Cohen:1987vi,Cohen:1988kt}, Cohen and Kaplan proposed that the angular velocity of the condensate could act as an effective chemical potential for a thermal bath, generating a baryon asymmetry for the quarks using a baryon number violating interaction. Conversion of lepton asymmetry into baryon asymmetry by electroweak sphaleron processes was utilized in leptogenesis~\cite{Fukugita:1986hr,Davidson:2008bu}. Possibilities for generating the baryon asymmetry by electroweak sphaleron processes at a first order electroweak phase transition, called electroweak baryogenesis, were investigated in~\cite{Kuzmin:1985mm, Cohen:1990it, Cohen:1990py}.

In later work, it was realized that baryogenesis could result from a condensate carrying charge $Q$ other than baryon number \cite{Chiba:2003vp, Takahashi:2003db}, although these papers required an interaction that violates both $Q$ and $B$ to be in thermal equilibrium. Baryogenesis by the $\ell\ell H^\dagger H^\dagger$ interaction and the coupling of the angular velocity of the condensate with weak gauge bosons was investigated in~\cite{Kusenko:2014uta}. The angular direction was assumed to oscillate rather than orbit. The oscillation was caused by a large mass for the angular direction and the connection with the QCD axion was not obvious; however, Ref.~\cite{Takahashi:2015waa} proposed a way to identify the angular direction with the QCD axion by giving it a large mass only in the early universe. For oscillations, the baryon asymmetry is dominantly produced at the beginning of the oscillation, while for rotations, considered in this paper, the production may be dominated at a much later time, qualitatively changing the physical picture.

Axiogenesis~\cite{Co:2019wyp} uses the condensate of a PQ field and $B+L$ number violation from the electroweak sphaleron process, and is closer to our work. The baryon asymmetry is produced near the weak scale, well after the radial component of the PQ field has settled to $f_a$. Although the observed baryon asymmetry can be explained by the mechanism, the kinetic misalignment mechanism overproduces axion dark matter, unless a new ingredient is added, for example by raising the temperature of the electroweak phase transition above the weak scale.
In the setup we discuss in this paper, the baryon asymmetry is produced much before the electroweak phase transition with aid from the $\ell\ell H^\dag H^\dag$ interaction. We call the mechanism lepto-axiogenesis.

Since the radial component of the PQ symmetry breaking field evolves, approximate PQ charge conservation implies that $\dot{\theta}$ decreases slower than the case with a constant radial component, or even stays constant. When the baryon asymmetry is dominantly  produced depends on the potential of the PQ symmetry breaking field.

We first investigate the simplest potential of the PQ symmetry breaking field with a negative quadratic term and a positive quartic term. We show that the observed baryon asymmetry can be explained by lepto-axiogenesis, and the reheat temperature after inflation may be as low as $10^9 \GeV$.

We next investigate supersymmetric theories where the PQ symmetry breaking has a nearly quadratic potential given by soft supersymmetry breaking, and is naturally flat. Assuming the oscillation of $P$ is initiated by the zero-temperature mass, the observed baryon asymmetry determines the scale of supersymmetry breaking to be $30-700 \TeV$ and $300-7000 \TeV$ for degenerate and hierarchical neutrino masses, respectively. This should be compared with Affleck-Dine baryogenesis from squarks and sleptons, where the soft mass scale is not restricted. The reheat temperature after inflation may be as low as $10^7 \GeV$. We also discuss the case where the oscillation is initiated by a thermal potential and show that TeV scale supersymmetry can be consistent with baryogenesis from the rotation of $P$.

In these two theories, one with supersymmetry and the other without, we investigate how the parameter space is restricted when we require cogenesis of the baryon asymmetry and axion dark matter.

This paper is organized as follows. In Sec.~\ref{sec:leptoaxiogenesis}, we review the mechanism of axiogenesis and introduce the new lepto-axiogenesis scenario. Secs.~\ref{sec:quartic} and \ref{sec:susy} investigate the PQ symmetry breaking field $P$ for quartic and nearly quadratic potentials, respectively.
Finally, Sec.~\ref{sec:con} is devoted to a summary of the results and presents our conclusions.

\section{Axiogenesis and Majorana neutrino masses}
\label{sec:leptoaxiogenesis}
In this section, we first review axiogenesis introduced in~\cite{Co:2019wyp} as a mechanism of baryogenesis involving the QCD axion. We then introduce lepto-axiogenesis.

\subsection{Axiogenesis}
\label{sec:axiogenesis_min}
\begin{figure}[!ht]
	\includegraphics[width=\linewidth]{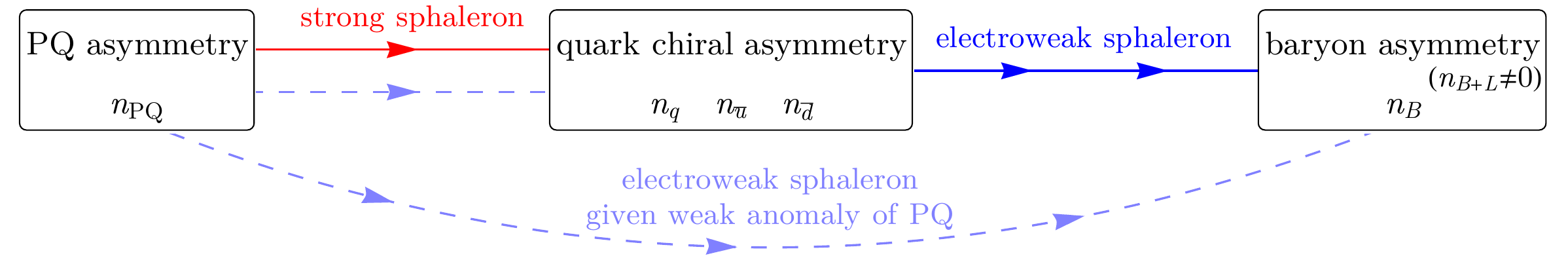}
	\caption{Transfer of asymmetries for axiogenesis.}
	\label{fig:asym_axio}	
\end{figure}

Let us assume that the PQ symmetry is explicitly broken in the early universe, and the explicit breaking induces a rotation in the phase direction of the PQ symmetry breaking field $P$, which has a radial component $S$ and an angular component $\theta$,
\begin{align}
\label{eq:P}
 P = \frac{1}{\sqrt{2}}(f_a N_{\rm DW} + S) e^{i \theta/N_{\rm DW}}.
\end{align}
Here $N_{\rm DW}$ is the domain wall number and we define $\theta$ so that it receives a potential with a periodicity of $2\pi$ from QCD.
The rotation corresponds to PQ charge asymmetry,
\begin{align}
\label{eq:PQ charge}
n_{\rm PQ} = &  \frac{1}{N_{\rm DW}}\left(i\dot{P^*}P- i\dot{P}P^*\right)= \dot{\theta}f_{\rm eff}^2,\nonumber \\
f_{\rm eff}^2 \equiv & \left(f_a + \frac{S}{N_{\rm DW}}\right)^2,
\end{align}
where $f_{\rm eff}$ takes into account that the effective decay constant may be different in the early universe from today's value $f_a$.
Since the PQ symmetry has a QCD anomaly, the PQ asymmetry is partially converted into chiral asymmetry of quarks via strong sphaleron transitions. The chiral asymmetry is further converted into $B+L$ asymmetry via electroweak sphaleron transitions. If the PQ symmetry also has weak anomaly, the PQ asymmetry is directly converted into $B+L$ asymmetry. The flow of the asymmetries is shown in Fig.~\ref{fig:asym_axio}.

The sphaleron transition is effective for low enough temperature.
The electroweak sphaleron transition rate per unit time and volume is given by~\cite{DOnofrio:2014rug}
\begin{align}
\Upsilon_{\rm ws} \simeq 20\times \alpha^{5}_{2} T^{4}.
\end{align}
The transitions creating and destroying baryon asymmetry occur almost at the same rate, but asymmetry of fermion numbers induces small bias to the transitions of the two directions. $B+L$ asymmetry is produced with a rate $\Gamma_{\rm ws} \equiv \Upsilon_{\rm ws}/T^3 $~\cite{Dine:1989kt}.
The asymmetry of quark doublets $q$ and lepton doublets $\ell$ evolve as
\begin{align}
\dot{n}_\ell = \Gamma_{\rm ws} (- n_q - n_\ell) + \cdots \nonumber \\
\dot{n}_q = 3\Gamma_{\rm ws} (- n_q - n_\ell) + \cdots,
\end{align}
and reach equilibrium value if $\Gamma_{\rm ws} > H$. For radiation domination, $\Gamma_{\rm ws} > H$ is satisfied for
\begin{align}
\label{eq:Tws_RD}
T  < T_{\rm ws} \equiv 20 \left( \frac{90}{\pi^2 g_*} \right)^{ \scalebox{1.01}{$\frac{1}{2}$} } \alpha_2^5 M_{\rm Pl}  \simeq 10^{11}\GeV  \left(\frac{g_{\rm SM}}{g_*} \right)^{ \scalebox{1.01}{$\frac{1}{2}$} } \left( \frac{\alpha_2}{1/40} \right)^5 ,
\end{align}
where $g_{\rm SM} = 106.75$ the full Standard Model degrees of freedom and $M_{\rm Pl} = 2.4 \times 10^{18} \GeV$ the reduced Planck mass.

The strong sphaleron transition rate per unit time and volume is given by~\cite{Moore:2010jd}
\begin{align}
\Upsilon_{\rm ss} \simeq 200 \times \alpha^{5}_{3} T^{4},
\end{align}
which is in equilibrium when $\Gamma_{\rm ss} \equiv \Upsilon_{\rm ss}/T^3 > H$, corresponding to the temperature 
\begin{equation}
\label{eq:Tss_RD}
T  < T_{\rm ss} \equiv 200 \left( \frac{90}{\pi^2 g_*} \right)^{ \scalebox{1.01}{$\frac{1}{2}$} } \alpha_3^5 M_{\rm Pl} \simeq 3 \times 10^{12}\GeV  \left(\frac{g_{\rm SM}}{g_*} \right)^{ \scalebox{1.01}{$\frac{1}{2}$} } \left( \frac{\alpha_3}{1/35} \right)^5 ,
\end{equation}
in a radiation-dominated universe.

The equilibrium values of baryon and lepton asymmetry are given by
\begin{align}
\label{eq:nB+L}
n_{B} = n_{L} = c_B \dot{\theta} T^2 = c_B \frac{T^2}{f_{\rm eff}^2} n_{\rm PQ},
\end{align}
where $c_B$ is a constant whose natural value is $\mathcal{O}(0.1)$, and we assume $f_{\rm eff} \gg T$. Minimizing the free energy leads to most of the PQ charge asymmetry remaining in the form of condensate rotation, rather than as asymmetry of particles in the thermal bath~\cite{Co:2019wyp}, explaining why the baryon and lepton asymmetries are much smaller than the PQ charge asymmetry. Similarly, other particle asymmetries such as quark/lepton chiral asymmetries and the Higgs number asymmetry are also much smaller than the PQ charge asymmetry.

The baryon asymmetry is fixed around a temperature $T_{\rm ws,FO}$, below which the electroweak sphaleron transition becomes ineffective. The baryon asymmetry normalized by the entropy density $s$ is then given by
\begin{align}
Y_B \equiv \frac{n_B}{s} = c_B \frac{T_{\rm ws,FO}^2}{f_a^2} \frac{n_{\rm PQ}}{s} = 8\times 10^{-11} \left( \frac{c_B}{0.1} \right) \left( \frac{T_{\rm ws,FO}}{130~{\rm GeV}} \right)^2  \left( \frac{10^8~{\rm GeV}}{f_a} \right)^2 \left( \frac{Y_{\rm PQ}}{500} \right) ,
\end{align}
where it is assumed $f_{\rm eff} = f_a$ by the time $T = T_{\rm ws,FO}$.

The non-zero value of $\dot{\theta}$ affects the axion dark matter abundance. 
Usually the QCD axion begins oscillation around the minimum once the axion mass exceeds the Hubble expansion rate $H$ around the QCD phase transition.
For large enough $Y_{\rm PQ}$, the axion moves rapidly even around the QCD phase transition, and the beginning of the oscillation is delayed~\cite{Co:2019jts}. The resultant dark matter abundance is enhanced in comparison with the conventional misalignment mechanism. The yield $Y_a$ of the axion oscillation is as large as the PQ charge asymmetry.
The yield of the axion to explain the observed dark matter abundance is
\begin{align}
\label{eq:yaDM}
Y_{a,{\rm DM}} = \frac{\rho_{\rm DM}/s}{m_a} =  70 \left( \frac{f_{a}}{10^{9}\GeV} \right). 
\end{align}
Using this, the axion abundance is 
\begin{align}
\Omega_a h^2 \simeq \Omega_{\rm DM} h^2 \left( \frac{10^8~{\rm GeV}}{f_a} \right) \left( \frac{Y_{\rm PQ}}{4} \right).
\end{align}
We review this kinetic misalignment mechanism in Appendix~\ref{app:KMM}.

For the SM prediction $T_{\rm ws,FO}\simeq 130$ GeV~\cite{DOnofrio:2014rug}, $c_B\sim 1$ and the decay constraint satisfying the astrophysical lower bound $f_a \gtrsim 10^8$ GeV~\cite{Ellis:1987pk,Raffelt:1987yt,Turner:1987by,Mayle:1987as,Raffelt:2006cw,Chang:2018rso,Carenza:2019pxu}, axion dark matter is overproduced. In Ref.~\cite{Co:2019wyp}, this problem is avoided by 1) an early electroweak phase transition from a new scalar coupled to the Higgs or 2) $c_B \gg 1$ by large coupling between the axion and the weak gauge boson.  

\subsection{Lepto-Axiogenesis}
\label{sec:lepto-axio}

As seen in axiogenesis, the non-zero velocity of the axion field $\dot{\theta}$ induces a quark chiral asymmetry as well as Higgs number and lepton chiral asymmetries via the anomaly of the PQ symmetry and sphaleron transitions. More generally, the axion velocity can also directly generate Higgs number and lepton chiral asymmetries depending on the UV theory of the axion. If $B-L$ symmetry is explicitly broken, these asymmetries may give rise to a $B-L$ asymmetry. Since the $B-L$ asymmetry is not washed-out by electroweak sphalerons, the baryon asymmetry may be generated far above the electroweak phase transition, evading the problem of overproducing axion dark matter.
If neutrino masses are Majorana, the explicit breaking of lepton symmetry can convert the asymmetries generated from the axion velocity into $B-L$ asymmetry at high temperatures. We call this scenario lepto-axiogenesis. 

To be concrete, we generate Majorana neutrino masses from the dimension-five operator~\cite{Weinberg:1979sa}
\begin{align}
{\cal L}_\nu = \frac{1}{2M_{5}} \ell \ell \, H^\dagger H^\dagger = \frac{m_\nu}{2 v_{\rm EW}^2} \ell \ell \, H^\dagger H^\dagger,
\end{align}
which may arise from the seesaw mechanism~\cite{Yanagida:1979as,GellMann:1980vs,Minkowski:1977sc,Mohapatra:1979ia}. Here $\ell$ and $H$ are lepton and Higgs doublets. In this paper, we assume that the states which generate the dimension-five operator, e.g.~right-handed neutrinos, are heavier than the energy scale of interest.
This operator converts the Higgs number asymmetry and/or the lepton chiral asymmetry into $B-L$ asymmetry at a rate 
\begin{align}
\Gamma_{L} \simeq
\frac{1}{4\pi^3}\frac{\bar{m}^2}{v^{4}_{\rm EW}} T^3,
\end{align}
where $v_{\rm EW} = 174 \GeV$ and $\bar{m}^2 = \Sigma_{i}m^{2}_{i}$ is the sum of active neutrino masses squared.
The bound from cosmology, $\sum m_i < 0.3$ eV (TT,TE,EE+lowE)~\cite{Aghanim:2018eyx}, and neutrino oscillation data require
\begin{align}
0.0024~{\rm eV}^2 < \bar{m}^2 < 0.03~{\rm eV}^2.
\end{align}
The interaction is in thermal equilibrium, $\Gamma_L > H$, if 
\begin{align}
\label{eq:TL_RD}
T >  T_{L} \equiv \left( \frac{\pi^2 g_*}{90} \right)^{ \scalebox{1.01}{$\frac{1}{2}$} } \frac{4\pi^3 v_{\rm EW}^4}{\bar{m}^2 M_{\rm Pl} } \simeq 5\times 10^{12}\GeV \left(\frac{g_*}{g_{\rm SM}}\right)^{ \scalebox{1.01}{$\frac{1}{2}$} } \left(\frac{0.03~{\rm eV}^{2}}{\bar{m}^2}\right),
\end{align}
where radiation domination is assumed.

For $T > T_L$ (or more generically $\Gamma_L > H$), since the $B-L$ symmetry breaking by the Majorana mass term is in thermal equilibrium, any $B-L$ asymmetry produced at $T > T_L$ is continually re-equilibrated as the temperature falls. Hence, the final $B-L$ asymmetry is dominantly produced at $T\leq T_L$. At any temperature $T < T_L$, the $B-L$ asymmetry is produced at a rate
\begin{align}
\label{eq:B-L_dot}
    \dot{n}_{B-L} = \Gamma_L \left(n_\ell - \frac{n_H}{2} \right),
\end{align}
where $n_\ell$ and $n_H$ are the lepton doublet and Higgs asymmetries, respectively. If all interactions converting the PQ charge asymmetry into $n_\ell$ or $n_H$ are in thermal equilibrium, one finds that $n_\ell - n_H/2 \sim \dot{\theta}T^2$.
Generically at a given temperature, some of the interactions are out of thermal equilibrium, suppressing the production of $B-L$ asymmetry,
\begin{align}
\label{eq:n_B-L_dot}
\dot{n}_{B-L} = \Gamma_L \times c_{B-L} \, \dot{\theta}T ^2 \times \prod_i \min \left( 1, \frac{\Gamma_i}{H} \right) ,
\end{align}
where $i$ runs over the interactions necessary for the production of $B-L$, e.g.~strong and electroweak sphaleron processes and Yukawa interactions. Here $c_{B-L}$ is an $\mathcal{O}(0.01-0.1)$ coefficient whose value depends on the set of interactions that are in thermal equilibrium.
For the cosmological era with $\Gamma_L \leq H$, the $B-L$ asymmetry produced per Hubble time is
\begin{align}
\label{eq:nB-L_generic}
\Delta n_{B-L} = \frac{\Gamma_L}{H} \times c_{B-L} \, \dot{\theta}T ^2 \times \prod_i \min \left( 1, \frac{\Gamma_i}{H} \right).
\end{align}
The final baryon asymmetry is then given by
\begin{equation}
\label{eq:c_B}
Y_B = C \times Y_{B-L} .
\end{equation}
Throughout this work, we assume only Standard Model particles are present during the electroweak phase transition so $C = 28/79$ and we define $c_B \equiv C \times c_{B-L}$ to be consistent with Eq.~(\ref{eq:nB+L}).

In lepto-axiogenesis, the production of $n_{B-L}$ occurs at high temperatures, where the UV completion of the QCD axion becomes crucial in understanding the efficiency of the production. There exist four possible scenarios for lepto-axiogenesis based on the KSVZ~\cite{Kim:1979if,Shifman:1979if} and DFSZ~\cite{Dine:1981rt,Zhitnitsky:1980tq} UV completions of the QCD axion. The KSVZ quarks can be either heavy or light compared to the temperature and the possible transfers of the asymmetries are described in \ref{model:KSVZ_heavy} and \ref{model:KSVZ_light}, while the DFSZ two Higgs doubles can be heavy or light as well, whose possible asymmetry routes are described in \ref{model:DFSZ_light} and \ref{model:DFSZ_heavy}. We describe \ref{model:KSVZ_heavy} below in detail and others in Appendix~\ref{app:models}.

\begin{enumerate}[label=\textbf{KSVZ-heavy},ref=\underline{KSVZ-heavy}, wide, labelindent=0pt]
\item -- KSVZ model with heavy quarks 

\label{model:KSVZ_heavy}
We consider a KSVZ model and assume that the KSVZ fermions have a mass much larger than the temperature of the universe, so that among the particles in the thermal bath, the axion couples only to gauge bosons. The flow of the asymmetry production via various Standard Model processes and the Majorana neutrino mass is shown in Fig.~\ref{fig:asym_leptoaxio}.
\begin{figure}[!ht]
	\includegraphics[width=\linewidth]{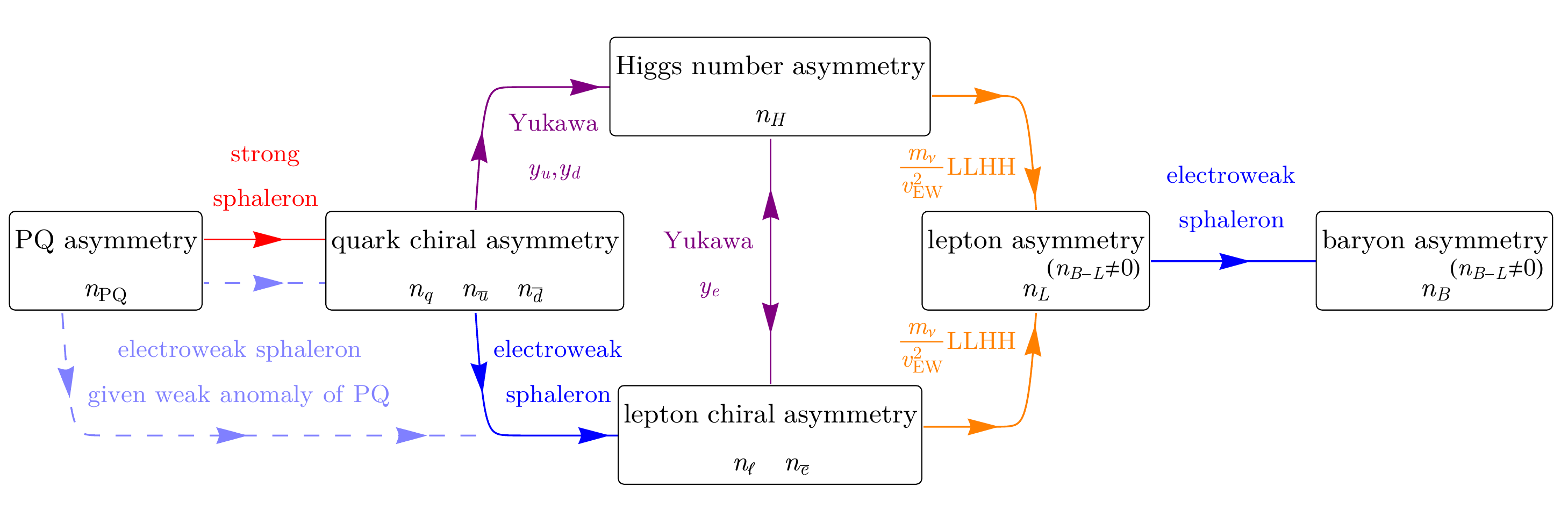}
	\caption{Transfer of asymmetries for lepto-axiogenesis for the scenario \ref{model:KSVZ_heavy}.}
	\label{fig:asym_leptoaxio}	
\end{figure}
The interaction rates for the top Yukawa interaction $\sim \alpha_3 y_t^2 T$ and the strong sphaleron transition are largest among Standard Model interactions. From Fig.~\ref{fig:asym_leptoaxio}, one can see that the suppression occurs if either or both are not in thermal equilibrium.
 
The $B-L$ asymmetry produced per Hubble time is
\begin{align}
\label{eq:nB-L}
\Delta n_{B-L} = \frac{\Gamma_L}{H} \times c_{B-L} \, \dot{\theta}T ^2 \times \min \left( 1, \frac{\Gamma_{\rm ss}}{H} \right) \times \min \left(1, \frac{\alpha_3 y_{t}^2 T}{H}\right).
\end{align}
In the cosmological evolution we consider in this paper, we find that the suppression by the top Yukawa does not enter for the final $B-L$ asymmetry.
\begin{table}[!ht]
\begin{center}
\begin{tabular}{|c|c||c|c|c|c|c|c|c|c|}
 \hline
  \multicolumn{6}{|c|}{Equilibrium} &  \multicolumn{3}{c|}{\ Conservation \ } &  \multirow{2}{*}{$c_{B-L}$}     \\   \cline{1-9} 
\ $\Gamma_{\rm ss}$ \ & \ $\alpha_3 y_t^2 T$ \ & \ $\Gamma_L$ \ & \ $\Gamma_{\rm ws}$ \ & \  $\alpha_3 y_\tau^2 T$ \ & \ $\alpha_2 y_b^2 T$ \ & \  $n_{B-L}$ \  & \  $n_{B+L}$ \  & \ $n_{\bar{\tau}}$ \  &   \\
\hline
\checkmark & \checkmark & \checkmark & \checkmark & \checkmark & \checkmark & \ding{55} &  \ding{55} & \ding{55} & $\frac{7 - 4c_W}{26}$  \\ \hline 
\checkmark & \checkmark & \ding{55}  & \checkmark & \checkmark & \checkmark & \checkmark & \ding{55} & \ding{55} & \ $\frac{7 - 4 c_W}{35}$ \ \\ \hline
\checkmark & \checkmark & \ding{55} & \ding{55} & \checkmark & \checkmark & \checkmark & \checkmark & \ding{55} & $\frac{1}{8}$  \\ \hline 
\checkmark & \checkmark & \ding{55} & \ding{55} & \ding{55} & \checkmark & \checkmark & \checkmark & \checkmark & $\frac{1}{10}$  \\ \hline 
\checkmark & \checkmark & \ding{55} & \ding{55} & \ding{55} & \ding{55} & \checkmark & \checkmark & \checkmark & $\frac{1}{10}$  \\ \hline
\end{tabular}
\end{center}
\caption{Values of $c_{B-L}$ for different sets of interactions in equilibrium using simplified Boltzmann equations presented in Appendix~\ref{app:Boltz_eq}.}
\label{tab:cB_L}
\end{table}
In Table~\ref{tab:cB_L}, we show the values of $c_{B-L}$ whether the electroweak sphaleron, the bottom Yukawa, and the tau Yukawa are in thermal equilibrium or not\footnote{The values of $c_{B-L}$ given in Table~\ref{tab:cB_L} are presented in the approximation $y_t \gg y_b$ for the cases where the bottom Yukawa interaction is in equilibrium. A source of an $\mathcal{O}(10)$ error arises in this approximation when $\alpha_3$ is small as in the Standard Model at high energies.}. We include the possibility that the PQ symmetry has a weak anomaly with $c_W$ the anomaly coefficient normalized to that of QCD. Here we only consider the third generation fermions for simplicity. We expect $c_{B-L}$ to be of the same order for the actual cases with three generations.
\end{enumerate}

The resultant baryon asymmetry in Eq.~(\ref{eq:c_B}) depends on the cosmological evolution of $\dot{\theta}$, $f_{\rm eff}$ $T$ and $H$ as can be seen from Eqs.~(\ref{eq:nB-L}), (\ref{eq:nB-L_light}), (\ref{eq:nB-L_light2}), and (\ref{eq:nB-L_DFSZ}).
In the following sections, we investigate concrete scenarios and show that the observed baryon asymmetry $Y_B^{\rm obs} = 8.7 \times 10^{-11}$~\cite{Aghanim:2018eyx} can be explained by lepto-axiogenesis.
We essentially focus on a KSVZ model with heavy quarks as described in~\ref{model:KSVZ_heavy} and comment on the connection to other cases in passing and in Appendix~\ref{app:models}.

In the concrete scenarios we consider, the rotation of $P$ is generically not completely circular, and during a cycle $\dot{\theta}$ may change rapidly. In Appendix~\ref{app:ave_thetadot}, we show that in most cases the baryon asymmetry produced per Hubble time is simply given by replacing $\dot{\theta}$ with a cycle averaged one, $\vev{\dot{\theta}}$.

\section{Models with a quartic potential}
\label{sec:quartic}

In this and the next section, we consider a scenario where the rotation of the PQ symmetry breaking field $P$ is initiated by a higher-dimensional operator of $P$ in a similar manner to the Affleck-Dine mechanism~\cite{Affleck:1984fy,Dine:1995uk,Dine:1995kz}.
We assume that the radial direction $S$ of the field $P$ defined in Eq.(\ref{eq:P}), which we call the ``saxion" following supersymmetric terminology, has a flat potential. Then in the early universe, the saxion may have a large field value. We consider the scenario where the large field value is developed during inflation and is therefore homogenized by inflation. For a sufficiently large field value, a higher-dimensional potential of $P$ which explicitly breaks the PQ symmetry,
\begin{align}
\label{eq:explicit_PQV}
V_{\cancel{\rm PQ}}(P) = A \frac{P^n}{M^{n-3}} + {\rm h.c.},
\end{align}
may not be negligible. Here $M$ is a cut-off scale and $A$ is a dimensionful parameter. (The parametrization by $A$ is motivated by a supersymmetric theory discussed later.) The explicit breaking gives a potential to the angular direction $\theta$ and drives a homogeneous angular motion. As can be seen from Appendix~\ref{app:epsilon}, a substantial rotation results even when the saxion initial field value is smaller than $M$, maintaining the validity of the effective theory.

The rotation also produces axion dark matter via parametric resonance or kinetic misalignment. We estimate the abundance and show the constraint on the parameter space.

\subsection{Rotation in a quartic potential}

In this section, we study the simplest potential for spontaneous PQ symmetry breaking,
\begin{align}
\label{eq:VQ_Q}
V_{\rm PQ}(P) =\lambda^2 \left( |P|^2 - \frac{f_a^2 N^{2}_{\rm DW}}{2} \right)^2,
\end{align}
which leads to a saxion vacuum mass $m_S =\sqrt{2} \lambda N_{\rm DW} f_a$.
When $|P| \gg f_a$, the potential is dominated by the quartic term. The saxion mass around the initial large field value $S_i$ is
\begin{align}
m_S (S_i) = 
\lambda S_i \, \simeq \, 
10^6 \GeV \left( \frac{\lambda }{10^{-10}} \right) \left( \frac{S_i}{10^{16}~{\rm GeV}} \right) .
\end{align}
The saxion begins to oscillate when $m_S(S_i) \simeq 3H$.
For a radiation-dominated universe, this occurs at a temperature 
\begin{equation}
\label{eq:Tosc_Si}
T_{\rm osc} \, \simeq \, 
5  \times 10^{11}~{\rm GeV} \left(\frac{g_{\rm SM}}{g_*} \right)^{ \scalebox{1.01}{$\frac{1}{4}$} } \left( \frac{\lambda }{10^{-10}} \right)^{ \scalebox{1.01}{$\frac{1}{2}$} } \left( \frac{S_i}{10^{16}~{\rm GeV}} \right)^{ \scalebox{1.01}{$\frac{1}{2}$} }.
\end{equation}

As the saxion begins to oscillate, the explicit PQ symmetry breaking potential~(\ref{eq:explicit_PQV}) kicks $P$ in the angular direction, inducing a non-zero angular velocity. During the rotation, $S$ oscillates.
We define $\bar{S}$ as
\begin{align}
 \bar{S}^2 \equiv \vev{S^2},
\end{align}
which is roughly the amplitude of the oscillation of $S$.
After the beginning of the rotation, the energy density and $\bar{S}$ scale as $R^{-4}$ and $R^{-1}$, respectively. As $\bar{S}$ decreases, the explicit PQ symmetry breaking soon becomes ineffective and the PQ asymmetry is conserved.
We parameterize the resultant PQ charge asymmetry $n_{\rm PQ}$ defined in Eq.~(\ref{eq:PQ charge}) using a parameter $\epsilon \leq 1$ defined by
\begin{align}
\label{eq:epsilon_def}
     n_{\rm PQ}  &\equiv  \dfrac{\epsilon }{N_{\rm DW}}\, \omega S_{\rm max}^2, &  n_S   &\equiv \frac{1}{\omega}\left(\frac{d|P|}{d t}\right)_{\max }^{2},  & \omega^2  &\equiv  \left. \frac{V'(S)}{S} \right|_{\rm max}, \\
\label{eq:epsilon_q}
    n_{\rm PQ}  &\simeq  \dfrac{4\epsilon }{N_{\rm DW}} n_S,    & n_S  &\simeq \frac{1}{4} \lambda S_{\rm max}^3,  & \omega   &\simeq  \lambda S_{\rm max},   &  \text{for $\epsilon \ll 1$},
\end{align}
where max denotes the maximum value during a cycle. Here $n_S$ and $\omega$ can be approximately understood as the number density of $S$ and the frequency of the motion. A detailed estimation of $\epsilon$ is given in Appendix~\ref{app:epsilon}.

The rotation is generically not circular, and $\dot{\theta}$ oscillates in time. Its time-average is
\begin{align}
 \vev{\dot{\theta}} = \frac{1}{\Delta t} \int_t^{t + \Delta t} {\rm d}t' \dot{\theta}(t')= \frac{\theta(t + \Delta t) - \theta(t)}{\Delta t}.
\end{align}
For coherently rotating $P$ (i.e.~$\epsilon \neq 0$), this is given by the frequency of the rotation,
\begin{align}
\label{eq:dtheta_ave}
\vev{\dot{\theta}} \, \simeq \, N_{\rm DW} \,  m_S(\bar{S}),
\end{align}
whose order of magnitude value does not depend on $\epsilon$ and scales as $R^{-1}$. For $\epsilon \ll 1$, the frequency of the rotation approaches that of the oscillation with $\epsilon =0$. As a result, for $\epsilon \ll 1$, the precise value of $\vev{\dot{\theta}}$ becomes almost independent of $\epsilon$. The independence from $\epsilon$ can suppress baryon isocurvature perturbations arising from lepto-axiogenesis. The dependence of $\vev{\dot{\theta}}$ on $\epsilon$ is discussed in Appendix~\ref{app:ave_thetadot} in detail.

During the rotation of $P$,  fluctuations of $P$ are produced by parametric resonance unless the rotation is very close to a circular one, $\epsilon > 0.8$~\cite{Co:2020dya}. By a numerical computation, we find that $\epsilon>0.8$ cannot be achieved by the rotation from higher dimensional operator unless $n=5$. As we show in Sec.~\ref{sec:exPQ_quartic}, $n=5$ gives too large an axion mass at the vacuum and reintroduce the strong CP problem. Once the amplitude of the fluctuations becomes comparable to the amplitude of the rotation, parametric resonance is terminated by the back-reaction, and the field value of $P$ is randomized. For the quartic potential, this occurs around $\bar{S} \sim 10^{-2}/10^{-4}~S_i$ during a radiation/matter dominated era~\cite{Kasuya:1998td,Kasuya:1999hy,Kawasaki:2013iha}. The average value of $\dot{\theta}$ now depends on $\epsilon$. From the PQ charge conservation,
\begin{align}
    \epsilon n_S \simeq N_{\rm DW}n_{\rm PQ} = \frac{1}{N_{\rm DW}}\vev{\dot{\theta} S^2} \simeq \frac{1}{N_{\rm DW}} \vev{\dot{\theta}}\bar{S}^2.
\end{align}
Here the averages of $\dot{\theta}$ and $S^2$ are separated, since $P$ is randomized and $\theta$ and $S$ move independently. Using $n_S \simeq m_S(\bar{S}) \bar{S}^2$, we obtain
\begin{align}
\label{eq:dtheta_ave_PR}
    \vev{\dot{\theta}} \simeq \epsilon N_{\rm DW} m_S(\bar{S}).
\end{align}
Parametric resonance reduces the source term driving asymmetries by order $\epsilon$. 

\subsection{Baryon asymmetry}

We investigate the parameter space where lepto-axiogenesis explains the observed baryon asymmetry. We consider scenarios where the onset of field rotation occurs after or before the end of reheating in the following subsections. In spite of the assumption of \ref{model:KSVZ_heavy}, the results still apply identically to some models in \ref{model:KSVZ_light} and \ref{model:DFSZ_light} as long as the oscillation occurs when the non-thermal mass assumed in Eq.~(\ref{eq:Tosc_Si}) dominates over the thermal masses from the light PQ-charged particles and the production of $n_{B-L}$ matches the form given in Eq.~(\ref{eq:nB-L}). For example, the additional interactions relevant for $\Delta n_{B-L}$ in Eqs.~(\ref{eq:nB-L_light}), (\ref{eq:nB-L_light2}), and (\ref{eq:nB-L_DFSZ}) should be in thermal equilibrium in order to match Eq.~(\ref{eq:nB-L}). These additional conditions can be straightforwardly examined but we will not investigate the applicability further to avoid obscuring the discussion.

\subsubsection{Rotation during a radiation dominated era}

In a radiation-dominated universe for $(T < T_{\rm ss}, \ T_{\rm ss} < T < T_L, \ T_L < T )$, $Y_B \equiv n_B/s \propto (T, T^0, T^{-1})$ respectively based on the scaling in Eq.~(\ref{eq:nB-L}). For $T_{\rm osc} < T_{\rm ss}$, i.e.~$m_S(S_i) \lesssim 3 \times 10^7 \GeV \, ( 35 \alpha_3)^{10} (g_{\rm SM}/g_*)^{1/2}$, the production of the baryon asymmetry from the PQ asymmetry is UV-dominated and peaks at the beginning of the oscillation. For $ T_{\rm osc} > T_{\rm ss}$, the baryon asymmetry is dominantly produced at $T = \min (T_L,T_{\rm osc})$.
For $m_S(S_i) \gtrsim 10^8 \GeV (0.03~{\rm eV}^2 / \bar{m}^2)^2 (g_*/g_{\rm SM})^{3/2}$, we have $T_{\rm osc} > T_L$. Assuming the saxion is thermalized before dominating the energy density, the produced baryon asymmetry is
\begin{align}
\label{eq:YB_quartic1}
Y_{B} & \simeq \left. \frac{n_B}{s} \right|_{T = \min(T_L, T_{\rm osc})} \simeq \left. \frac{c_B \, \dot\theta T^2}{s}  \right|_{T = T_{\rm osc}} \times \left. \min \left( 1, \frac{\Gamma_{\rm ss}}{H} \right) \times \frac{\Gamma_L}{H} \right|_{T = \min(T_L, T_{\rm osc})} \\
& \simeq 10^{-10} \, N_{\rm DW} \left( \frac{c_B}{0.1} \right)  \left( \frac{\bar{m}^2}{0.03~{\rm eV}^2} \right)  \times 
\begin{cases}
0.8  \left(\frac{g_{\rm SM}}{g_*} \right)^{ \scalebox{0.9}{$\frac{3}{2}$} } \left( \frac{m_S (S_i) }{200 \TeV} \right)  & {\rm for} \ \ T_{\rm osc} < T_{\rm ss} \\
200 \left(\frac{g_{\rm SM}}{g_*} \right)^{ \scalebox{0.9}{$\frac{7}{4}$} } \left( \frac{m_S (S_i) }{10^8 \GeV} \right)^{\scalebox{0.9}{$\frac{1}{2}$}} \left(\frac{\alpha_3}{1/35} \right)^5 &  {\rm for} \ \ T_{\rm osc} > T_{\rm ss} \nonumber
\end{cases},
\end{align}
where we use $\dot\theta(T_{\rm osc}) = N_{\rm DW} m_S(S_i)$ from Eq.~(\ref{eq:dtheta_ave}). The observed value $Y_B^{\rm obs}$ is explained with $m_S(S_i) \simeq 200 \TeV~(0.1/c_B) (0.03~{\rm eV}/\bar{m}^2)$, while larger values of $m_S(S_i)$ require dilution especially if $\alpha_3$ is larger than SM value in the case $T_{\rm osc} > T_{\rm ss}$.
If $T_{\rm osc} \gtrsim 100~T_L$, parametric resonance becomes effective at $T > T_L$, and the $B-L$ asymmetry produced at $T\simeq T_L$ is suppressed by $\epsilon$ as explained below Eq.~(\ref{eq:dtheta_ave}).

\begin{figure}[!t]
	\includegraphics[width=0.495\linewidth]{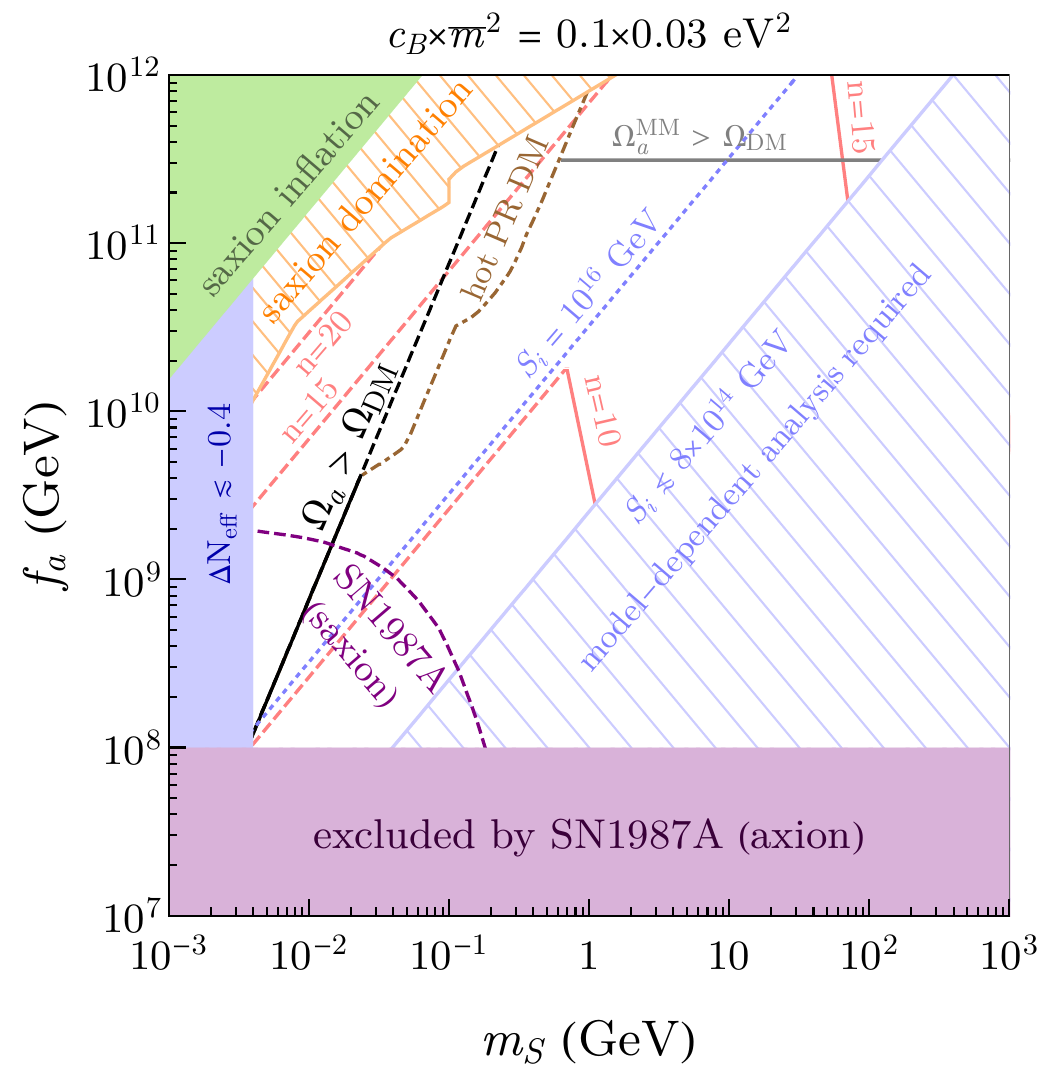}
	\includegraphics[width=0.495\linewidth]{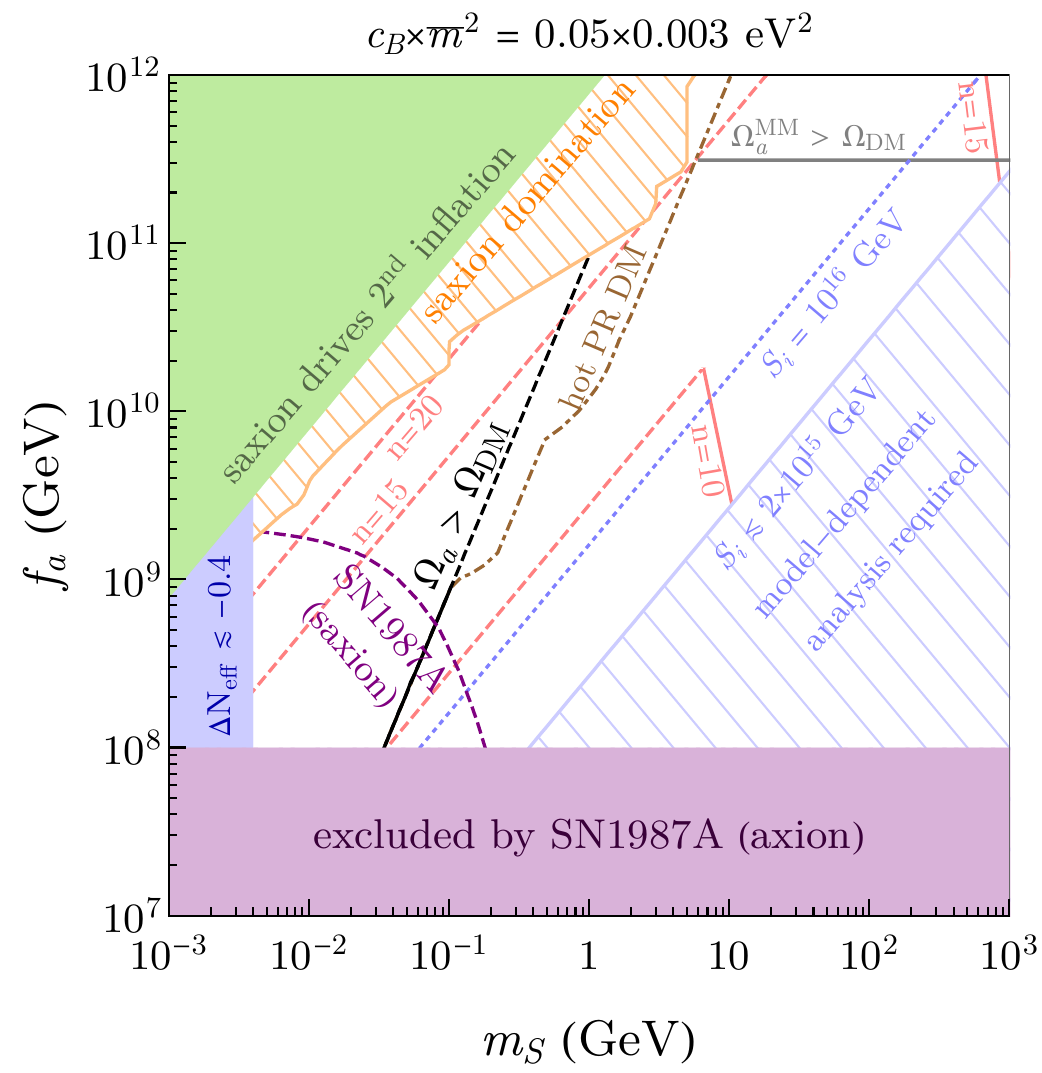}
	\includegraphics[width=0.495\linewidth]{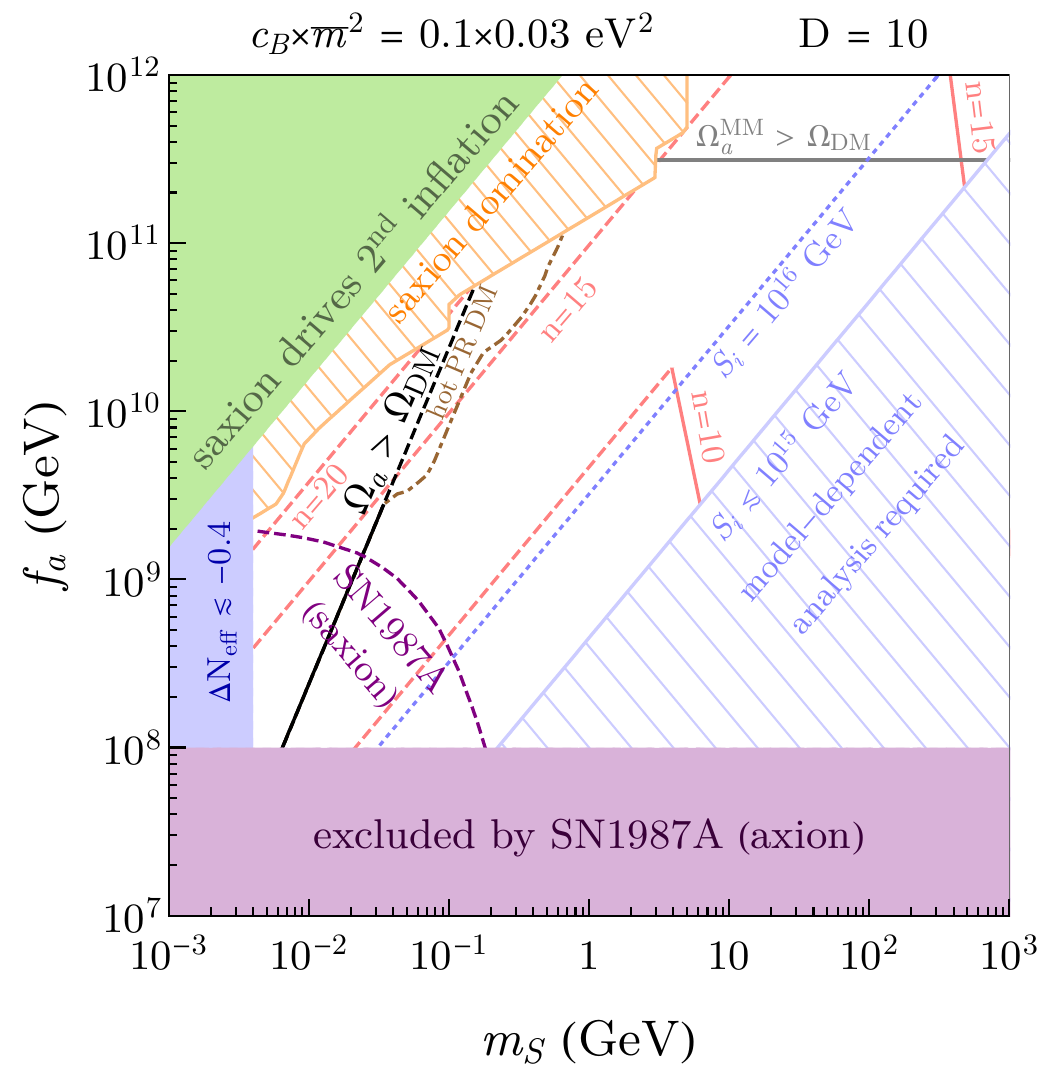}
	\includegraphics[width=0.495\linewidth]{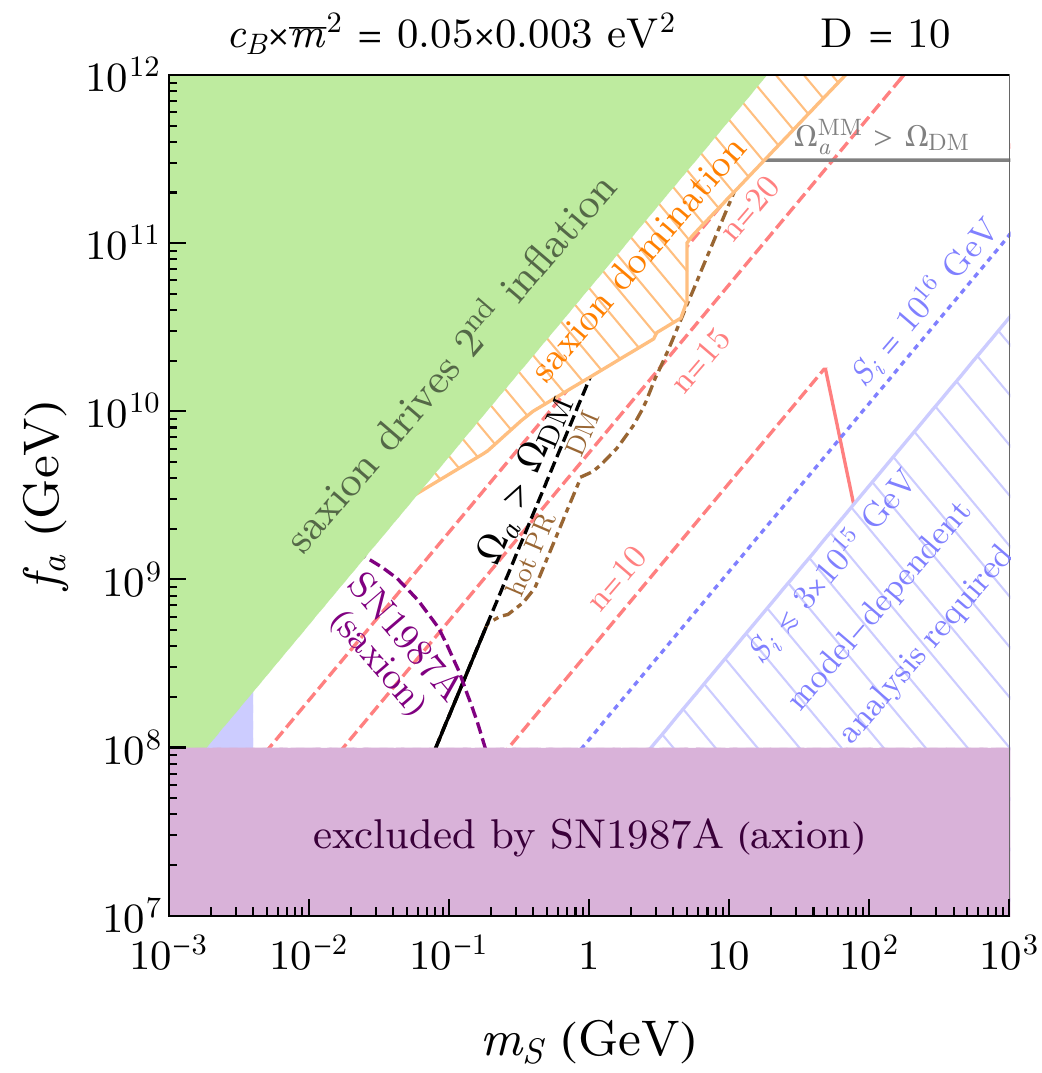}
	\caption{Parameter space compatible with the observed baryon asymmetry for the quartic potential for $N_{\rm DW} = 1$. The upper panels assume a radiation-dominated universe throughout the evolution, while the lower panels include a dilution factor of $D = 10$ that can arise from late-time entropy production. Left (right) panels are for larger (smaller) $c_B \times \bar{m}^2$ as labeled. We fix
	$\epsilon = 0.125$, which affects the black and pink solid lines.}
	\label{fig:quartic}	
\end{figure}

In Fig.~\ref{fig:quartic}, the observed baryon asymmetry can be generated throughout the unshaded region of the $(m_S,f_a)$ plane. In the green-shaded region, $S_i$ exceeds the Planck scale, and a secondary inflation by the saxion occurs; the baryon asymmetry is diluted and lepto-axiogenesis cannot explain the observed baryon asymmetry. Contours of the required initial saxion field $S_i$ are shown by blue dotted lines, with values varying from the Planck scale to $10^{14}$ GeV and below. 

Due to assumptions about the heavy KSVZ quarks (see \ref{model:KSVZ_heavy}), the thermal logarithmic potential~\cite{Anisimov:2000wx} becomes important and the oscillation occurs earlier than in Eq.~(\ref{eq:Tosc_Si}) when $S_i \lesssim 3 \times 10^{16} \GeV (35 \alpha_3) (g_* / g_{\rm SM})^{1/4}$. We continue the computation to lower values of $S_i$ because the results may still be applicable to other cases in \ref{model:KSVZ_light} and \ref{model:DFSZ_light} with conditions mentioned earlier. However, if $S_i$ becomes too small the estimation of the baryon asymmetry necessarily changes. For example, let us consider a coupling with a light particle $y P Q\bar{Q}$ as in \ref{model:KSVZ_light}. For the oscillation to begin in the zero-temperature potential rather than in the thermal potential, requires
\begin{align}
    y < 2\times 10^{-6} \left(\frac{g_*}{g_{\rm SM}}\right)^{\scalebox{0.9}{$\frac{1}{4}$}} 
    \left( \frac{m_S(S_i)}{10^3~{\rm TeV}}\right)^{\scalebox{0.9}{$\frac{1}{2}$}}.
\end{align}
On the other hand, the charge of rotating $P$ must be efficiently transferred into the chiral asymmetry of $Q\bar{Q}$ in order for the baryon asymmetry to reproduce the case of \ref{model:KSVZ_heavy} with $T_{\rm osc} < T_{\rm ss}$. This requires $\Gamma \simeq \alpha_3 m_Q(S_i)^2 / T_{\rm osc} > H(T_{\rm osc})$ as discussed in \ref{model:KSVZ_light}, which translates to
\begin{align}
    y > 2 \times 10^{-6}
    \left(\frac{1/35}{\alpha_3}\right)^{\scalebox{0.9}{$\frac{1}{2}$}}
    \left(\frac{g_{\rm SM}}{g_*}\right)^{\scalebox{0.9}{$\frac{1}{8}$}}
    \left( \frac{m_S(S_i)}{10^3~{\rm TeV}}\right)^{\scalebox{0.9}{$\frac{3}{4}$}} 
    \left( \frac{10^{15}~{\rm GeV}}{S_i} \right).
\end{align}
These two conditions on the coupling can be compatible with each other if
\begin{align}
    S_i > 10^{15}~{\rm GeV} 
    \left(\frac{1/35}{\alpha_3}\right)^{\scalebox{0.9}{$\frac{1}{2}$}}
    \left(\frac{g_{\rm SM}}{g_*}\right)^{\scalebox{0.9}{$\frac{3}{8}$}}
    \left( \frac{m_S(S_i)}{10^3~{\rm TeV}}\right)^{\scalebox{0.9}{$\frac{1}{4}$}}.
\end{align}
A similar analysis for \ref{model:DFSZ_light} will reveal the same condition but with $\alpha_3$ replaced by $\alpha_2$. The violation of this condition is shown as the blue hatched region in Fig.~\ref{fig:quartic}, inside which the evaluation of $Y_B$ is highly model-dependent and beyond the scope of this work.

In the purple shaded regions, extra cooling of supernovae cores by the emission of axions suppresses the neutrino emission, in contradiction with the observed neutrino spectrum from SN1987A~\cite{Ellis:1987pk,Raffelt:1987yt,Turner:1987by,Mayle:1987as,Raffelt:2006cw,Chang:2018rso,Carenza:2019pxu}. The region below the purple dashed line is similarly excluded by saxion emission due to the coupling with gluons~\cite{Ishizuka:1989ts}. This can be avoided, however, by introducing a large coupling between the saxion and the Higgs so that saxions are trapped inside supernova cores. Such large couplings can be probed by observations of rare kaon decays at KLEVER~\cite{Ambrosino:2019qvz, Beacham:2019nyx}.

The rotation of $P$ includes an angular motion and a radial motion. The energy density of the radial motion must be dissipated into radiation. In the blue shaded region, the dissipation causes problems. If the coupling with the SM particles is small, the saxion decays into axions and creates too much dark radiation compared to the experimental bound~\cite{Aghanim:2018eyx}. For larger couplings to the SM particles, the saxion is in thermal equilibrium with the SM bath, and a significant portion of its energy is dumped into electrons and photons after neutrinos decouples, leading to an excessively negative amount of dark radiation, which spoils Big Bang Nucleosynthesis (BBN)~\cite{Fields:2019pfx}.

Axion dark matter, discussed in Secs.~\ref{sec:PR_quartic}--\ref{sec:kmm_quartic}, can be obtained from parametric resonance or kinetic misalignment on the solid black lines of Fig.~\ref{fig:quartic} but, in the case of parametric resonance, is constrained by warmness of DM as indicated by the dashed brown lines.   

We require the explicit PQ symmetry breaking potential of (\ref{eq:explicit_PQV}) to preserve the PQ solution to the strong CP problem, leading to the solid pink contours labeled by $n$. The dashed pink lines labelled by $n$ follow from vacuum requirements.  For any given $n$, these constraints are satisfied by the area below the wedge formed by the corresponding solid and dashed lines, as detailed in Sec.~\ref{sec:exPQ_quartic}.
In the orange hatched regions of Fig.~\ref{fig:quartic}, simple possibilities to thermalize the saxion oscillation fail, leading to saxion domination, as discussed in Sec.~\ref{sec:quartic_therm}.

In the lower panels of Fig.~\ref{fig:quartic}, we show the parameter space when some entropy production after the $B-L$ asymmetry production leads to dilution by a factor $D$ of $Y_B$ and axion dark matter from $Y_{\rm PQ}$. This dilution can result from a generic moduli field or thermalized saxions. The required initial field value $S_i$ becomes larger correspondingly. The lower right panel belongs to the second case in Eq.~(\ref{eq:YB_quartic1}), i.e.~$T_{\rm osc} > T_{\rm ss}$, with $\alpha_3 = 1/35$, while $T_{\rm osc} < T_{\rm ss}$ for the other three panels.

\subsubsection{Rotation before the completion of reheating}

If the reheat temperature after inflation $T_R$ is small enough, $P$ starts to rotate before reheating completes, $T_{\rm osc} \ge T_R$, when the universe is matter dominated. During the matter domination, we have the scaling laws for temperature $T \propto R^{-3/8}$ and for the Hubble scale $H \propto T^4$. Here we only consider the most extreme case $T_R < T_{\rm ss}$ to explore the lowest possible $T_R$; investigation of all possible cases is beyond the scope of this paper. For $\bar{S} > f_a$, using Eq.~(\ref{eq:nB-L}), one can show that $n_B / \rho_{\rm inf} \propto (T^{-13/3}, T^{-22/3})$ for $(T_R < T < T_{\rm ss}, \ T_{\rm ss} < T < T_L)$ respectively.
Once $\bar{S} \simeq \sqrt{2} f_a$ at the temperature $T_S$, the potential is dominated by the quadratic term and the scaling changes to $n_B / \rho_{\rm inf} \propto (T, T^{-2})$, after which the production of the $B-L$ asymmetry is negligible ($T_S < T_{\rm ss}$ in the parameter of interest). 
Therefore, the baryon asymmetry is produced dominantly at $T= {\rm max }(T_R,T_S)$,
\begin{align}
\label{eq:YB_quartic_lowTRH}
Y_B & \simeq \left. \frac{n_B}{\rho_{\rm inf}} \right|_{T = {\rm max }(T_R,T_S)} \times \left. \frac{\rho_{\rm inf}}{s} \right|_{T = T_R} \simeq  \left. \frac{c_B \dot{\theta} T^2}{\rho_{\rm inf}} \frac{\Gamma_L}{H} \right|_{T = {\rm max }(T_R,T_S)} \times \frac{3}{4}T_R \nonumber  \\
  & \simeq 8 \times 10^{-11} N_{\rm DW} \left( \frac{c_B}{0.1} \right) \left( \frac{\bar{m}^2}{0.03~{\rm eV}^2} \right) \nonumber \\ 
  & \hspace{0.5cm} \times  \begin{cases}
  \left(\frac{g_{\rm SM}}{g_*} \right)^{ \scalebox{0.9}{$\frac{7}{6}$} } \left( \frac{T_R}{10^{9} \GeV} \right)^{ \scalebox{0.9}{$\frac{4}{3}$} } \left( \frac{m_S(S_i)}{5 \times 10^{14} \GeV} \right)^{ \scalebox{0.9}{$\frac{1}{3}$} } & {\rm for} \ \ S_i > S_0  \\
  \left(\frac{g_{\rm SM}}{g_*} \right)^{ \scalebox{0.9}{$\frac{5}{8}$} } \left( \frac{T_R}{10^{9} \GeV} \right)^{ \scalebox{0.9}{$\frac{7}{2}$} } \left( \frac{5 \times 10^{14} \GeV}{m_S(S_i)} \right)^{ \scalebox{0.9}{$\frac{3}{4}$} } \left( \frac{10^9 \GeV}{f_a} \right)^{ \scalebox{0.9}{$\frac{13}{8}$} } \left( \frac{S_i}{8 \times 10^{17} \GeV} \right)^{ \scalebox{0.9}{$\frac{13}{8}$} } & {\rm for} \ \ S_i < S_0 
  \end{cases} 
  \\ \nonumber 
  S_0 & \equiv 8 \times 10^{17} \GeV \left( \frac{f_a}{10^{9}\GeV} \right) \left(\frac{g_{\rm SM}}{g_*}\right)^{ \scalebox{1.01}{$\frac{1}{3}$} } \left(\frac{m_S(S_i)}{6 \times 10^{13} \GeV} \right)^{ \scalebox{1.01}{$\frac{2}{3}$} } \left( \frac{10^9 \GeV}{T_R} \right)^{ \scalebox{1.01}{$\frac{4}{3}$} } , \nonumber 
\end{align}
where $S_0$ is the critical value of $S_i$ below which $S$ reaches $\sqrt{2} f_a$ after $T_R$.

\begin{figure}[!t]
	\includegraphics[width=0.495\linewidth]{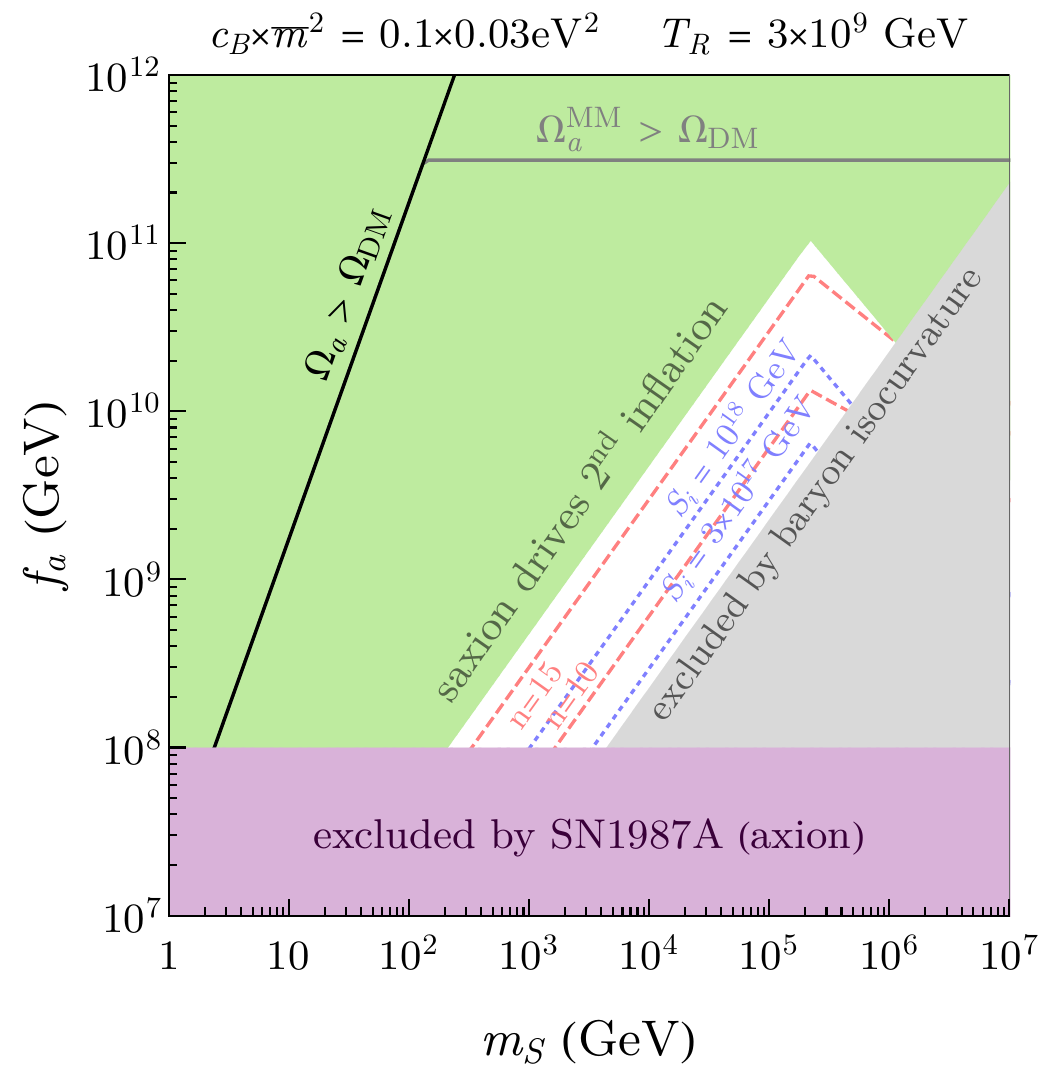}
	\includegraphics[width=0.495\linewidth]{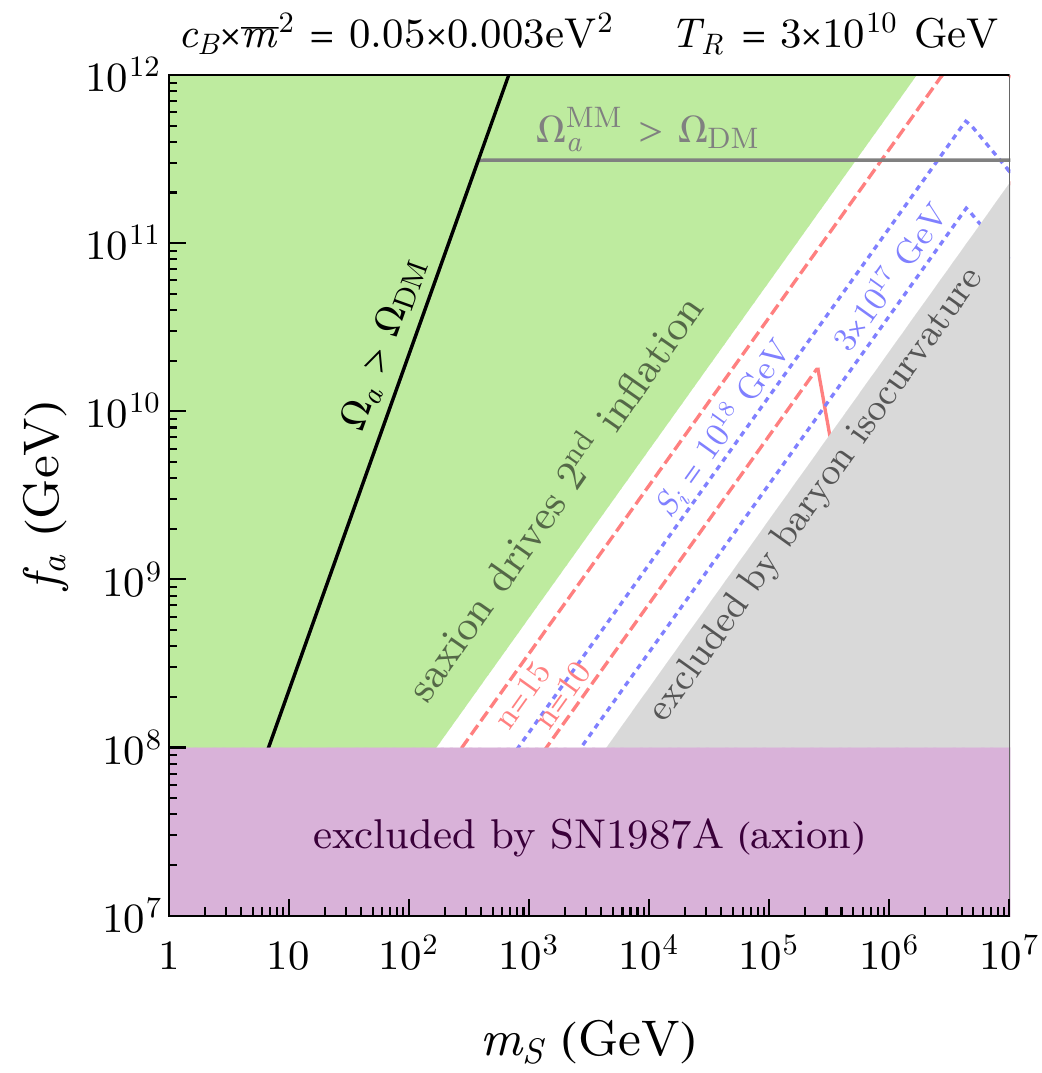}
	\includegraphics[width=0.495\linewidth]{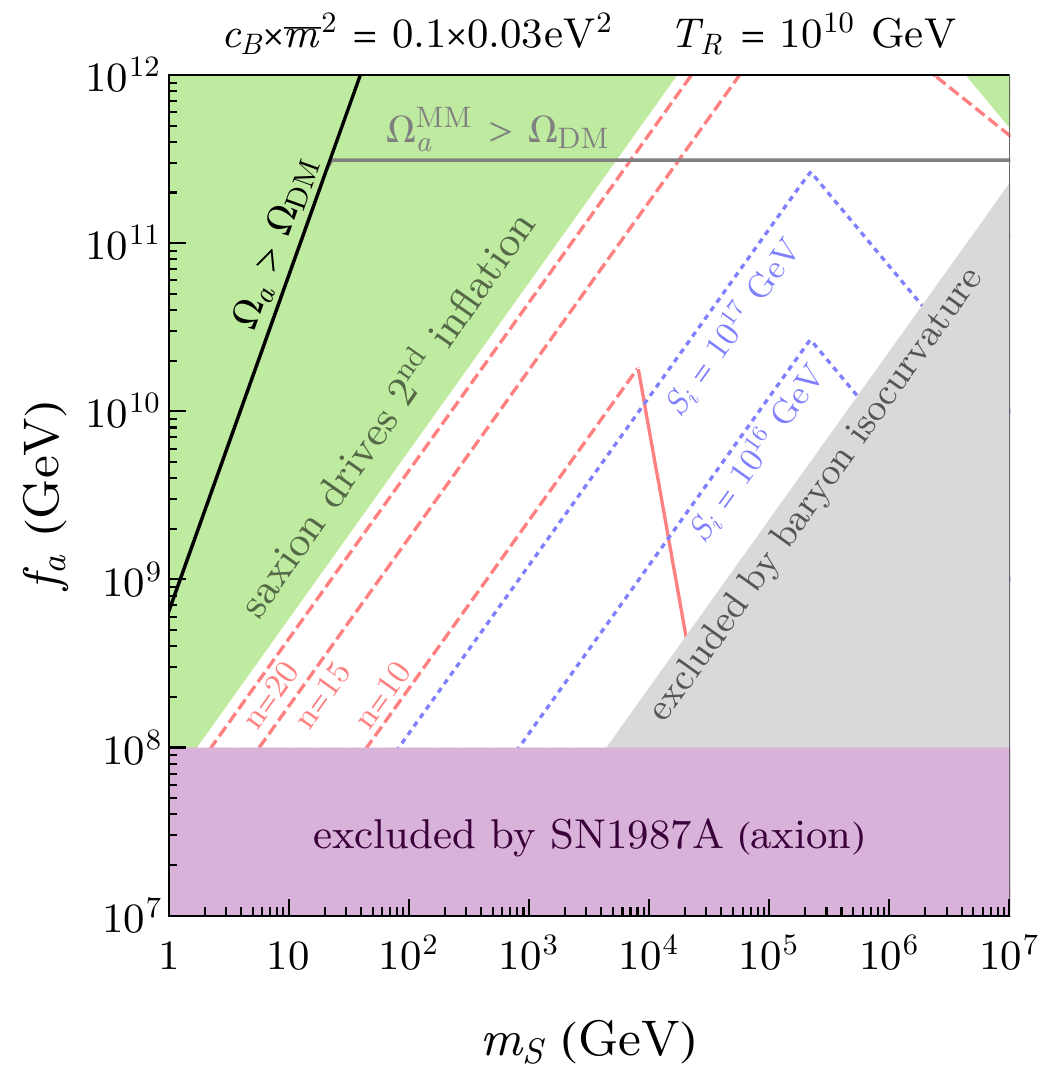}
	\includegraphics[width=0.495\linewidth]{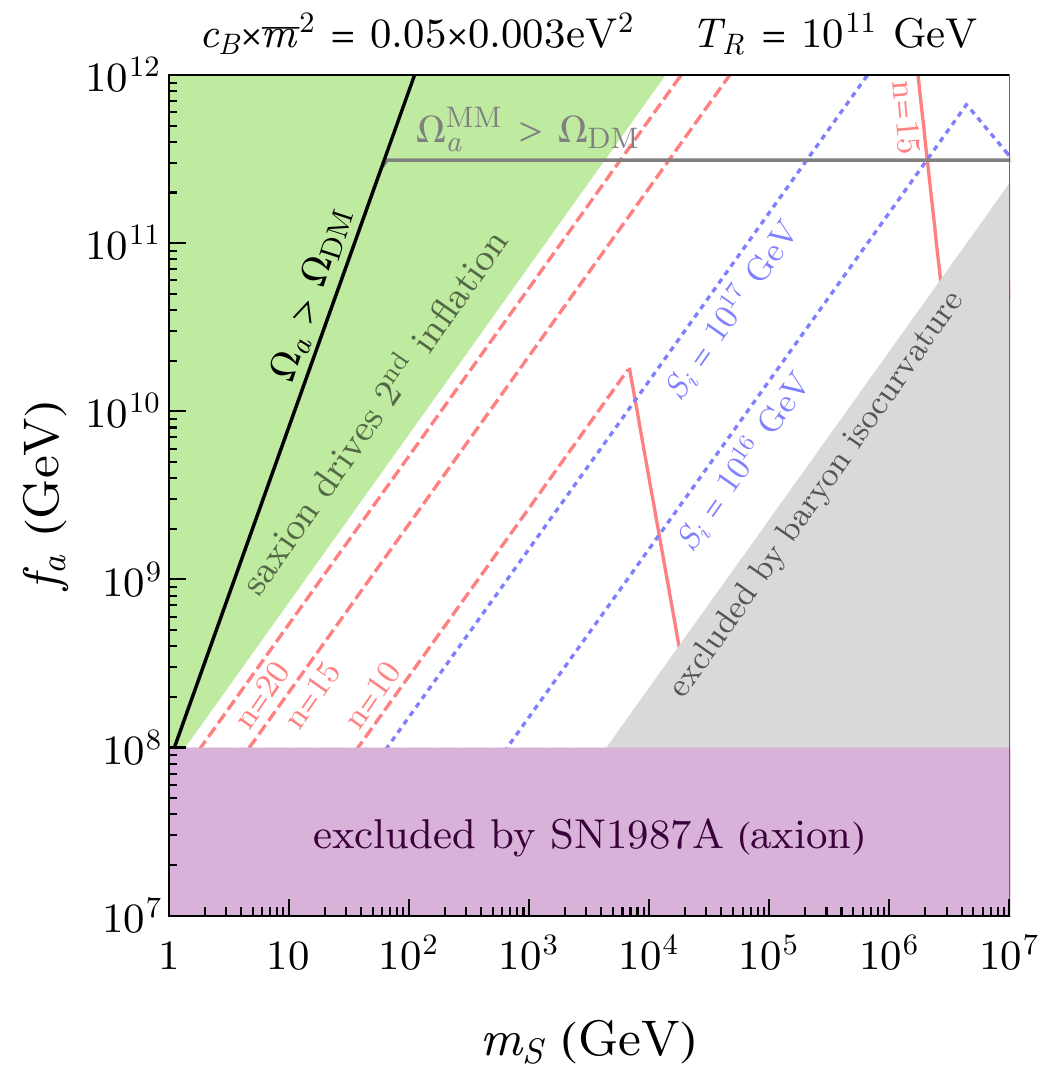}
	\caption{Parameter space compatible with the observed baryon asymmetry for the quartic potential for $N_{\rm DW} = 1$. Here the production of $Y_B$ is dominated during inflationary reheating with a different reheat temperature for each panel. Left (right) panels are for larger (smaller) $c_B \bar{m}^2$ as labeled. We fix $\epsilon = 0.125$, which affects the black and pink solid lines.}
	\label{fig:quartic_lowTRH}	
\end{figure}

In Fig.~\ref{fig:quartic_lowTRH}, we show the constraints on the parameter space for several reheat temperatures. Here $m_S(S_i)$ and $S_i$ are constrained by Eq.~(\ref{eq:YB_quartic_lowTRH}). Due to entropy production from the inflationary reheating, the required values of $m_S$ and $S_i$ are larger than in Fig~\ref{fig:quartic}. The color scheme of various constraints is the same as in Fig.~\ref{fig:quartic}. The additional gray shaded region is excluded because of the large isocurvature perturbation of the baryon asymmetry. Since the $B-L$ asymmetry is dominantly produced at $T={\rm max}(T_{R}, T_S)$ long after the rotation begins, parametric resonance becomes effective and $\vev{\dot{\theta}}$ depends on $\epsilon$. If saxion thermalization occurs before parametric resonance becomes effective, parametric resonance is avoided since the rotation becomes circular, but then $\dot{\theta}$ depends on $\epsilon$ after the thermalization. In both cases, the isocurvature perturbation comes from the fluctuation of $\epsilon$,
\begin{align}
    \frac{\delta Y_B}{Y_B} \simeq n \frac{H_{\rm inf}}{2\pi S_i}.
\end{align}
To derive the constraints conservatively, we assume $H_{\rm inf} = m_S (S_i)$. This is naturally the case if the saxion field during inflation is determined by a balance between the quartic potential and a negative Hubble induced mass $- H^2 |P|^2$. We then obtain the upper bound $\lambda < 2 \times 10^{-5} (10/n)$, which is illustrated by the gray region in Fig.~\ref{fig:quartic_lowTRH} using $n = 10$. The reheat temperature can be as small as a few times $10^9$ GeV, which is comparable to the lower bound on $T_R$ from successful thermal leptogenesis from right-handed neutrinos~\cite{Fukugita:1986hr,Buchmuller:2004nz,Giudice:2003jh}. Importantly, in thermal leptogenesis, right-handed neutrinos are produced from the thermal bath so their mass must be no larger than $T_R$~\cite{Croon:2019dfw}, while for lepto-axiogenesis right-handed neutrinos can be much heavier. As $T_R$ is increased from the minimal value, the parameter space for lepto-axiogenesis rapidly opens up. 

\subsection{Axion dark matter from parametric resonance}
\label{sec:PR_quartic}
In this subsection we consider the possibility that axions produced by parametric resonance explain the observed dark matter density.
Since the axion couples to the saxion $S$, oscillations of the saxion field produce axions non-perturbatively through PR.

The production continues until the energy density of the axions becomes comparable to the saxion oscillations, at which point the back-reaction from the axions stops PR. For the quartic potential, this occurs around $\bar{S} \sim 10^{-2}~S_i$ (or $10^{-4}~S_i$) during a radiation (or a matter) dominated era~\cite{Kasuya:1998td,Kasuya:1999hy,Kawasaki:2013iha}.

The axions produced through parametric resonance are not necessarily cold at matter radiation equality and the parameter space is subject to warm dark matter constraints. Assuming the axion makes up all of the dark matter, the warm dark matter constraint requires~\cite{Irsic:2017ixq, Lopez-Honorez:2017csg}
\begin{align}
\label{eq:warmness}
v_{a}|_{T= 1\rm eV} \equiv \frac{k_a|_{T= 1\rm eV}}{m_{a}} \lesssim 2\times 10^{-4}.
\end{align}
Cosmic 21-cm lines can probe $v_{a}|_{T= 1\rm eV} \gtrsim 10^{-5}$~\cite{Sitwell:2013fpa}.
After axions are produced, their number density and the momentum are approximately conserved up to cosmic expansion~\cite{Micha:2002ey}.
Since $k_a^{3}$ and the axion number density scale as $1/R^{3}$, their ratio remains invariant and is given by $n_{a}/k_a^{3} = 1/(4 \lambda^{2})$. Replacing $k_a$ with $(4 \lambda^{2}n_{a})^{1/3}$ at $T = 1 {\rm~eV}$, requiring $Y_{a}\equiv n_{a}/s = Y_{a,{\rm DM}}$ in Eq.~(\ref{eq:yaDM}), and imposing the constraint in Eq.~(\ref{eq:warmness}), we obtain an upper bound on $m_S$,
\begin{align}
\label{eq:warmconstraint}
m_S \lesssim  
80 \MeV \; N_{\rm DW} \left(\frac{10^9 \GeV}{f_a}\right)\left(\frac{v_{a}|_{T= 1\rm eV}}{2\times 10^{-4}}\right)^{\scalebox{1.01}{$\frac{3}{2}$}}.
\end{align}
For larger masses, the PR axions are too warm, assuming they make up all of dark matter. This bound is satisfied by the solid black line in Fig.~\ref{fig:quartic}. Note that this constraint is generically applicable to axion dark matter produced by parametric resonance in a quartic potential, and is independent of the cosmological evolution, such as possible entropy production.

Assuming that no entropy is produced after PR occurs, the axion yield is given by\footnote{Note that Eqs~(\ref{eq:warmconstraint}) and (\ref{eq:yaPR}) differ from Ref.~\cite{Co:2017mop} due to our change of the estimation of $n_S$.}
\begin{align}
\label{eq:yaPR}
Y_{a} \equiv \frac{n_a}{s} \simeq \frac{n_S(S_{i})}{s} \simeq 
60 N_{\rm DW}^{ \scalebox{0.9}{$\frac{1}{2}$} }
\left(\frac{g_{\rm SM}}{g_{*}}\right)^{ \scalebox{1.01}{$\frac{1}{4}$} } \left(\frac{S_{i}}{10^{17} \GeV} \right)^{ \scalebox{1.01}{$\frac{3}{2}$} } \left(\frac{{\rm GeV}}{m_S} \right)^{ \scalebox{1.01}{$\frac{1}{2}$} } \left(\frac{f_{a}}{10^9 \GeV} \right)^{ \scalebox{1.01}{$\frac{1}{2}$} },
\end{align}
where we have used $m_S(S_i) = \lambda S_{i}$ and $m_S(S_i) = 3 H_{\rm osc}$. 
Using Eqs.~(\ref{eq:YB_quartic1}), (\ref{eq:yaDM}) and (\ref{eq:yaPR}), axion dark matter by parametric resonance and the baryon asymmetry by lepto-axiogenesis require
\begin{align}
\scalebox{0.9}{$m_S$}&\simeq 
\begin{cases}
\scalebox{0.9}{$10 \MeV$}
     N_{\rm DW}^{ \scalebox{0.9}{$\frac{1}{4}$} } 
    \left(\frac{f_a}{10^{9} \GeV}\right)^{\scalebox{0.9}{$\frac{1}{2}$} }
    \left(\frac{g_{*}}{g_{\rm SM}}\right)
    \left(\frac{0.1}{c_{B}}\right)^{ \scalebox{0.9}{$\frac{3}{4}$} } 
    \left(\frac{0.03~\rm eV^{2}}{\bar{m}^{2}}\right)^{ \scalebox{0.9}{$\frac{3}{4}$} }  & {\rm for} \ \ T_{\rm osc} < T_{\rm ss} \\
\scalebox{0.9}{$250 \MeV$}
     N_{\rm DW}^{ -\scalebox{0.9}{$\frac{1}{2}$} } 
    \left(\frac{D}{10}\right)
    \left(\frac{f_{a}}{10^{9} \GeV}\right)^{\scalebox{0.9}{$\frac{1}{2}$}}
    \left(\frac{g_{*}}{g_{\rm SM}}\right)^{\scalebox{0.9}{$\frac{5}{2}$}}
    \left(\frac{0.1}{c_B}\right)^{\scalebox{0.9}{$\frac{3}{2}$}} 
    \left(\frac{0.03~\rm eV^{2}}{\bar{m}^{2}}\right)^{\scalebox{0.9}{$\frac{3}{2}$}}
    \left(\frac{1/35}{\alpha_{3}}\right)^{\scalebox{0.9}{$\frac{15}{2}$}} &  {\rm for} \ \ T_{\rm osc} > T_{\rm ss}
\end{cases}.
\end{align}
For $T_{\rm osc} > T_{\rm ss}$, since the baryon asymmetry is overproduced, we introduce a dilution factor $D$.
The prediction for $m_S$ is shown as the black solid lines in Fig.~\ref{fig:quartic}. Above (below) the black lines, axion dark matter is overproduced (underproduced) by parametric resonance. Therefore, while dark matter can only be explained by parametric resonance below the brown dot-dashed lines with small $m_S$ and $f_a$, most of the parameter space is free from overproduction of dark matter-- a major obstacle faced by the minimal theory of axiogenesis presented in~\cite{Co:2019wyp}.

Lyman-$\alpha$ constraints \cite{Boyarsky:2008xj} analyzed by VHS \cite{Viel:2004bf} and SDSS \cite{McDonald:2004eu,McDonald:2004xn} surveys limit the fraction of warm dark matter to be $\mathcal{O}(30\%)$ and hot dark matter to be $\mathcal{O}(10\%)$. Below the brown dot-dashed line in Fig.~\ref{fig:quartic} is the region where this constraint is satisfied i.e. $\rho_{a}/\rho_{\rm DM} \lesssim 0.1$, whereas the region above is ruled out if the PR axions are not successfully thermalized.

To avoid too warm and/or an excessive amount of axions from PR above the brown dot-dashed lines and black lines, it is required that the axion fluctuations from PR be thermalized when the saxion is thermalized~\cite{Co:2020dya}. This thermalization needs to occur before $S$ is relaxed to $f_a$ so that saxions and axions are sufficiently mixed with each other via the PQ symmetry restoration. Otherwise, the axion thermalization rate becomes suppressed by its momentum due to its derivative coupling when $S = N_{\rm DW}f_a$. This assumption is essential for the kinetic misalignment mechanism discussed below in Sec.~\ref{sec:kmm_quartic} to give enough dark matter in regions above the same black lines in Fig.~\ref{fig:quartic}.

\subsection{Axion dark matter from kinetic misalignment}
\label{sec:kmm_quartic}
The kinetic misalignment mechanism can also produce axion dark matter. This contribution dominates when the axions produced by parametric resonance are thermalized. Using Eq.~(\ref{eq:N_PQ_epsilon}), the axion dark matter abundance by kinetic misalignment is

\begin{align}
Y_a = 2 \times  \frac{4 \epsilon}{N_{\rm DW}} \frac{n_S}{s} \simeq 10 \epsilon \, \scalebox{0.9}{$N_{\rm DW}^{-\frac{1}{2}}$} \left( \frac{g_{\rm SM}}{g_*}\right)^{ \scalebox{1.01}{$\frac{1}{4}$} } \left(\frac{S_{i}}{10^{16} \GeV} \right)^{ \scalebox{1.01}{$\frac{3}{2}$} } \left(\frac{{\rm GeV}}{m_S} \right)^{ \scalebox{1.01}{$\frac{1}{2}$} } \left(\frac{f_{a}}{10^9 \GeV} \right)^{ \scalebox{1.01}{$\frac{1}{2}$} } \,,
\end{align}
where $\epsilon$ is defined in Eq.~(\ref{eq:epsilon_def}) and the factor of 2 is a deviation from the analytical estimation (see Appendix~\ref{app:KMM}). Axion dark matter by kinetic misalignment and the baryon asymmetry by lepto-axiogenesis require
\begin{align}
\scalebox{0.9}{$m_S$} & \scalebox{0.9}{$\simeq $}
\begin{cases}
\scalebox{0.9}{$10 \MeV$} 
    \scalebox{0.8}{$N_{\rm DW}^{-\frac{1}{4}}$}
    \left(\frac{\epsilon }{0.125}\right)^{\scalebox{0.8}{$\frac{1}{2}$}}
    \left(\frac{f_a}{10^{9} \GeV}\right)^{\scalebox{0.8}{$\frac{1}{2}$}}
    \left(\frac{g_{*}}{g_{\rm SM}}\right)
    \left(\frac{0.1}{c_{B}}\right)^{ \scalebox{0.8}{$\frac{3}{4}$} } 
    \left(\frac{0.03~\rm eV^{2}}{\bar{m}^{2}}\right)^{ \scalebox{0.8}{$\frac{3}{4}$}}  & \scalebox{0.8}{{\rm for} \ $T_{\rm osc} < T_{\rm ss}$}\\
 \scalebox{0.9}{$250 \MeV$} 
    \scalebox{0.8}{$N_{\rm DW}^{-1}$}
    \left(\frac{\epsilon }{0.125}\right)^{\scalebox{0.8}{$\frac{1}{2}$}}
    \left(\frac{D}{10}\right)
    \left(\frac{f_{a}}{10^{9} \GeV}\right)^{\scalebox{0.8}{$\frac{1}{2}$}}
    \left(\frac{g_{*}}{g_{\rm SM}}\right)^{\scalebox{0.8}{$\frac{5}{2}$}}
    \left(\frac{0.1}{c_B}\right)^{\scalebox{0.8}{$\frac{3}{2}$}} 
    \left(\frac{0.03~\rm eV^{2}}{\bar{m}^{2}}\right)^{\scalebox{0.8}{$\frac{3}{2}$}}
    \left(\frac{1/35}{\alpha_{3}}\right)^{\scalebox{0.8}{$\frac{15}{2}$}}& 
    \scalebox{0.8}{{\rm for} \ $T_{\rm osc} > T_{\rm ss}$}.
\end{cases}
\end{align}
The black lines in Fig.~\ref{fig:quartic} show the prediction for $m_{S}$, where the lower right  panel is for the $T_{\rm osc} > T_{\rm ss}$ case. Kinetic misalignment can explain axion dark matter in the regions above the black lines by smaller $\epsilon$. In addition, the produced axions are cold and not subject to the warmness constraint.

Lastly, on the horizontal gray lines in Figs.~\ref{fig:quartic} and \ref{fig:quartic_lowTRH}, the observed dark matter abundance is explained by the conventional misalignment mechanism in the regions where kinetic misalignment is inefficient, i.e., larger $f_a$. Due to non-thermal PQ symmetry restoration by parametric resonance, the misalignment angle $\theta_i$ is randomized and averaged to $\pi/\sqrt{3}$. Axion emission from cosmic strings gives similar amount of axions~\cite{Davis:1986xc,Kawasaki:2014sqa,Klaer:2017ond,Gorghetto:2018myk,Kawasaki:2018bzv,Martins:2018dqg,Hindmarsh:2019csc,Buschmann:2019icd}. In the lower panels of Fig.~\ref{fig:quartic}, the gray lines are unaffected despite the dilution factor $D=10$ because entropy production is assumed to occur before the axion oscillations near the QCD phase transition. If entropy is produced after instead, the gray line will be shifted upward by a factor of $\simeq 7 \, (D/10)^{6/7}$. 

In Fig.~\ref{fig:quartic_lowTRH}, the black lines also represent the axion dark matter contribution from KMM (for $\epsilon = 0.125$) and PR. It is evident that this constraint is irrelevant in the case of rotation during inflationary reheating and we thus do not present the derivations of the formulae.

\subsection{Constraints on explicit PQ symmetry breaking}
\label{sec:exPQ_quartic}

To achieve rotations with a parameter $\epsilon \simeq 3  V'_{\cancel{\rm PQ}} / V'_{\rm PQ}$ from Eq.~(\ref{eq:epsilon_quartic}), the required coupling $A$ is
\begin{align}
\label{eq:A_approx}
A \simeq  \frac{2^{n/2} \lambda^2 \epsilon}{6 n} \frac{M^{n-3}}{S_i^{n-4}}\,,
\end{align}
whose exact form is given in Eq.~(\ref{eq:epsilon_quartic}) for a radiation-dominated universe.
The axion mass given by the explicit PQ symmetry breaking $V_{\cancel{\rm PQ}}$ is
\begin{align}
 \Delta m_a^2 \simeq \frac{2 n^2}{2^{n/2}}\frac{A \, N^{n-2}_{\rm DW} f_a^{n-2}}{M^{n-3}} \simeq  \frac{\lambda^2 n \epsilon}{3} \frac{N^{n-2}_{\rm DW} f_a^{n-2}}{S_i^{n-4}}.
\end{align}
To obey the experimental bound on the neutron electric dipole moment~\cite{Baker:2006ts, Afach:2015sja, Abel:2020gbr}, we require
\begin{align}
\label{eq:deltama_max}
    \Delta m_a^2  < 10^{-10} m_a^2.
\end{align}
For a given $(m_S,f_a)$, we fix $S_i$ to reproduce the observed baryon asymmetry, and then obtain a lower bound on $n$ from Eq.~(\ref{eq:deltama_max}). The contours of the lower bound are shown as pink solid lines in Figs.~\ref{fig:quartic} and \ref{fig:quartic_lowTRH} for the labeled values of $n$.

It is required that $\epsilon < 1$; otherwise, the angular direction will start oscillating before the radial mode and get trapped into a false vacuum at a large field value created by $V_{\cancel{\rm PQ}}$. Therefore, we impose
\begin{align}
\label{eq:Si_max_quartic}
S_{i} < \sqrt{2} \left(\frac{2\lambda^2}{3 n} \frac{M^{n-3}}{A}\right)^{\frac{1}{n-4}}\,.
\end{align}
The contours of the lower bound on $S_i$ are shown as pink dashed lines in Figs.~\ref{fig:quartic} and \ref{fig:quartic_lowTRH} with $A = M = 4\pi M_{\rm Pl}$ and the labeled values of $n$. The constraint is mildly relaxed if $A \ll M$.

\subsection{Saxion thermalization}
\label{sec:quartic_therm}
We first consider the case where the rotations begin during a radiation dominated era.
In order to avoid entropy production from saxion domination we need the saxion to thermalize before it dominates the energy density at $T_{M}$. In the quartic potential the saxion field value evolves as $S \propto T$ until $S \sim \sqrt{2}f_{a}N_{\rm DW}$ at temperature $T_{S}$ which can be written as 
\begin{align}
\label{eq:Tfaquartic}
T_{S} \simeq 20 \TeV \,
N_{\rm DW}^{ \scalebox{0.9}{$\frac{1}{2}$} }
\left(\frac{g_{SM}}{g_{*}}\right)^{ \scalebox{1.01}{$\frac{1}{4}$}}
\left(\frac{m_{S}}{{10 \MeV}}\right)^{ \scalebox{1.01}{$\frac{1}{2}$}}
\left(\frac{f_{a}}{{10^{9}\GeV}}\right)^{ \scalebox{1.01}{$\frac{1}{2}$}}
\left(\frac{10^{16}\GeV}{S_{i}}\right)^{ \scalebox{1.01}{$\frac{1}{2}$}} .
\end{align}
After $T_{S}$ the saxion potential is dominated by the quadratic term and saxion field value scales $S \propto T^{3/2}$. The temperature at which saxion enters matter domination is given by
\begin{align}
\label{eq:Tfaquartic_MD}
T_{M} \simeq 200 \MeV \,
N_{\rm DW}^{ \scalebox{0.9}{$\frac{1}{2}$} }
\left(\frac{g_{SM}}{g_{*}}\right)^{ \scalebox{1.01}{$\frac{1}{4}$}}
\left(\frac{m_{S}}{{10 \MeV}}\right)^{ \scalebox{1.01}{$\frac{1}{2}$}}
\left(\frac{f_{a}}{{10^{9}\GeV}}\right)^{ \scalebox{1.01}{$\frac{1}{2}$}}
\left(\frac{S_{i}}{10^{16}\GeV}\right)^{ \scalebox{1.01}{$\frac{3}{2}$}}.
\end{align}

The saxion can thermalize by scattering with gluons, non-KSVZ fermions, or through its couplings with the Higgs boson. In the case where the KSVZ quarks are heavier than the temperature of the universe when the saxion begins oscillating, as is the case for \ref{model:KSVZ_heavy}, they stay heavier since $S \propto T$ for quartic potential. In this case the saxion can thermalize by scattering with gluons. The thermalization rate for $T \geq T_{S}$ is given by~\cite{Bodeker:2006ij, Laine:2010cq, Mukaida:2012qn}
\begin{align}
\label{eq:gtherm}
\Gamma_{g} &= b N_{\rm DW}^2 \frac{T^{3}}{S^{2}(T)} ,
\end{align}
where $b\simeq 10^{-2} \alpha^{2}_{3} \simeq 10^{-5}$. Once $S \simeq N_{\rm DW} f_{a}$, $\Gamma_{g} = b T^{3}/f_{a}^{2}$ which drops faster than the Hubble scale. Thus, gluons can only thermalize before $T_{S}$. Requiring that $\Gamma_{g} > 3 H(T_{S})$ from gluon scattering we obtain
\begin{align}
\label{eq:gthermbound}
f_{a} \lesssim 4\times 10^{9} \GeV \, 
N_{\rm DW}^{ \scalebox{0.9}{$\frac{1}{3}$} }
\left(\frac{g_{SM}}{g_{*}}\right)^{ \scalebox{1.01}{$\frac{1}{2}$}}
\left(\frac{m_{S}}{10\MeV}\right)^{ \scalebox{1.01}{$\frac{1}{3}$}}
\left(\frac{10^{16}\GeV}{S_{i}}\right)^{ \scalebox{1.01}{$\frac{1}{3}$}} .
\end{align}

Thermalization through gluons is not sufficient to thermalize the saxion in most of the parameter space of interest in Fig.~\ref{fig:quartic}. In the absence of thermalization through gluons, the saxion can thermalize via its coupling to the Higgs or a non-KSVZ fermion. We will begin by  discussing constraints that arise from coupling to the non-KSVZ fermions.

The saxion-fermion coupling can be written as $z P \psi \bar{\psi}$ and the thermalization rate is
\begin{align}
\label{eq:quartthermq}
\Gamma_{\psi} \simeq b_{\psi} z^{2} T,
\end{align}
where $b_{\psi}\simeq 0.1$. In order to avoid entropy production by the saxion, the saxion needs to be thermalized at $T_{\rm th} > T_{M}$. Once the saxion settles at its minimum, the fermion mass remains constant at $z N_{\rm DW}f_{a}$ while the temperature keeps decreasing.   This implies $T_{\rm th} > \max [T_{M},z N_{\rm DW}f_{a}]$. 
However, $z$ cannot be arbitrarily large. The quantum correction to the quartic coupling of $P$ coming from $z$ needs to be less than $\lambda^{2}$, giving $z \lesssim (16\pi^{2}\lambda^{2})^{1/4}$. The thermal corrections from non-KSVZ fermions to the saxion mass should also be smaller than the saxion mass at the beginning of oscillations. This gives $z \lesssim \lambda S_{i}/T_{\rm osc}$. We also require the mass of the fermion to be lower than the temperature at the beginning of oscillations, so that fermions are in the thermal bath at thermalization $z \lesssim T_{\rm osc}/S_{i}$. Imposing the above constraints, we find that the fermion scattering can thermalize the saxion for $f_a \lesssim 10^{10} $ GeV.

Once the temperature drops below the mass of the fermion, the number density of the fermion is Boltzmann-suppressed and the thermalized saxions decouple from the thermal bath. If there is no subsequent thermalization, the saxion decays into axions and produce dark radiation with an amount given by
\begin{align}
\label{eq:DNeff}
\Delta N_{\rm eff} \simeq   \, 0.25 \left(\frac{10~{\rm MeV}}{m_{S}}\right)^{ \scalebox{1.01}{$\frac{1}{2}$} }  \left( \frac{f_a}{10^8~{\rm GeV}} \right)  \left( \frac{g_{\rm SM}}{g_* (T_{\rm D})} \right) \left(\frac{10.75}{g_* (T_{\rm dec})}\right)^{ \scalebox{1.01}{$\frac{1}{12}$} },
\end{align}
where $g_*(T_{\rm D})$ and $g_*(T_{\rm dec})$ are the effective degrees of freedom when the saxions decouple and decay respectively.
The present upper bound is $\Delta N_{\rm eff} < 0.3$~\cite{Aghanim:2018eyx}, and future observations of the cosmic microwave background can probe $\Delta N_{\rm eff}>0.02$~\cite{Wu:2014hta,Abazajian:2016yjj}.
$f_a > 10^8\GeV (m_S/10\MeV)^{1/2}$ is excluded, but this constraint can be avoided if the saxion couples to the Higgs and keeps thermalized until the temperature drops below $m_S$.

We next consider the coupling of the saxion to Higgs bosons, $\xi^{2}S^{2}H^{\dagger}H$.
The Higgs can thermalize the saxion via scatterings $S\,H \rightarrow H\,Z/W$ with a rate given by
\begin{align}
\label{eq:hthermrate}
\Gamma_{\rm th, H} &= \alpha_{2}(T)\xi^{4}\frac{\left(S(T) + N_{\rm DW}f_{a} \right)^2}{T} ,
\end{align}
where $S(T) = S_{i}(T/T_{\rm osc})$.
During radiation domination, $\Gamma_{\rm th,H}(T)/H(T)$ always increases as temperature decreases, and hence thermalization is IR-dominated.
If we demand that the Higgs mass from the field value of $P$ does not exceed the observed Higgs mass at the vacuum, i.e.~$\xi N_{\rm DW} f_a < m_{H,0}$, we find that in most of the parameter space of interest in Fig.~\ref{fig:quartic}, the saxion cannot be thermalized. Hence, we consider the case where the Higgs mass is fine-tuned to give the observed value,
\begin{align}
\label{eq:tunedmh}
m^{2}_{H,0} = -m^{2} + \xi^{2}f^{2}_{a},~~~\xi^2 N^{2}_{\rm DW}f_a^{2},m^2 \gg m_{H,0}^2.
\end{align}

We find that, after taking into account the constraints outlined below, in most of the parameter space thermalization happens at $T < T_S$, the temperature where $S$ reaches  $\sqrt{2} N_{\rm DW} f_{a}$. We thus set $S = N_{\rm DW}f_a$ in the following. Typically the saxion-Higgs mixing is parameterized by a mixing angle $\theta_{SH}$ given by
\begin{align}
\label{eq:thSH}
\theta_{SH}\simeq 2\sqrt{2}\xi^{2}\frac{N_{\rm DW}f_{a}v}{m^{2}_{H,0}}\simeq 3\times 10^{-5} \, N_{\rm DW}
\left(\frac{\xi}{10^{-6}}\right)^{2}
\left(\frac{f_{a}}{10^{9}\GeV}\right).
\end{align}

In order to avoid entropy production, thermalization needs to occur before saxion domination at temperature $T_{M}$, giving the lower bound
\begin{align}
\label{eq:xilbTM}
\xi \gtrsim 7 \times 10^{-10} \, 
N_{\rm DW}^{ \scalebox{0.9}{$\frac{3}{8}$} }
\left(\frac{1/40}{\alpha_{2}}\right)^{ \scalebox{1.01}{$\frac{1}{4}$}}
\left(\frac{g_{SM}}{g_{*}}\right)^{ \scalebox{1.01}{$\frac{1}{16}$}}
\left(\frac{{10\MeV}}{m_{S}}\right)^{ \scalebox{1.01}{$\frac{3}{8}$}}
\left(\frac{10^{9} \GeV}{f_{a}}\right)^{ \scalebox{1.01}{$\frac{1}{8}$}}
\left(\frac{S_{i}}{10^{16}\GeV}\right)^{ \scalebox{1.01}{$\frac{9}{8}$}}.
\end{align}
We also need $T_{\rm th} > m_{H,0}$, since for lower $T_{\rm th}$ the Higgs will not be present in the thermal bath. Other lower bounds on $\xi$ come from the BBN, which requires its lifetime to be shorter than one second if the saxion decays primarily through the Higgs portal, and from the saxion emission from supernova cores which requires large enough coupling to trap the saxion inside the cores~\cite{Turner:1987by,Frieman:1987ui,Burrows:1988ah, Essig:2010gu}. However, $\xi$ cannot be arbitrarily large. For $T < T_S$, $S(T) \simeq \sqrt{2}N_{\rm DW}f_{a}(T/T_S)^{3/2}$ and the Higgs mass is $m^{2}_{H}(T) \simeq 2\xi^{2}N^{2}_{\rm DW}f_{a}S(T)$.
For Higgs thermalization to be efficient we need $m_{H}(T_{\rm th}) < T_{\rm th}$,
\begin{align}
\label{eq:xiubTS_H}
\xi \lesssim 3\times 10^{-5} \, 
N_{\rm DW}^{- \scalebox{0.9}{$\frac{11}{16}$} }
\left(\frac{\alpha_{2}}{1/40}\right)^{ \scalebox{1.01}{$\frac{1}{8}$}}
\left(\frac{g_{SM}}{g_{*}}\right)^{ \scalebox{1.01}{$\frac{11}{32}$}}
\left(\frac{m_{S}}{10\MeV}\right)^{ \scalebox{1.01}{$\frac{9}{16}$}}
\left(\frac{10^{9} \GeV}{f_{a}}\right)^{ \scalebox{1.01}{$\frac{11}{16}$}}
\left(\frac{10^{16}\GeV}{S_{i}}\right)^{ \scalebox{1.01}{$\frac{9}{16}$}}.
\end{align}
Other upper bounds on the saxion-Higgs mixing come from the branching ratio for $\rm{K} \rightarrow (\pi\,+\,{\rm invisible})$~\cite{Artamonov:2008qb} and from LHCb constraints on visible decays of B mesons through a scalar mediator \cite{Aaij:2015tna,Aaij:2016qsm}. We require that corrections to the quartic coupling from $\xi$ are smaller than $\lambda^{2}$ i.e. $\xi \lesssim (16\pi^{2}\lambda^{2})^{1/4}$. We also require the thermal corrections to the mass of the saxion at the beginning of saxion oscillations to be lower than the zero-temperature saxion mass, giving $\xi \lesssim \lambda S_{i}/T_{\rm osc}$. Comparing the upper and lower bounds on $\xi$, we obtain the region that cannot be successfully thermalized through the saxion-Higgs coupling. A summary of the astrophysical and experimental constraints we use in terms of the mixing angle $\theta_{SH}$ can be found in Ref.~\cite{Beacham:2019nyx}.

Combining the constraints from the gluons, fermions, and Higgs thermalization processes mentioned in this section, we can determine the region where the saxion is not thermalized before dominating the energy density. This is shown by the orange hatched region in Fig.~\ref{fig:quartic}, which shows that most of the parameter space is consistent with the assumption of a radiation-dominated universe. The jagged nature of the lines shown in Fig.~\ref{fig:quartic} comes from satisfying the LHCb and $\rm{K} \rightarrow (\pi\,+\,{\rm invisible})$ constraints.

Some of the parameter space in Fig.~\ref{fig:quartic} is excluded by the brown dot-dashed lines because of the hot axions produced by PR. The bound can be evaded if the PR axions are thermalized. For this the thermalization must occur at $T > T_{S}$, since for $T< T_S$, where the PQ symmetry is broken, the axion direction is derivatively coupled to the thermal bath and the thermalization rate is suppressed. For $T>T_S$, the saxion and the axion mix with each other and thermalization of saxions automatically lead to that of axions. As we have seen, thermalization through the Higgs and fermions occurs predominantly when $T_{\rm th} < T_{S}$ and does not allow for the axion to thermalize. A thermalization process more efficient than considered thus far is needed to allow for KMM dark matter to the left of the PR warmness constraint. However, we do not pursue this direction further.

For the case where the rotation begins before the completion of reheating after inflation, since the amplitude of the rotation of $P$ is much smaller than the case discussed above, possible thermalization rates are larger. Also, since the quartic coupling of $P$ is larger, upper bound on the thermalization rate from the quantum correction to the quartic coupling is relaxed. As a result, the thermalization is more effective than the case discussed above.

\subsection{Domain wall problem}
PR grows the fluctuations, leading to the restoration of the PQ symmetry and randomization of the field values~\cite{Tkachev:1995md,Kasuya:1996ns,Kasuya:1997ha,Kasuya:1998td,Tkachev:1998dc,Kasuya:1999hy}. Once the amplitude of the fluctuations becomes smaller than $f_a$ by the cosmic expansion, the PQ symmetry is broken again. Domain walls are produced around the QCD phase transition. If the domain walls are stable, they eventually dominate the energy density of the universe.
Avoiding the problem requires one of the following:
\begin{enumerate}
    \item Domain wall number is unity so that the domain walls are unstable~\cite{Sikivie:1982qv}.
    \item $P$ thermalizes before the magnitude of the fluctuations become as large as the amplitude of the rotation. The rotation then becomes circular and PR no longer occurs.
    \item Explicit PQ symmetry breaking lets the domain walls decay early enough~\cite{Sikivie:1982qv}.
\end{enumerate}

The option 2 requires the coupling of $P$ to particles in the thermal bath to be strong enough. We find that in the allowed parameter region in Fig.~\ref{fig:quartic}, such a strong coupling generates too large $\lambda$ by radiative corrections.

In the option 3, domain walls decay into axions. To make this axion population less abundant than the observed DM abundance requires the decay to occur early enough by a large amount of explicit PQ symmetry breaking. The large PQ symmetry breaking spoils the axion solution to the strong CP problem unless $f_a< 3-8\times 10^8$ GeV, depending on the domain wall number~\cite{Hiramatsu:2010yn,Hiramatsu:2012sc,Kawasaki:2014sqa,Ringwald:2015dsf}. The constraint is relaxed by a factor of few by diluting the axions produced from the domain walls~\cite{Harigaya:2018ooc}.

\section{Supersymmetric models}
\label{sec:susy}

For efficient lepto-axiogenesis, the saxion oscillations after inflation must start from a large initial field value, requiring a flat saxion potential. A flat potential is natural in supersymmetric theories, where the saxion is the scalar superpartner of the axion; the saxion potential is essentially flat in the supersymmetric limit.

For example, spontaneous PQ symmetry breaking can be induced by running of the soft mass of the field $P$~\cite{Moxhay:1984am},
\begin{align}
\label{eq:dim_trans}
V_{\rm PQ}(P) = m_S^2 |P|^2 \left( {\rm ln} \frac{2 |P|^2}{f_a^2 N_{\rm DW}^{2}} -1 \right).
\end{align}
The curvature of the saxion potential is given by the supersymmetry breaking soft mass $m_S$, and the potential is sufficiently flat for a low enough scale of supersymmetry breaking.
Alternatively, in a two-field model with soft masses,
\begin{align}
\label{eq:two_field}
W = X( P \bar{P} - V_{\rm PQ}^2 ),~~V_{\rm soft} = m_P^2 |P|^2 +  m_{\bar{P}}^2 |\bar{P}|^2,
\end{align}
the stabilizer field $X$ fixes the PQ symmetry breaking fields $P$ and $\bar{P}$ on the moduli space $P \bar{P} = V_{\rm PQ}^2$, and the moduli space is lifted by the soft masses. For $P \gg V_{\rm PQ}$ or $\bar{P} \gg V_{\rm PQ}$, the potential of the saxion is dominated by the soft masses $m_P$ and $m_{\bar{P}}$, respectively.

Motivated by the above two examples, we consider the case where the saxion potential is well approximated by a quadratic term $m_S^2 |P|^2$ at large field values. In this case, the rotation of $P$ is initiated in the same manner as the rotation of the squark and slepton fields in the Affleck-Dine baryogenesis scheme~\cite{Affleck:1984fy,Dine:1995uk,Dine:1995kz}.

\subsection{Rotation in a nearly quadratic potential}
\label{sec:rot_quadratic}

In supersymmetric theories, scalar fields in general obtain squared masses proportional to $H^2$, called Hubble-induced masses. We assume that in the early universe, $P$ obtains a negative Hubble-induced mass,
\begin{align}
V(P) = - c_H H^2 |P|^2\,,
\end{align}
where $c_H$ is an $\mathcal{O}(1)$ positive constant. This negative Hubble-induced mass drives the saxion to a large field value. Although lepto-axiogenesis works for generic origins of a large field value, this origin simplifies the discussion of the dynamics of $P$ as we will see.

We assume that the PQ symmetry is explicitly broken by a superpotential term
\begin{align}
\label{eq:PQVW}
W = \frac{1}{n} \frac{P^{n}}{M^{n-3}}\,.
\end{align}
The $F$-term potential given by this superpotential stabilizes the saxion against the negative Hubble-induced mass term. The supersymmetry breaking $A$-term potential associated with the PQ breaking superpotential~(\ref{eq:PQVW}) is
\begin{align}
\label{eq:explicit_PQV_S}
V_{\cancel{\rm PQ}} =  A \frac{P^{n}}{M^{n-3}} + {\rm h.c.}\,,
\end{align}
and drives the angular motion  of $P$. (The superpotential term alone does not do this, since a linear combination of the $R$ symmetry and the PQ symmetry remains unbroken without the $A$ term). Under these conditions, the entire potential is  
\begin{align}
V(P) = (m_S^{2} - c_H H^2) |P|^2  +\left( A \frac{P^{n}}{M^{n-3}}  + {\rm h.c.} \right) + \frac{|P|^{2n-2}}{M^{2n-6}}\,.
\end{align}
After inflation, for $3H > m_S$ the saxion tracks the time-dependent minimum of the potential given by~\cite{Dine:1995kz,Harigaya:2015hha} 
\begin{align}
\label{eq:saxion_minimum}
S(H) \simeq \left(H M^{n-3}\right)^{ \frac{1}{n-2}} \left(\frac{2^{n-2}}{n-1} \right)^{\frac{1}{2n-4}},
\end{align}
where we take $c_H \sim 1$.%
\footnote{Even if the Hubble induced mass term becomes negligible at some point, as is the case during radiation dominated eras~\cite{Kawasaki:2011zi,Kawasaki:2012qm,Kawasaki:2012rs}, $S$ still follows the value around Eq.~(\ref{eq:saxion_minimum}) because of the balance between Hubble friction and the potential gradient.}

When the Hubble scale becomes comparable to $m_S$, $3H_{\rm osc} \equiv m_S$ and $S_i \equiv S(H_{\rm osc})$, $P$ starts to oscillate. The $A$ term drives the angular motion. Subsequently, as $\bar{S}$ decreases by redshifting, explicit PQ symmetry breaking becomes negligible and the PQ asymmetry is conserved. Following the definitions in Eq.~(\ref{eq:epsilon_def}), we parameterize the PQ charge asymmetry by a parameter $\epsilon \leq 1$ and obtain
\begin{align}
\label{eq:nPQ_def}
    n_{\rm PQ} =  \frac{\epsilon}{N_{\rm DW}} m_S S_{\rm max}^2 \simeq  2 \frac{\epsilon}{N_{\rm DW}} n_S , \hspace{1cm} n_S \simeq \frac{1}{2} m_S S_{\rm max}^2,
\end{align}
for $\epsilon \ll 1$. The typical size of $\epsilon$ is given by
\begin{align}
\label{eq:npq_SUSY}
    \epsilon \simeq \frac{A}{m_S}.
\end{align}
A detailed estimation of $\epsilon$ is given in Appendix~\ref{app:epsilon}.
Here we assume that the initial phase of $P$ is not accidentally aligned with the minimum of the potential given by the $A$-term. If the soft supersymmetry breaking terms of $S$ are dominantly given by gravity mediation, $A\sim m_S$, and hence $\epsilon = \mathcal{O}(1)$; in other schemes $\epsilon$ can be small. The time-average of $\dot{\theta}$ is approximately $N_{\rm DW} m_S$, as in Eq.~(\ref{eq:dtheta_ave}), and nearly independent of $\epsilon$. If parametric resonance becomes effective and randomizes the field value of $P$, $\vev{\dot{\theta}}$ depends on $\epsilon$ and is given by $\epsilon N_{\rm DW} m_S$, as in Eq.~(\ref{eq:dtheta_ave_PR}).

Once the PQ symmetry breaking field $P$ starts to rotate and oscillate, it evolves in the following way. The energy density of $P$ redshifts as matter ($R^{-3}$) and, likewise, $n_{\rm PQ}$ redshifts in the same manner due to conservation of PQ charge. $\vev{\dot\theta}$ remains constant as long as $S \gg N_{\rm DW} f_a$. Since we are taking large initial field values, there is the possibility that the saxion could end up dominating the energy density of the universe over the radiation bath. Therefore, in the next subsection we study two scenarios: in the first, saxions are thermalized before they dominate, while, in the second, there is a period of saxion domination before thermalization with consequent dilution of the baryon asymmetry and axion dark matter. After thermalization of the $P$ field, the orbit becomes circular with the energy density of the radial motion transferred into the thermal bath. After thermalization, the energy density of the circular rotation $\rho_{\theta}$ decreases as matter ($R^{-3}$) until $S$ drops to $N_{\rm DW}f_{a}$, when a period of kination ensues with $\rho_{\theta}$ redshifting as $R^{-6}$. Notice that the kination era due to the axion rotation is conceptually different from the known scenarios~\cite{Spokoiny:1993kt, Joyce:1996cp, Salati:2002md} where a scalar field \emph{rolls down} the potential, while in our case the axion \emph{revolves around} it. An early period of kination era could leave a distinctive imprints, and possible experimental signals, in a wide range of different cosmological phenomena compared to the standard cosmology: from modifying the relic abundance of DM~\cite{DEramo:2017gpl,Redmond:2017tja,DEramo:2017ecx,Visinelli:2017qga}, increasing the signal of primordial gravitational waves~\cite{Cui:2018rwi, Bernal:2019lpc, Ramberg:2019dgi, Gouttenoire:2019kij}, to boosting the matter power spectrum, enhancing small-scale structure formation~\cite{Redmond:2018xty,Visinelli:2018wza}.

\subsection{Baryon asymmetry}

The baryon asymmetry $Y_B$ generated at temperature $T$ from the rotation of $P$ is always proportional to $\dot{\theta}/ T$. With a quadratic potential we have shown that there are periods when $\dot{\theta}(T)$ is constant, whereas with a quartic potential $\dot{\theta}(T)$ always falls. Hence, relative to the quartic case, baryogenesis with a quadratic potential is dominated more by lower temperatures. Even if $P$ starts to oscillate early, the baryon asymmetry is dominantly produced at $T < T_{\rm ss}$ where the strong sphaleron transition is in thermal equilibrium. The baryon number produced per unit Hubble time is generically given by
\begin{align}
\label{eq:nBquad}
\Delta n_{B} \, \simeq \, c_B \, \dot{\theta}T^{2} \, \frac{\Gamma_{L}}{H}.
\end{align}

We first consider the case where the saxion is thermalized before it dominates the universe, so that no entropy is produced after the onset of the rotation. Due to PQ charge conservation, the energy density associated with the rotation of $P$ can dominate even after the saxion is thermalized without subsequently producing entropy, as explained in Sec.~\ref{sec:rot_quadratic} and Appendix~\ref{app:scaling}. We define $S_M$ as the saxion field value at the beginning of this matter-like domination. Similarly, $S_{\rm ss}$ is the field values of the saxion at $T_{\rm ss}$. We find 
\begin{align}
\label{eq:yB_quad}
Y_{B} & \simeq \frac{n_B}{s}  \simeq c_B \, \frac{\dot\theta T^2}{s} \, \frac{\Gamma_L}{H} \, \ln \left( \frac{t_2}{t_1} \right) \\
& \simeq 10^{-10} N_{\rm DW}  \left( \frac{c_B}{0.1} \right) \left(\frac{g_{\rm MSSM}}{g_*} \right)^{ \scalebox{1.01}{$\frac{3}{2}$} } \left( \frac{\bar{m}^2}{0.03~{\rm eV}^2} \right) \left( \frac{m_S}{600 \TeV} \right) \ln \left( \frac{\min \left( S_i, S_{\rm ss}\right)}{ \max \left(N_{\rm DW} f_a, S_M \right) } \right) , \nonumber
\end{align}
where we assume a radiation dominated era with $g_{\rm MSSM} = 228.75$. Here the logarithmic dependence appears because $\dot n_B$ in Eq.~(\ref{eq:TL_RD}) is inversely proportional to time when the universe is radiation-dominated, $\dot\theta$ is constant ($S > N_{\rm DW} f_a$), and $T < T_{\rm ss}$. Here $t_1$ and $t_2$ denote the initial and final times of this period, whereas $\ln(t_2/t_1) = 4 \ln(S(t_1)/S(t_2))/3$ with $S(t_1) = \min \left[S_i, S_{\rm ss}\right]$ and $S(t_2) = \max\left[N_{\rm DW}f_a,S_M\right]$, where $S_{M}$ is the average vev of the radial mode when saxion matter domination begins. The observed baryon asymmetry is explained for
\begin{equation}
\label{eq:mS_quadratic}
m_S \simeq 30 \TeV \ N_{\rm DW}^{-1} \left( \frac{0.1}{c_B} \right) \left(\frac{g_*}{g_{\rm MSSM}} \right)^{ \scalebox{1.01}{$\frac{3}{2}$} } \left( \frac{0.03~{\rm eV}^2}{\bar{m}^2} \right) \left( \frac{17}{\ln \left( \frac{\min \left( S_i, S_{\rm ss}\right)}{ \max \left(N_{\rm DW}f_a, S_M \right) } \right)} \right) .
\end{equation}
This value of $m_S$ gives $T_{\rm EW} < T_{\rm osc} < T_{\rm ss}$.   We find $\min \left( S_i, S_{\rm ss}\right) = S_i$ provided $m_S \lesssim 6 \times 10^9 \GeV (20 \alpha_3)^{10} (g_{\rm SM}/g_*)^{1/2}$. Eqs.~(\ref{eq:yB_quad}) and (\ref{eq:mS_quadratic}) are valid if $S_{\rm ss} > N_{\rm DW} f_a$, for which we need $S_i > N_{\rm DW} f_a~(m_S / 6 \times 10^9 \GeV)^{3/4} (g_*/g_{\rm SM})^{3/8} (20 \alpha_3)^{15/2}$, and this is easily satisfied. 

Remarkably, the order of magnitude of $m_S$, the scale of supersymmetry breaking, is determined to be of order 30 TeV by the observed baryon asymmetry and neutrino masses. The dependence on the neutrino spectrum, $N_{\rm DW}$, and the parameters inside the log is mild.  In particular, $m_S$ cannot be of order the TeV scale unless $N_{\rm DW} \sim (30, 300)$ for neutrino masses with (near degeneracy, a normal hierarchy).

\begin{figure}[!t]
	\includegraphics[width=0.495\linewidth]{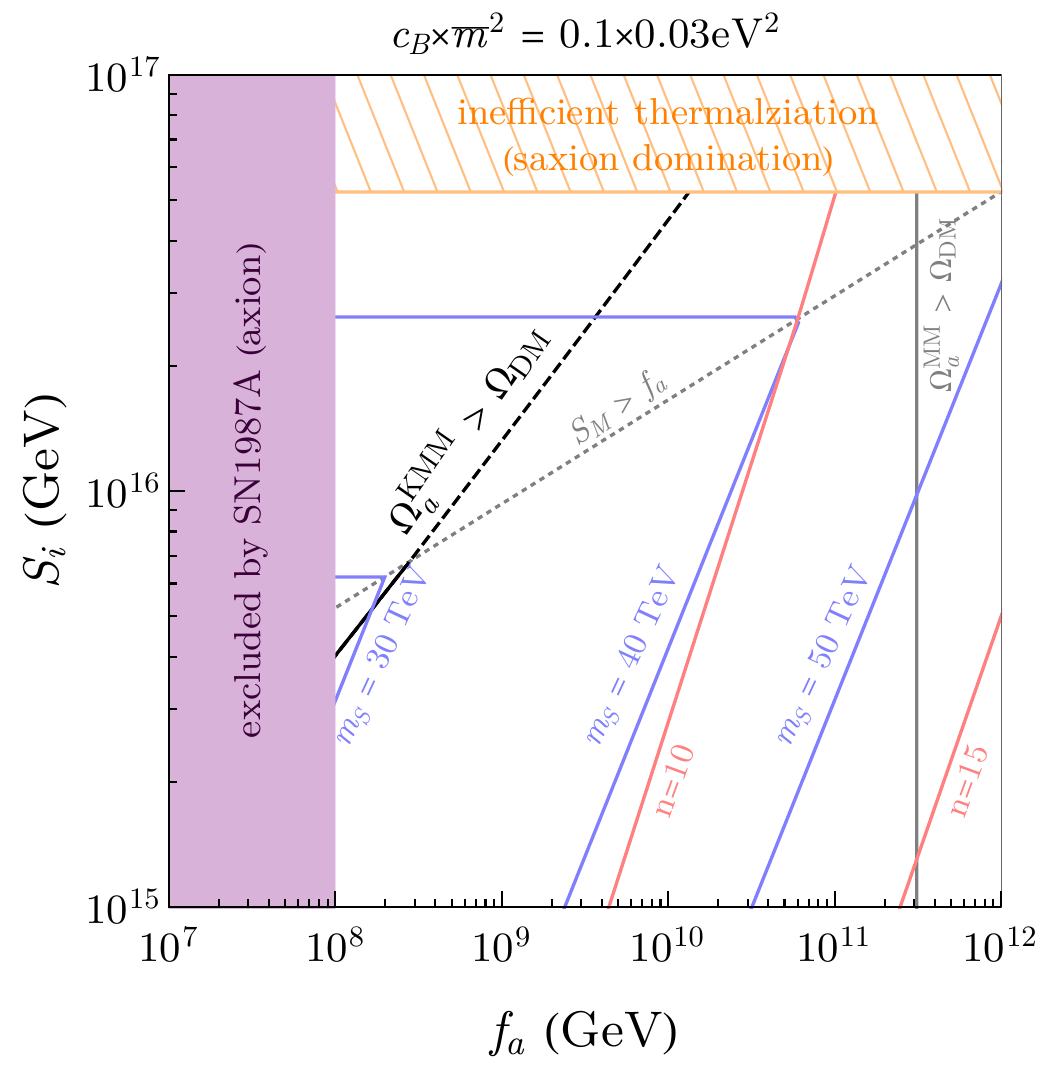}
	\includegraphics[width=0.495\linewidth]{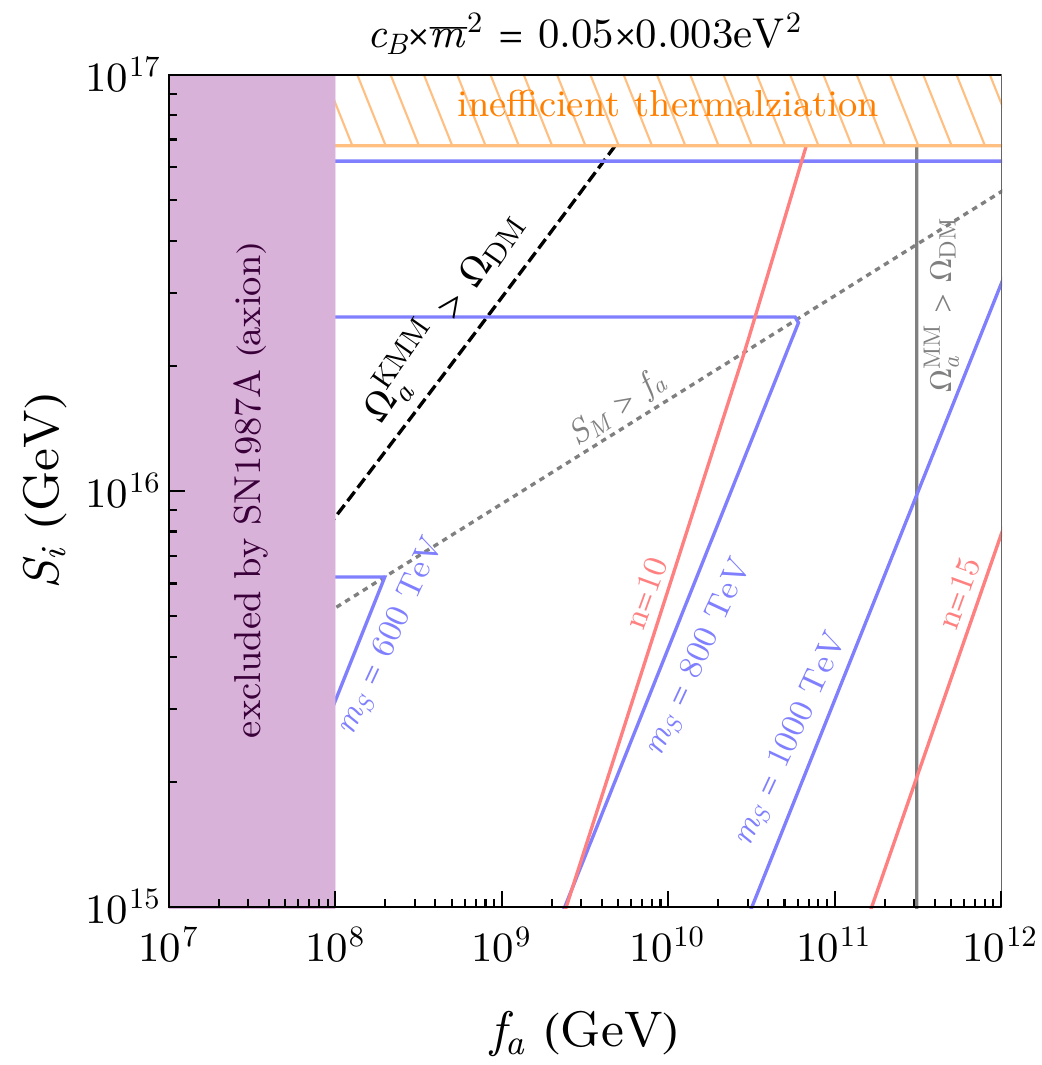}
	\caption{Parameter space compatible with the observed baryon asymmetry for nearly quadratic potentials in the case without entropy production for $N_{\rm DW} = 1$. 
	Values of $m_S$ that yield the observed baryon asymmetry are shown by blue contours.  Axions account for the observed dark matter via either parametric resonance or kinetic misalignment along the black solid line, while only kinetic misalignment is viable along the black dashed line.  We fix $\epsilon = 0.25$, which affects the black (both solid and dashed) and pink solid lines.}
	\label{fig:quadratic}	
\end{figure}

The blue contours in Fig.~\ref{fig:quadratic} show the values of $m_S$ required to explain the observed baryon asymmetry. Relevant for the determination of $Y_B$ using Eq.~(\ref{eq:yB_quad}), $\min \left( S_i, S_{\rm ss}\right) = S_i$ throughout the parameter space of interest as noted above, while above the gray dotted lines, we have $S_M > N_{\rm DW} f_a$, in which case $Y_B$ and thus $m_S$ become independent of $f_a$. The purple region is the same as in Fig.~\ref{fig:quartic}, while other constraints will be discussed in the following subsections.

We now consider an era of saxion domination. From the scaling laws presented in Appendix~\ref{app:scaling} and summarized in Table~\ref{tab:scalings}, one can see that the baryon asymmetry is dominantly produced around the beginning of the matter dominated era or at the end of reheating\footnote{After the saxion is thermalized at $T_{\rm th}$, an era dominated by the rotation labeled as MD$_{\rm A}^{\rm rot}$ in Appendix~\ref{app:scaling} is present unless $\epsilon$ is  sufficiently small. However, we expect $\epsilon \simeq A/m_S \simeq 1$ for gravity mediation and thus do not consider small $\epsilon$ here.}.
At both temperatures, $\Delta n_B/s$ are of the same order, but the former contribution experiences entropy production by the PQ symmetry breaking field.
Thus, the production is actually dominated at $T_{\rm th}$. The baryon asymmetry is given by
\begin{align}
\label{eq:YB_quadratic_MD}
Y_{B} \simeq \left. \frac{c_B \, \dot\theta T^2}{s} \frac{\Gamma_L}{H} \right|_{T = T_{\rm th}} \simeq 10^{-10} N_{\rm DW} \left( \frac{c_B}{0.1} \right) \left( \frac{\bar{m}^2}{0.03~{\rm eV}^2} \right) \left( \frac{m_S}{800 \TeV} \right),
\end{align}
which is valid as long as $S_{\rm th}$, the saxion field value at thermalization, is still larger than $N_{\rm DW}f_a$. It is remarkable that the cases of thermalization during radiation and matter domination lead to such similar results for $Y_B$; the only essential difference between (\ref{eq:yB_quad}) and (\ref{eq:YB_quadratic_MD}) is that the latter lacks the logarithm. The absence of this logarithm implies that, unlike with radiation domination, $Y_{B}$ is completely independent of the initial field value of the saxion, giving a sharp prediction for $m_S$ from the observed baryon asymmetry 
\begin{align}
\label{eq:mS_YB_quad_DM}
m_S & \simeq  700 \TeV \,
    N_{\rm DW}^{\scalebox{0.9}{$-1$}}
    \left(\frac{g_{*}}{g_{\rm MSSM}}\right)^{\scalebox{0.9}{$\frac{3}{2}$}}
    \left(\frac{0.1}{c_{B}}\right) 
    \left(\frac{0.03~\rm eV^{2}}{\bar{m}^{2}}\right) \,.
\end{align}

\begin{figure}[!t]
	\includegraphics[width=0.495\linewidth]{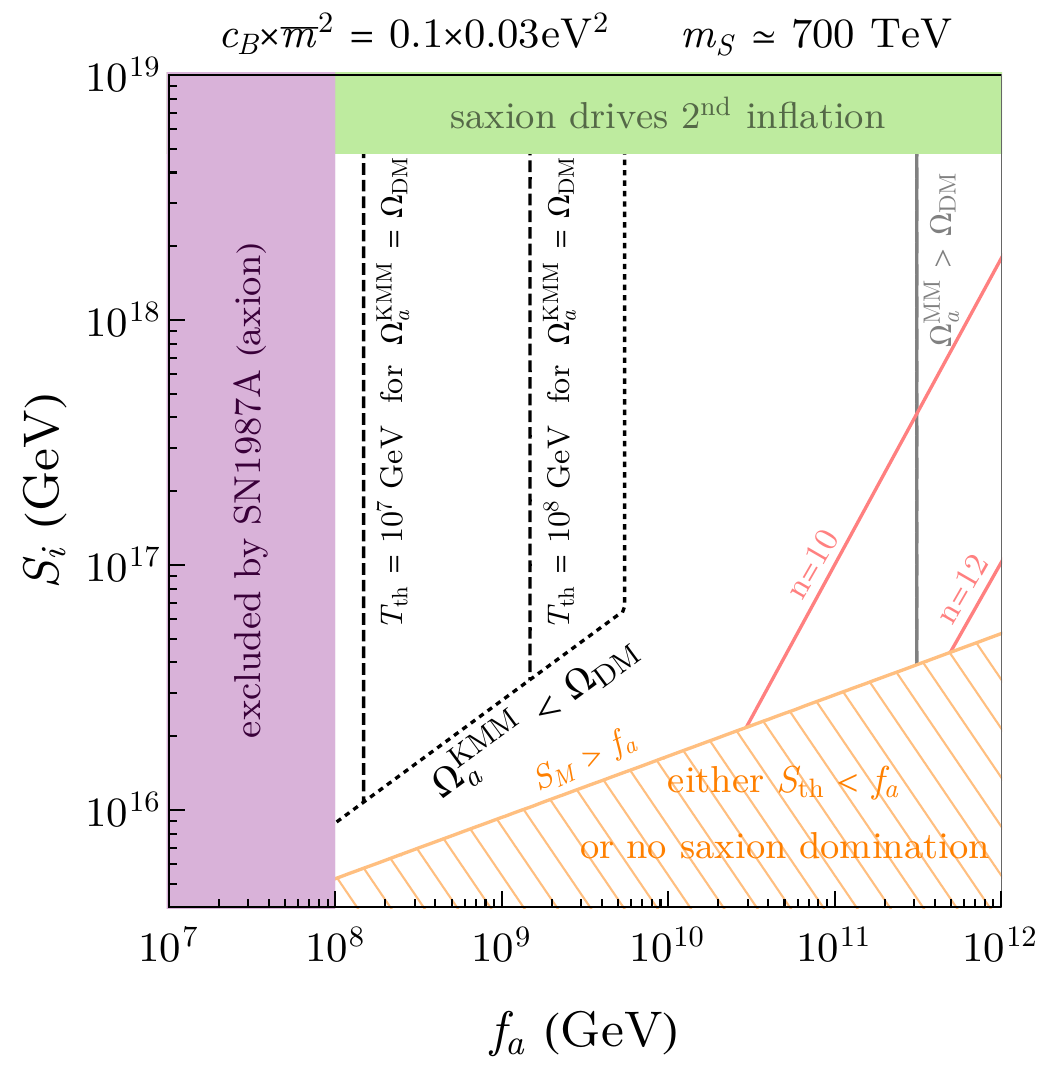}
	\includegraphics[width=0.495\linewidth]{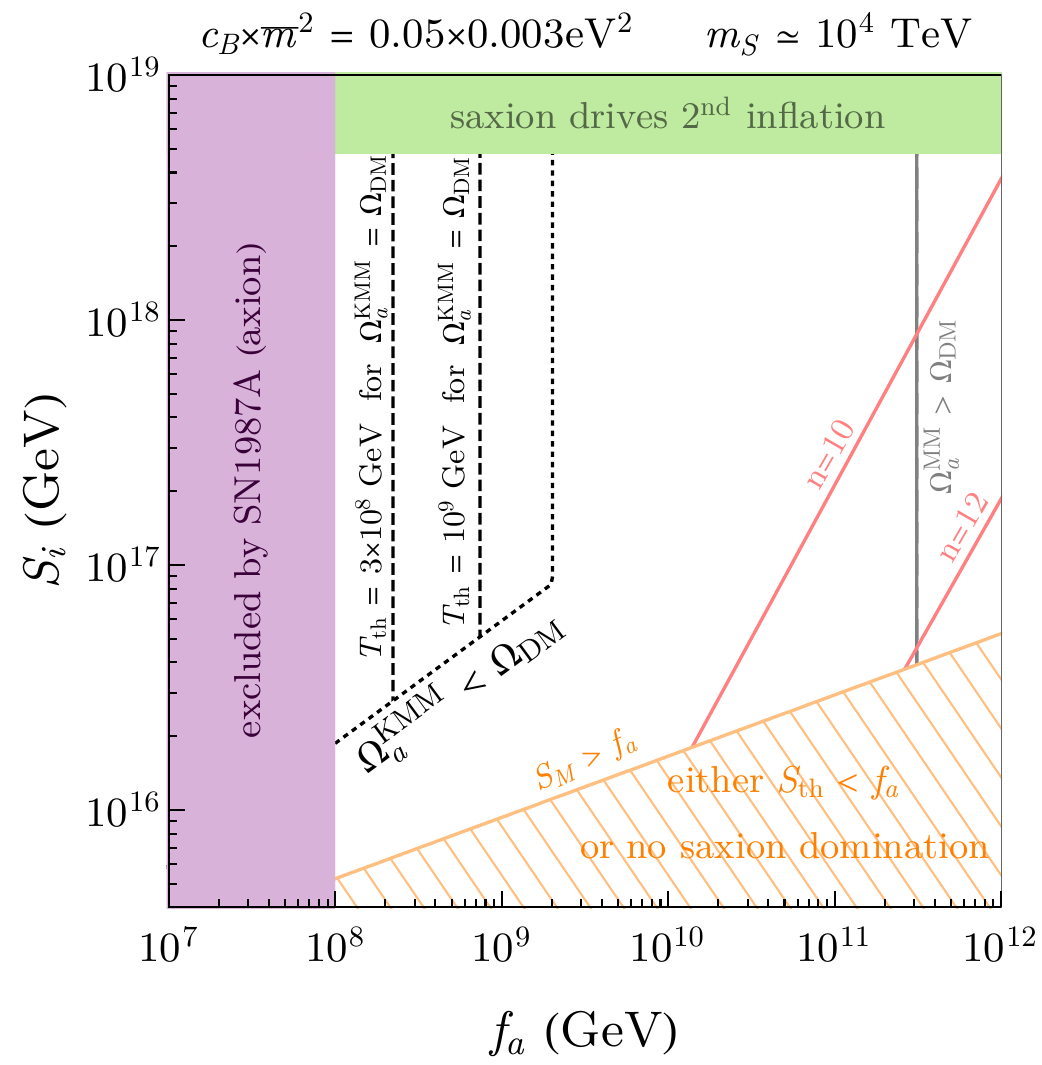}
	\caption{Parameter space compatible with the observed baryon asymmetry for nearly quadratic potentials in the case with entropy production from saxion domination for $N_{\rm DW} = 1$. Values of $m_S$ that yield the observed baryon asymmetry are shown at top.
	We fix $\epsilon = 0.25$, which affects the black (both solid and dotted) and pink lines.}
	\label{fig:quadratic_MD}	
\end{figure}

In the unshaded region of Fig.~\ref{fig:quadratic_MD}, the observed baryon asymmetry is explained with a uniquely-determined $m_S$ in Eq.~(\ref{eq:mS_YB_quad_DM}) shown at the top of each panel along with the values of $c_B$ and $\bar{m}^2$. Above the orange hatched region, we have $S_M > N_{\rm DW} f_a$ so $S_{\rm th} < S_M$ can be larger than $N_{\rm DW} f_a$ to be consistent with the assumptions made in Eq.~(\ref{eq:YB_quadratic_MD}). Inside the orange hatched, the observed baryon asymmetry can in principle be explained but the analysis becomes model-dependent due to different scaling behaviors shown in Table~\ref{tab:scalings} for different saxion thermalization channels. The purple and green regions are the same as in Fig.~\ref{fig:quartic}, while other constraints and contours will be discussed in the following subsections.

Remarkably, for the cases considered in Figs.~{\ref{fig:quadratic}} and {\ref{fig:quadratic_MD}}, the baryon asymmetry is proportional to the mass of the PQ symmetry breaking field, and requires $m_S = \mathcal{O}(10-10^4) \TeV$. Such a large scalar mass is consistent with the scenario of high scale supersymmetry breaking~\cite{Giudice:1998xp,Wells:2004di,Ibe:2006de,Acharya:2007rc,Hall:2011jd,Ibe:2011aa,Arvanitaki:2012ps,ArkaniHamed:2012gw}, which has the following features:
\begin{itemize}
\item
The observed Higgs mass is explained by quantum correction from stops~\cite{Okada:1990gg,Okada:1990vk,Ellis:1990nz,Haber:1990aw}.
\item
Because of anomaly mediated contribution to the gaugino mass~\cite{Randall:1998uk,Giudice:1998xp,Bagger:1999rd,DEramo:2012vvz,Harigaya:2014sfa}, singlet SUSY breaking fields are not required. The Polonyi problem~\cite{Coughlan:1983ci} is absent.
\item
With anomaly mediated gaugino masses, the wino is the light supersymmetric particle (LSP) with a mass around the TeV scale. The thermal freeze-out abundance of the wino can explain the observed dark matter abundance~\cite{Hisano:2006nn,Cirelli:2007xd}.
\item
Gravitinos decay before BBN and there is no serious gravitino problem~\cite{Pagels:1981ke,Weinberg:1982zq,Khlopov:1984pf,Kawasaki:2008qe}.
\item
Due to the large squark and slepton masses, the SUSY flavor/CP problems are solved.
\end{itemize}

We next consider the case where the universe is inflaton-dominated when the PQ symmetry breaking field begins rotation. As long as there is a period when the universe is radiation-dominated and $S > N_{\rm DW}f_a$, the baryon asymmetry is given by Eq.~(\ref{eq:YB_quadratic_MD}). This is possible when reheating completes early enough, i.e.,
\begin{align}
    T_{R} > 4\times 10^7~{\rm GeV}\times
    N_{\rm DW}^{\scalebox{0.8}{$\frac{1}{2}$} }
    \left(\frac{g_{\rm MSSM}}{g_*} \right)^{\scalebox{0.9}{$\frac{1}{4}$}} 
    \left(\frac{m_S}{10^6~{\rm GeV}} \right)^{\scalebox{0.9}{$\frac{1}{2}$}}
    \left(\frac{f_a}{10^9~{\rm GeV}} \right)^{\scalebox{0.9}{$\frac{1}{2}$}} 
    \left(\frac{10^{17}~{\rm GeV}}{S_i} \right)^{\scalebox{0.9}{$\frac{1}{2}$}}.
\end{align}
Lower reheat temperatures may still explain the observed baryon asymmetry but require a different evaluation of $Y_B$ and we do not pursue this further.

We comment on the LSP production from gravitinos.
If the reheat temperature of the universe is large, gravitinos are abundantly produced at reheating and later decay to LSPs.  Avoiding too much stable LSP dark matter then requires~\cite{Kawasaki:1994af,Kawasaki:2004qu}
\begin{align}
\label{eq:TRsusy}
    T_{\rm R} < 2\times 10^{9}~{\rm GeV} \left( \frac{m_{\rm LSP}}{\rm TeV} \right).
\end{align}
This bound is violated if the rotation begins during a radiation-dominated era, requiring $R$-parity violation or entropy production from the saxion. If the rotation begins during a matter dominated era, since $T_R$ may be as low as $10^7 \GeV$, the bound can be satisfied.

\subsection{Axion dark matter from parametric resonance}
\label{sec:quadraticPR}

As described in Sec.~\ref{sec:PR_quartic}, the oscillations of the saxion field produce axions non-perturbatively through parametric resonance. For the quadratic potential with a logarithmic correction in Eq.~(\ref{eq:dim_trans}) this happens for $\epsilon \lesssim 0.5 $. For the theory with the superpotential and the soft masses in Eq.~(\ref{eq:two_field}) with $\epsilon = {\cal O}(0.1-1)$, PR is ineffective unless $\bar{S} \lesssim 100 f_a$. See Appendix~\ref{app:PR_quadratic}. We thus consider the potential in Eq.~(\ref{eq:dim_trans}) in this subsection.

The time at which PR becomes effective can be written as $t_{\rm PR} \equiv N_{\rm PR}/m_S$ where we find $N_{\rm PR} \simeq  \mathcal{O}(10^3)$ from a numerical calculation in Appendix~\ref{app:PR_quadratic}. Assuming radiation domination during PR and no further entropy production, e.g.~from saxion domination, the axion yield is given by 
\begin{align}
Y_{a} \equiv \frac{n_a}{s} \simeq \frac{n_S(S_{i})}{s} \simeq  20 \left(\frac{g_{\rm MSSM}}{g_{*}}\right)^{ \scalebox{1.01}{$\frac{1}{4}$} } \left(\frac{S_{i}}{10^{16} \GeV} \right)^{2}\left(\frac{100 \TeV}{m_S} \right)^{ \scalebox{1.01}{$\frac{1}{2}$} } .
\end{align}
If the axions produced by PR do not thermalize and contribute to more than ${\cal O}(10)\%$ of DM, they are subject to warm DM constraints.

When the PR axions explain the entire DM, using $n_{a}/k_a^{3}$ as a red-shift invariant quantity, we obtain the warm dark matter bound, 
\begin{align}
\label{eq:warmquad}
S_{i} &\lesssim 8\times 10^{15}\GeV 
\left( \frac{g_{*}}{g_{\rm MSSM}} \right)^{\scalebox{1.01}{$\frac{1}{6}$}}
\left( \frac{0.1}{k_a/m_S} \right)^{\scalebox{1.01}{$\frac{1}{2}$}} 
\left(\frac{10^3}{N_{\rm PR}}\right)^{\scalebox{1.01}{$\frac{1}{4}$}}
\left(\frac{v_{a}|_{T= 1\rm eV}}{2\times 10^{-4}}\right)^{\scalebox{1.01}{$\frac{1}{2}$}}.
\end{align}
This constraint is satisfied by the solid black line in Fig.~\ref{fig:quadratic}. For the parameter space that does not satisfy Eq.~(\ref{eq:warmquad}), the observed dark matter abundance can be explained by kinetic misalignment discussed in the next subsection on the black dashed lines, as long as the axions from parametric resonance are thermalized.

In order to obtain both the correct baryon and DM densities, the initial field value is required to be
\begin{align}
\label{eq:Si_quadratic_PR_RD}
S_{i} \simeq  10^{16} \GeV
    \, N_{\rm DW}^{\scalebox{0.8}{$-\frac{1}{4}$}}
    \left(\frac{g_{*}}{g_{\rm MSSM}}\right)^{\scalebox{0.9}{$\frac{1}{2}$}}
    \left(\frac{0.1}{c_{B}}\right)^{\scalebox{0.9}{$\frac{1}{4}$}}
    \left(\frac{0.03~\rm eV^{2}}{\bar{m}^{2}}\right)^{\scalebox{0.9}{$\frac{1}{4}$}}  
    \left(\frac{f_a}{10^{9}~\rm GeV}\right)^{\scalebox{0.9}{$\frac{1}{2}$}} \nonumber \\
\hspace{-1.0 cm} \times
\begin{cases}
    \left(\frac{20}{- \log(\xi_3) + \log(-\log(\xi_3))}\right)^{\scalebox{0.9}{$\frac{1}{4}$}} & {\rm for} \,\, \scalebox{0.9}{$ S_{M} > N_{\rm DW} f_{a}$} \\ 
    \left(\frac{65}{\log(\xi_4) - \log(\log(\xi_4))}\right)^{\scalebox{0.9}{$\frac{1}{4}$}} & {\rm for} \,\, \scalebox{0.9}{$ S_{M} <  N_{\rm DW}f_{a}$} \\
\end{cases}\,,
\end{align}
where
\begin{align}
\xi_{1} &= 4 \times 10^{-8}
    N_{\rm DW}^{-1}
    \left(\frac{g_{*}}{g_{\rm MSSM}}\right)^2
    \left(\frac{0.1}{c_{B}}\right) 
    \left(\frac{0.03~\rm eV^{2}}{\bar{m}^{2}}\right)  
    \left(\frac{f_a}{10^{9}~\rm GeV}\right)^2  \\
\xi_{2} &= 2 \times 10^{30}
    N_{\rm DW}^{-5}
    \left(\frac{g_{*}}{g_{\rm MSSM}}\right)^2
    \left(\frac{0.1}{c_{B}}\right) 
    \left(\frac{0.03~\rm eV^{2}}{\bar{m}^{2}}\right)  
    \left(\frac{10^{9}~\rm GeV}{f_a}\right)^2 \,.
\end{align}
The logarithmic dependence on $\xi_{1,2}$ reflects the weak dependence of the predicted saxion mass on $f_a$, shown in Eq.~(\ref{eq:mS_quadratic}).

If the saxion thermalizes at a temperature $T_{\rm th}$ after saxion domination, assuming no further entropy production, the axion yield is given by 
\begin{align}
\label{eq:Ya_PR_quad_MD}
Y_{a} \simeq \frac{3}{4}\frac{T_{\rm th}}{m_S}.
\end{align}
However, in this paper, when the saxion dominates, we consider the case where the saxion is thermalized when $\bar{S}>f_a$, so that the estimation in Eq.~(\ref{eq:YB_quadratic_MD}) is valid. Since the PQ symmetry breaking field $P$ is not at the minimum of the potential, the saxion and the axion mix with each other, and hence the saxion thermalization necessarily leads to the thermalization of the PR axions. The KMM mechanism explained below dominates.

\subsection{Axion dark matter from kinetic misalignment}

KMM can produce the observed DM abundance regardless of the thermalization of the $P$ field. In a radiation dominated universe, the axion dark matter abundance from KMM is given by
\begin{align}
\label{eq:yaKMMquad}
Y_a = 2 \times \frac{2 \epsilon}{N_{\rm DW}} \frac{n_S}{s} \simeq 100  \frac{\epsilon}{N_{\rm DW}} \left( \frac{g_{\rm MSSM}}{g_*}\right)^{ \scalebox{1.01}{$\frac{1}{4}$} } \left(\frac{S_{i}}{10^{16} \GeV} \right)^{2}\left(\frac{100 \TeV}{m_S} \right)^{ \scalebox{1.01}{$\frac{1}{2}$} } \,.
\end{align}
Therefore, we explain axion dark matter by kinetic misalignment and the baryon asymmetry by lepto-axiogenesis if the initial field value of the saxion is
\begin{align}
\label{eq:Si_quadratic_KMM_RD}
S_{i} \simeq  10^{16} \GeV
    \, N_{\rm DW}^{\scalebox{0.8}{$\frac{1}{4}$}}
    \left(\frac{0.25}{\epsilon}\right)^{\scalebox{0.8}{$\frac{1}{2}$}}
    \left(\frac{g_{*}}{g_{\rm MSSM}}\right)^{\scalebox{0.9}{$\frac{1}{2}$}}
    \left(\frac{0.1}{c_{B}}\right)^{\scalebox{0.9}{$\frac{1}{4}$}}
    \left(\frac{0.03~\rm eV^{2}}{\bar{m}^{2}}\right)^{\scalebox{0.9}{$\frac{1}{4}$}}  
    \left(\frac{f_a}{10^{9}~\rm GeV}\right)^{\scalebox{0.9}{$\frac{1}{2}$}} \nonumber \\
\hspace{-1.0 cm} \times
\begin{cases}
    \left(\frac{20}{- \log(\xi_3) + \log(-\log(\xi_3))}\right)^{\scalebox{0.9}{$\frac{1}{4}$}} & {\rm for} \,\, \scalebox{0.9}{$ S_{M} > N_{\rm DW}f_{a}$} \\ 
    \left(\frac{65}{\log(\xi_4) - \log(\log(\xi_4))}\right)^{\scalebox{0.9}{$\frac{1}{4}$}} & {\rm for} \,\, \scalebox{0.9}{$ S_{M} < N_{\rm DW}f_{a}$} \\
\end{cases}\,,
\end{align}
where
\begin{align}
\xi_{3} &= 4 \times 10^{-8}
    N_{\rm DW}
    \left(\frac{0.25}{\epsilon}\right)^2
    \left(\frac{g_{*}}{g_{\rm MSSM}}\right)^2
    \left(\frac{0.1}{c_{B}}\right) 
    \left(\frac{0.03~\rm eV^{2}}{\bar{m}^{2}}\right)  
    \left(\frac{f_a}{10^{9}~\rm GeV}\right)^2  \\
\xi_{4} &= 2 \times 10^{30}
    N_{\rm DW}^{-3}
    \left(\frac{0.25}{\epsilon}\right)^2
    \left(\frac{g_{*}}{g_{\rm MSSM}}\right)^2
    \left(\frac{0.1}{c_{B}}\right) 
    \left(\frac{0.03~\rm eV^{2}}{\bar{m}^{2}}\right)  
    \left(\frac{10^{9}~\rm GeV}{f_a}\right)^2 \,.
\end{align}
The logarithmic dependence on $\xi_{3,4}$ reflects the weak dependence of the predicted saxion mass on $f_a$, shown in Eq.~(\ref{eq:mS_quadratic}).
These required values of $S_i$ as a function of $f_a$ are shown by the black line in Fig.~\ref{fig:quadratic}.

On the other hand, if the saxion thermalizes at a temperature $T_{\rm th}$ after saxion domination, assuming no further entropy production, the axion yield is given by 
\begin{align}
\label{eq:ya_KMM_quad_MD}
Y_a = 2 \times \frac{2\epsilon }{N_{\rm DW}} \frac{ n_S}{s} =  \frac{4\epsilon}{N_{\rm DW}} \frac{3 T_{\rm th}}{4 m_S}\,,
\end{align}
where the first factor of $2$ is explained in Appendix~\ref{app:KMM}, while the second is from Eq.~(\ref{eq:nPQ_def}).
The required thermalization temperature to obtain the correct dark matter abundance is
\begin{align}
\label{eq:Tthdil}
T_{\rm th} &\simeq 7 \times 10^{7}  \GeV N_{\rm DW} \left(\frac{0.25}{\epsilon}\right) \left(\frac{f_a}{10^{9}~\rm GeV} \right)\left(\frac{m_S}{700 \TeV}\right)\nonumber \\
 &\simeq 7 \times 10^{7}  \GeV
    \, 
    \left(\frac{0.25}{\epsilon}\right)
    \left(\frac{f_a}{10^{9}~\rm GeV}\right)
    \left(\frac{g_{*}}{g_{\rm MSSM}}\right)^{\scalebox{0.9}{$\frac{3}{2}$}}
    \left(\frac{0.1}{c_{B}}\right)
    \left(\frac{0.03~\rm eV^{2}}{\bar{m}^{2}}\right) \,,
\end{align}
where in the second line we have used the prediction on $m_S$ in Eq.~(\ref{eq:mS_YB_quad_DM}) to obtain the observed baryon asymmetry from lepto-axiogenesis.
The required thermalization temperature is shown in Fig.~\ref{fig:quadratic_MD} by the black dashed lines. As we discuss in Sec.~\ref{sec:saxthermquad}, it is difficult to achieve the required $T_{\rm th}$ for large $f_a$, and the KMM underproduces axion DM. The vertical black dotted line shows the upper bound on $f_a$.

The saxion dominates the energy density of the universe at the temperature
\begin{align}
\label{eq:Tmatterdomination}
T_{M} \simeq  2 \times 10^{6} \GeV  
\left(\frac{g_{\rm MSSM}}{g_{*}}\right)^{\scalebox{0.9}{$\frac{1}{4}$} }
\left(\frac{m_S}{50\TeV}\right)^{\scalebox{0.9}{$\frac{1}{2}$} }
\left(\frac{S_{i}}{10^{16}\GeV}\right)^{2}.
\end{align}
We impose $T_{M} > T_{\rm th}$ so that the saxion actually dominates the energy density of the universe before it thermalizes. 
From Eqs.~(\ref{eq:Tthdil}) and (\ref{eq:Tmatterdomination}), the condition translates into a lower bound on $S_{i}$,
\begin{align}
\label{eq:Si_min_MM}
S_{i} \gtrsim  3 \times 10^{16} \GeV
    \, N_{\rm DW}^{\scalebox{0.8}{$\frac{1}{4}$}}
    \left(\frac{0.25}{\epsilon}\right)^{\scalebox{0.9}{$\frac{1}{2}$}}
    \left(\frac{f_a}{10^{9}~\rm GeV}\right)^{\scalebox{0.9}{$\frac{1}{2}$}}
    \left(\frac{g_{*}}{g_{\rm MSSM}}\right)^{\scalebox{0.9}{$\frac{1}{2}$}}
    \left(\frac{0.1}{c_{B}}\right)^{\scalebox{0.9}{$\frac{1}{4}$}} 
    \left(\frac{0.03~\rm eV^{2}}{\bar{m}^{2}}\right)^{\scalebox{0.9}{$\frac{1}{4}$}}   ,
\end{align}
as shown by the sloped part of the black dotted lines in Fig.~\ref{fig:quadratic_MD}, below which the minimum thermalization temperature consistent with the assumption of saxion domination, $T_{\rm th} = T_M$, is too small to reproduce axion dark matter from kinetic misalignment. However, the conventional misalignment mechanism explains axion dark matter at $f_a \simeq 10^{11} \GeV$, denoted by the vertical gray lines in Figs.~\ref{fig:quadratic} and \ref{fig:quadratic_MD}, and excludes higher $f_a$ due to overproduction. 

\subsection{Constraints on explicit PQ symmetry breaking}
\label{sec:exPQ_quadratic}
As in Sec.~\ref{sec:exPQ_quartic}, the explicit PQ-breaking potential may shift the axion vacuum to give too large of a neutron electric dipole moment. 
The contribution to the mass of the axion from the $A$-term PQ breaking potential is
\begin{align}
 \Delta m_a^2 \simeq \frac{n \epsilon \, m_S^{2}}{3} \left(\frac{N_{\rm DW} f_a}{S_i}\right)^{n-2} \,,
\end{align}
where we made use of Eq.~(\ref{eq:epsilon_quadratic}). We require $\Delta m_a^2$ to satisfy the bound in Eq.~(\ref{eq:deltama_max}). 

Similar to Eq.~(\ref{eq:Si_max_quartic}), we require $\epsilon <1$ to avoid large curvature in the phase direction that traps the saxion in one of the minima of the $A$-term potential, preventing, or substantially damping, rotations of $P$. Since $\epsilon \simeq A/m_S$ from Eq.~(\ref{eq:epsilon_quadratic}), $\epsilon <1$ is automatically satisfied for the natural scenario where $A \lesssim  m_S$, which is true as long as $m_S$ is not accidentally small. 

\subsection{Saxion thermalization}
\label{sec:saxthermquad}
Just as in the case of the quartic potential described in Sec.~\ref{sec:quartic_therm}, the saxion in the quadratic potential can also thermalize through gluons or fermions. The coupling with the Higgs exists in DFSZ models, but since we find that the scattering with gluons or fermions can be efficient enough, we do not consider Higgs here.
To prevent entropy production, as assumed in Eq.~(\ref{eq:yB_quad}) and Fig.~\ref{fig:quadratic}, the saxion needs to thermalize while the universe is still in radiation domination, i.e., $T_{\rm th} > T_{M}$ with $T_{M}$ given by Eq.~(\ref{eq:Tmatterdomination}).

We first consider scattering through gluons with a rate given by Eq.~(\ref{eq:gtherm}). Since $\Gamma_{g}$ decreases faster than Hubble for $T < T_S$, successful thermalization by gluons is possible only at $T > \max[T_{M},T_S]$. Using values of $m_S$ determined by $Y_B$ in Eq.~(\ref{eq:yB_quad}), gluons successfully thermalize saxions before matter domination if
\begin{align}
\label{eq:Sboundquadg}
S_{i} & \lesssim 
\begin{cases}
10^{16} \GeV \,
N_{\rm DW}^{\scalebox{0.8}{$\frac{1}{4}$}}
\left(\frac{b}{10^{-5}}\right)^{\scalebox{0.9}{$\frac{1}{6}$}}
\left(\frac{0.1}{c_B}\right)^{\scalebox{0.9}{$\frac{1}{12}$}} 
\left(\frac{0.03~\rm eV^{2}}{\bar{m}^{2}}\right)^{\scalebox{0.9}{$\frac{1}{12}$}}
\left(\frac{62}{- \log(\xi_5) + \log(-\log(\xi_5))}\right)^{\scalebox{0.9}{$\frac{1}{12}$}} \\
4\times 10^{16} \GeV \,
N_{\rm DW}^{\scalebox{0.8}{$\frac{1}{4
}$}} 
\left(\frac{b}{10^{-5}}\right)^{\scalebox{0.9}{$\frac{3}{2}$}}
\left(\frac{10^{9}\GeV}{f_{a}}\right)^{2}
\left(\frac{0.1}{c_B}\right)^{\scalebox{0.9}{$\frac{3}{4}$}} 
\left(\frac{0.03~\rm eV^{2}}{\bar{m}^{2}}\right)^{\scalebox{0.9}{$\frac{3}{4}$}}
\left(\frac{23}{\log(\xi_6) - \log(\log(\xi_6))}\right)^{\scalebox{0.9}{$\frac{3}{4}$}} ,
\end{cases}
\end{align} 
where the first (second) case is for $S_{M} > f_{a}$ ($S_{M} < f_{a}$) respectively and 
\begin{align}
\xi_{5} &= 6 \times 10^{-26}
    N_{\rm DW}^{-1}
    \left(\frac{b}{10^{-5}}\right)^{2}
    \left(\frac{0.1}{c_{B}}\right) 
    \left(\frac{0.03~\rm eV^{2}}{\bar{m}^{2}}\right) 
     \\
\xi_{6} &= 3\times 10^{11}
    N_{\rm DW}^{-7/3}
    \left(\frac{b}{10^{-5}}\right)^{2}
    \left(\frac{10^{9}\GeV}{f_{a}}\right)^{4}
    \left(\frac{0.1}{c_{B}}\right) 
    \left(\frac{0.03~\rm eV^{2}}{\bar{m}^{2}}\right) 
      \,.
\end{align}
This shows that gluons thermalize the saxion in some but not all of the relevant parameter space in Fig.~\ref{fig:quadratic}. Therefore, we continue to explore more efficient thermalization processes.

Thermalization can also proceed through fermions, which couple to the saxion through $W = z S\psi \bar{\psi}$; the thermalization rate by the scalar part of $\psi$ is smaller. The thermalization rate by the fermions is given by $\Gamma_\psi = b_\psi z^2 T$ with $b_\psi \simeq 0.1$.
To be consistent with the assumption that the saxion begins oscillation by the zero-temperature mass, we require that $z S_i > T_{\rm osc} $. This condition is easily satisfied for large $S_i$. We may instead assume that $z T_{\rm osc} < m_S$, but this leads to less efficient thermalization.

The fermions need to be in the thermal bath by $T_{\rm th}$, implying $z \leq T_{\rm th} / S(T_{\rm th})$. With this condition, the scattering rate for any $T_{\rm th} \ge T_{M}$ is at the most
\begin{align}
\label{eq:qthermTM}
\Gamma_\psi = b_\psi \frac{T_{\rm th}^{3}}{S^{2}(T_{\rm th})} = b_\psi \frac{T^{3}_{\rm osc}}{S^{2}_{i}}.
\end{align}
This maximal rate takes the same form as that of the gluon scattering in Eq.~(\ref{eq:gtherm}) but is four orders of magnitude larger with $b_{\psi} \simeq 0.1$. The constraint $T_{\rm th} \ge T_M$ is given by Eq.~(\ref{eq:Sboundquadg}) with $b$ replaced by $b_\psi$.

The first case of Eq.~(\ref{eq:Sboundquadg}) with $b$ set to  $b_{\psi} = 0.1$ leads to the orange hatched region in Fig.~\ref{fig:quadratic}. The second case of Eq.~(\ref{eq:Sboundquadg}) is obtained from the requirement $T_{\rm th} > T_S$, which is a condition stronger than necessary ($T_{\rm th} > T_M$ with $T_M < T_S$) but already shows the irrelevance of the constraint. The mass $m_\psi = z f_a$ must be above the TeV ($100 \GeV$) scale for colored (uncolored) $\psi$, which can be satisfied in the parameter space in Fig.~\ref{fig:quadratic}.

In Fig.~\ref{fig:quadratic_MD}, it is assumed that the saxion thermalizes after matter domination. In this case, even though the baryon asymmetry is independent of the thermalization temperature as discussed around Eq.~(\ref{eq:YB_quadratic_MD}), the DM density depends on the thermalization temperature as shown in Eqs.~(\ref{eq:Ya_PR_quad_MD}) and (\ref{eq:ya_KMM_quad_MD}). To explain the thermalization temperature necessary for dark matter, we consider thermalization by the fermion scattering from $W= z P \psi \bar\psi$ with a rate given by $\Gamma_{\psi} = b_\psi z^{2} T$.%
\footnote{Since the baryon asymmetry is dominantly produced at $T_{\rm th}$, to be consistent with the assumption of~\ref{model:KSVZ_heavy}, $\psi \bar{\psi}$ should not be the KSVZ quark. We may still identify $\psi \bar{\psi}$ with the KSVZ quark and instead consider~\ref{model:KSVZ_light}. As long as the transfer of the charge asymmetry of $P$ into $\psi$ is effective, Eq.~(\ref{eq:YB_quadratic_MD}) is applicable.}
The required $T_{\rm th}$ in Eq.~(\ref{eq:Tthdil}) is obtained for
\begin{align}
\label{eq:z}
z = 7 \times 10^{-5}N_{\rm DW}^{\scalebox{0.8}{$-\frac{1}{2}$}}
\left(\frac{0.1}{b_{\psi}}\right)^{\scalebox{0.9}{$\frac{1}{2}$}}
\left(\frac{0.25}{\epsilon}\right)^{\scalebox{0.9}{$\frac{1}{2}$}}
\left(\frac{0.1}{c_{B}}\right)^{\scalebox{0.9}{$\frac{1}{2}$}}
\left(\frac{0.03~\rm eV^{2}}{\bar{m}^{2}}\right)^{\scalebox{0.9}{$\frac{1}{2}$}}
\left(\frac{g_{*}}{g_{\rm MSSM}}\right) 
\left(\frac{f_{a}}{10^{9}\GeV}\right)^{\scalebox{0.9}{$\frac{1}{2}$}}.
\end{align}

Several constraints on $z$ are in order. We require that $N_{\rm DW}f_{a} \le S_{\rm th} \le S_{M}$ to be consistent with Eq.~(\ref{eq:YB_quadratic_MD}), giving lower and upper bounds. An upper bound on $z$ arises from requiring the fermions to be in the thermal equilibrium at $T_{\rm th}$, i.e., $z S_{\rm th} < T_{\rm th}$. Collider bounds on the fermion mass requires $z f_{a} \gtrsim$ TeV (100 GeV) scale for colored (uncolored) $\psi$. An earlier assumption that the saxion oscillates because of the vacuum mass rather than the thermal mass requires that $z T_{\rm osc} < m_S$ or $z S_i \ge T_{\rm osc}$ so that the thermal mass is small or the fermions are not in the thermal bath when oscillations begin. We find that these constraints are satisfied in the parameter space enclosed by the black dotted lines in Fig.~\ref{fig:quadratic_MD}. The lower edge was previously  discussed in Eq.~(\ref{eq:Si_min_MM}) and the right edge is set by the condition $z S_{\rm th} < T_{\rm th}$. Therefore, an appropriate choice of $z$ will allow $T_{\rm th}$ to take the values required for dark matter shown in Fig.~\ref{fig:quadratic_MD}.

\subsection{Domain wall problem}

For the rotation in the logarithmic potential (\ref{eq:dim_trans}) with $\epsilon=\mathcal{O}(1)$, the growth rate of the fluctuations is about $10^{-2}m_S$. The amplitude of the fluctuations becomes as large as $\bar{S}$ when $H \sim 10^{-3} m_S$, restoring the PQ symmetry. The PQ symmetry is broken again once $\bar{S} < f_a$. Unless the domain wall number is unity, stable domain walls are produced around the QCD phase transition. This is avoided if the thermalization of $P$ occurs before the amplitude of the fluctuation becomes as large as $\bar{S}$. This is because, after thermalization, the rotation is circular and parametric resonance is absent. This is possible if
\begin{align}
    S_i \lesssim 8\times 10^{14}\GeV 
    \left(\frac{b}{0.1}\right)^{\scalebox{0.9}{$\frac{1}{2}$}}
    \left(\frac{g_{\rm MSSM}}{g_*}\right)^{\scalebox{0.9}{$\frac{3}{8}$}}
    \left(\frac{m_S}{10^5\GeV}\right)^{\scalebox{0.9}{$\frac{1}{4}$}} .
\end{align}
For the rotation in the potential (\ref{eq:two_field}) with $\epsilon= \mathcal{O}(1)$, parametric resonance does not occur for $\bar{S} \gtrsim 10^2 f_a$. Once $\bar{S}\lesssim 10^2 f_a$, a resonant band appears and parametric resonance quickly grows the fluctuations to be as large as $\bar{S}$. If the field value becomes random, this leads to the formation of domain walls.
Since the PQ symmetry is not restored, those domain walls do not have boundaries and are stable even if the domain wall number is unity. Ref.~\cite{Kawasaki:2017kkr} investigates if the field value becomes random by a lattice computation for $\epsilon = 0$, and finds that the field value does not become random for $S_i < 10^3 f_a$ because the gradient term tries to homogenize the field value. Although Ref.~\cite{Kawasaki:2017kkr} could not continue the computation to $S_i > 10^3 f_a$, parametric resonance for $S_i < 10^3f_a$ is already strong enough to make the fluctuations as large so $\bar{S}$, so it is expected that their result holds for $S_i > 10^3 f_a$. Our case is different from theirs since we have $\epsilon \neq 0$. Although parametric resonance is weaker for $\epsilon \neq 0$, the rotating background may make it easier for the angular direction to be randomized. Also, if kinetic misalignment occurs, the axion starts oscillations from near the hilltop. Even small fluctuations may set the axion to fall into different minima, producing domain walls. We thus provide a condition such that thermalization occurs before parametric resonance becomes effective so that the domain wall problem is for sure avoided. The PQ symmetry breaking field is thermalized before saxion domination and before $\bar{S}$ reaches $10^2 f_a$ if
\begin{align}
    S_i \lesssim 9 \times 10^{14}\GeV 
    \left(\frac{b}{0.1}\right)^{\scalebox{0.9}{$\frac{3}{2}$}}
    \left(\frac{g_{\rm MSSM}}{g_*}\right)^{\scalebox{0.9}{$\frac{9}{8}$}}
    \left(\frac{m_S}{10^5\GeV}\right)^{\scalebox{0.9}{$\frac{3}{4}$}} 
    \left(\frac{10^{11} \GeV}{f_a}\right)^2,
\end{align}
which applies when $S_M < 10^2 f_a$. The constraint becomes irrelevant when $S_M > 10^2 f_a$ in the parameter space in Fig.~\ref{fig:quadratic} so the constraint is not explicitly written here. If the saxion instead dominates before thermalization occurs, the constraint from thermalization before $\bar{S}$ reaches $10^2 f_a$ is 
\begin{align}
    f_a \lesssim 10^{11} \GeV
    \left(\frac{b}{0.1}\right)^{\scalebox{0.9}{$\frac{2}{3}$}}
    \left(\frac{m_S}{10^5\GeV}\right)^{\scalebox{0.9}{$\frac{1}{3}$}} ,
\end{align}
which is not stringent in Fig.~\ref{fig:quadratic_MD}. Since we assume $T_M > T_{\rm th}$ in Fig.~\ref{fig:quadratic_MD} and now impose $S (T_{\rm th}) > 10^2 f_a$, the region consistent with both of these conditions, i.e., $S_M > 10^2 f_a$, is above the boundary obtained from rescaling the orange boundary, corresponding to $S_M > f_a$, up in $S_i$ by a factor of $10^{1/2} \simeq 3$.

\subsection{Oscillation by a thermal mass and TeV scale SUSY}

So far we consider the case where the PQ symmetry breaking field $P$ rotates in a zero-temperature potential, and find that a large saxion mass $\gtrsim 10$ TeV is required. There may be an era where $P$ rotates in a thermal potential with a mass larger than the zero-temperature mass, so that a lower saxion mass $\sim$ TeV is consistent with lepto-axiogenesis.

Let us, for example, consider a coupling between $P$ and $Q\bar{Q}$, $W = y_Q P Q\bar{Q}$. If $y_Q S_i$ is smaller than the temperature, $Q\bar{Q}$ is in the thermal bath and gives a thermal mass term $y_Q^2 T^2 |P|^2$. This initiates the oscillations of $P$ if
\begin{align}
    y_Q > 8\times 10^{-8} \left(\frac{m_S}{1 \TeV}\right)^{1/2} \left(\frac{g_*}{g_{\rm MSSM}}\right)^{\scalebox{0.9}{$\frac{1}{4}$}}
\end{align}
at a temperature
\begin{align}
\label{eq:Tosc_TeV}
    T_{\rm osc} = 2\times 10^{11}\GeV \left( \frac{y_Q}{10^{-6}} \right) \left(\frac{g_{\rm MSSM}}{g_*}\right)^{\scalebox{0.9}{$\frac{1}{2}$}} 
\end{align}
with an initial thermal mass
\begin{align}
    m_{T,i}= y_QT_{\rm osc} = 10^{5}\GeV \left(\frac{y_Q}{10^{-6}}\right)^2\left(\frac{g_{\rm MSSM}}{g_*}\right)^{\scalebox{0.9}{$\frac{1}{2}$}}.
\end{align}
The thermal mass decreases in proportion to $R^{-1}$, so the $B-L$ asymmetry is dominantly produced at the beginning of the rotation. To explain the observed baryon asymmetry, the initial thermal mass must be as large as in Eq.~(\ref{eq:mS_YB_quad_DM}), which requires
\begin{align}
\label{eq:yQ_TeV}
    y_Q = 2\times 10^{-6} \times \, N_{\rm DW}^{-1/2} \left(\frac{g_{*}}{g_{\rm MSSM}}\right)
    \left(\frac{0.1}{c_{B}}\right)^{{\scalebox{0.9}{$\frac{1}{2}$}}}
    \left(\frac{0.03~\rm eV^{2}}{\bar{m}^{2}}\right)^{\scalebox{0.9}{$\frac{1}{2}$}}.
\end{align}
The parameter $\epsilon$ is suppressed since $A\sim m_S \ll m_{T,i}$,
\begin{align}
    \epsilon \simeq 10^{-3} \left( \frac{m_S}{1\TeV} \right) \left(\frac{2\times 10^{-6}}{y_Q} \right)^2.
\end{align}

In order for the estimation of the baryon asymmetry to be correct, the asymmetry of $P$ charge must be efficiently transferred into $Q \bar{Q}$ asymmetry. The transfer rate is $y_Q^2 |P|^2/T$. For this case, as is shown in Appendix~\ref{app:ave_thetadot}, the transfer must be efficient when $P$ is closest to the origin where $S \sim \epsilon S_i$, because this is when the contribution to $\vev{\dot{\theta}}$ is dominated.  This is impossible since 1) $S\sim \epsilon S_i$ only for a short time scale $\epsilon / m_{T,i}$ and 2) the transfer rate $\sim y_Q^2 m_Q^2/T_{\rm osc}$ is suppressed by small $\epsilon$ when $S\sim \epsilon S_i$. The observed baryon asymmetry cannot be reproduced because of those suppressions.
Increasing $m_{T,i}$ seems to solve the problem by compensating the suppression with a larger $\vev{\dot{\theta}}$, but that makes $\epsilon$ even smaller and suppresses the baryon asymmetry.

To remedy this difficulty, we may add a larger coupling to another pair of fields $y_\psi P \psi \bar{\psi}$. If $y_\psi S_i > T_{\rm osc}$, $\psi\bar{\psi}$ are not in the thermal bath and do not give a thermal mass to $P$. A possible thermal log potential~\cite{Anisimov:2000wx} does not initiate the oscillation if $S_i \gtrsim 10^{16}$ GeV. This term can, however, be effective in transferring the asymmetry of $P$ into that of $\psi \bar{\psi}$.
We require that $\psi\bar{\psi}$ are in the thermal bath when $S$ passes nearest to the origin, $y_\psi \epsilon S_i < T_{\rm osc}$. Once this condition is satisfied, when $P$ passes near the origin, the potential of $P$ is dominated by the thermal potential generated by $\psi \bar{\psi}$. Then from charge and energy conservation, the minimum $S$ and the maximum $\dot{\theta}$ during the rotation are given by
\begin{align}
    S_{\rm min} & \simeq \epsilon S_i \times \frac{m_{T,i}S_i}{T_{\rm osc}^2}, \nonumber \\
    \dot{\theta}_{\rm max} & \simeq \frac{m_{T,i}}{\epsilon} \times \frac{T_{\rm osc}^4}{m_{T,i}^2 S_i^2}.
\end{align}
We have $\dot{\theta}\sim \dot{\theta}_{\rm max}$ when $S\sim S_{\rm min}$ for a time scale $\dot{\theta}_{\rm max}^{-1}$.
An efficient transfer of the $P$ asymmetry into $\psi\bar{\psi}$ asymmetry requires $0.1 (y_\psi S_{\rm min})^2 / T_{\rm osc} > \dot{\theta}_{\rm max}$. The bounds on $y_{\psi}$ are summarized as
\begin{align}
  y_{\psi} &> 6\times 10^{-4}\times \left(\frac{1\TeV}{m_S}\right)^{\scalebox{0.9}{$\frac{3}{2}$}} \left( \frac{y_Q}{2\times 10^{-6}} \right)^{\scalebox{0.9}{$\frac{9}{2}$}} \left( \frac{10^{17}\GeV}{S_i} \right)^{3} \left(\frac{g_{\rm MSSM}}{g_*}\right)^{\scalebox{0.9}{$\frac{9}{4}$}} \nonumber \\ 
  y_{\psi} &< 2\times 10^{-3} \times \left( \frac{1\TeV}{m_S} \right)  \left( \frac{y_Q}{2\times 10^{-6}} \right)^3 \left( \frac{10^{17}\GeV}{S_i} \right)  \left( \frac{g_{\rm MSSM}}{g_*} \right) .
\end{align}
The asymmetry of $\psi \bar{\psi}$ needs to be transferred into $n_{\ell,H_u}$. This can be done by the sphaleron processes and/or a direct coupling between $\psi$ or $\bar{\psi}$ with MSSM fields. Unlike the $|P|$-dependent transfer rate, the rate of these transfers only has to be larger than $H$ in order for the time-average of $n_{\ell,H_u}$ to reach $\vev{\dot{\theta}}T^2$.

From Eqs.~(\ref{eq:Tosc_TeV}) and (\ref{eq:yQ_TeV}), one can see that the reheat temperature after inflation must be above $10^{11}$ GeV. If the gravitino mass is also ${\cal O}(1)$ TeV, the late-time decay of the gravitinos produced around reheating is excluded by BBN. The gravitino mass must be above the $100$ TeV scale, which requires sequestering~\cite{Randall:1998uk}. $R$-parity violation is also required to avoid the LSP overproduction from the gravitinos.

\section{Conclusions and discussion}
\label{sec:con}

In this paper, we introduced and studied the generation of the baryon asymmetry of the universe from rotation of the PQ symmetry breaking field and the dimension-five interaction $\ell \ell H^\dag H^\dag$-- a mechanism we call lepto-axiogenesis. The rotation of the PQ symmetry breaking field corresponds to the charge asymmetry of $P$, which is transferred into the asymmetry of the Higgs boson $H$ and/or the lepton chirality through Yukawa couplings and the electroweak sphalerons. These asymmetries are transferred via the dimension-five interaction, into a lepton asymmetry, which is transferred into a baryon asymmetry through the electroweak sphaleron process. The schematic diagram in Fig.~\ref{fig:asym_leptoaxio} shows the possible routes of the asymmetries.

The rotation of the PQ symmetry breaking field is driven by a mechanism analogous to the Affleck-Dine mechanism. If the radial direction of the PQ symmetry breaking field, the saxion, takes a large field value in the early universe, explicit PQ symmetry breaking by a higher-dimensional operator may be effective, driving the rotation.
The rotation also produces QCD axion dark matter through parametric resonance and/or kinetic misalignment for $f_a \sim 10^{10}$ GeV. The conventional misalignment mechanism and the axion emission from strings and domain walls explains QCD axion dark matter for $f_a \sim 10^{11} \GeV$. The contribution from parametric resonance may be warm enough to affect structure formation in the universe at an observable level. Larger $f_a$ requires entropy production after the QCD phase transition.

We investigated quartic and quadratic potentials of the PQ symmetry breaking field, and found that the observed baryon asymmetry can be produced by lepto-axiogenesis without producing too much QCD axion dark matter, a major obstacle faced by axiogenesis.

In the case of a quartic potential, $Y_B$ is proportional to $M_{\rm Pl}^2 \Sigma m_\nu^2/v_{\rm EW}^4$ and also to a power of $m_S(Si)/M_{\rm Pl}$.  The latter gives sensitivity to UV physics; nevertheless various constraints on the scheme narrow the allowed ranges of $f_a$ and $m_S$ as shown in Figs.~\ref{fig:quartic} and \ref{fig:quartic_lowTRH}. Even with these constraints, baryogenesis is successful in the large unshaded regions; however, successful cogenesis of the axion dark matter abundance greatly limits the parameter space.

Fig.~\ref{fig:quartic} assumes a large enough reheat temperature such that the rotation begins during a radiation dominated era. 
When the axions produced by parametric resonance are not thermalized and remain as (a part of) dark matter, warmness constraints exclude the region above the brown dashed lines. The observed dark matter abundance is explained by parametric resonance or kinetic misalignment mechanism on the black lines. The former is allowed only on the solid black lines because of the warmness constraints. Most of the solid black line is excluded by the constraint from the saxion emission at supernovae, which can be evaded by the mixing between the saxion and the Higgs which traps the saxion inside the supernova cores. Such a large mixing can be probed by rare decays of kaons.

The rotation of the PQ symmetry breaking field also involves a radial motion, which should be thermalized to avoid the over-closure by the radial mode. We investigated possible thermalization processes and found that thermalization can successfully occur. If the saxion couples to SM particles only through gauge bosons, the thermalized saxion dominantly decays into axions, producing dark radiation with an amount shown in Eq.~(\ref{eq:DNeff}). If the saxion couples to the Higgs, the amount of the dark radiation is suppressed.

Fig.~\ref{fig:quartic_lowTRH} assumes a low reheat temperature such that the rotation begins before the completion of inflationary reheating. A reheat temperature as low as $10^9$ GeV is compatible with lepto-axiogenesis. The abundance of the QCD axions produced by parametric resonance and kinetic misalignment is too small, and $f_a \sim 10^{11} \GeV$ is required to explain the observed dark matter abundance by the QCD axion. 

The model with a quadratic potential is realized in supersymmetric theories. The constraints on the parameter space are summarized in Figs.~\ref{fig:quadratic} and \ref{fig:quadratic_MD}. Again, baryogenesis is successful over a wide range of $f_a$ and for $S_i \sim 10^{15} - 10^{18} \GeV$. Simultaneous production of axion dark matter via parametric resonance or kinetic misalignment occurs for lower values of $f_a$, or via conventional misalignment and the axion emission from strings and domain walls for $f_a \sim  10^{11} \GeV$.  

Since the angular velocity, $\dot{\theta}$, is constant in time, the baryon asymmetry produced per Hubble time is independent of temperature during radiation dominated eras. As a result, the baryon asymmetry is not sensitive to the cosmological evolution such as reheating, and is essentially determined by the mass of the PQ symmetry breaking field and hence by the soft mass scale. Indeed, the parametric behavior of the baryon asymmetry is given by
\begin{align}
\label{eq:YBpara}
    Y_B \sim   \left( \frac{m_S \, M_{\rm Pl} \; \Sigma m_\nu^2}{v_{\rm EW}^4} \right),
\end{align}
demonstrating insensitivity to initial conditions and the details of UV physics.  The observed baryon asymmetry is reproduced if the soft mass scale is $\mathcal{O}(10-10^{4})$ TeV, where the range reflects the difference in the numerical coefficient in Eq.~(\ref{eq:YBpara}) for saxion thermalization in the radiation and matter dominated eras. This naturally fits into scenarios with high scale supersymmetry, a well-motivated framework to address the observed Higgs mass, the Polonyi problem, the gravitino problem, and the SUSY flavor/CP problems. Alternatively, we have shown that, if extra fermions are coupled to $P$, axion rotations may occur in a thermal potential, allowing $m_S$ to be decreased to the TeV scale.
It is remarkable that, even though the baryon asymmetry is generated at a very high temperature, in minimal models the result depends only on parameters associated with much lower energies; contrasting, for example, with grand unification, leptogenesis and Affleck-Dine schemes.

\vspace{1cm}

\noindent
{\it Note added.}
While finalizing our manuscript, an interesting paper appeared on the arXiv~\cite{Domcke:2020kcp}. This has an overlap with our paper as it discusses the production of $B-L$ asymmetry from a non-zero velocity of a generic axion field and the dimension-five interaction. However, there is no further overlap as the physical setup they present is for a very heavy oscillating axion rather than a circulating QCD axion initiated by a higher-dimensional operator. Although they compute the baryon asymmetry arising from a constant $\dot{\theta} /T$, the physical origin is not specified and they do not view it as realistic. As we have seen in this paper, initiation of the velocity by a higher-dimensional operator involves rich physics such as dominant production of the baryon asymmetry at a low temperature much after the dimension-five interaction freezes out, kinetic misalignment, parametric resonance, thermalization of the radial mode, domain wall production, and oscillations or even rapid changes of $\dot{\theta}$ as discussed in Appendix~\ref{app:ave_thetadot}.

\section*{Acknowledgments}
The work was supported in part by the DoE Early Career Grant DE-SC0019225 (R.C.) and DE-SC0017840 (N.F), the DoE grants DE-AC02-05CH11231 (L.H.) and DE-SC0009988 (K.H.), the NSF grant NSF-1915314 (L.H.), NSF CAREER grant PHY-1915852 (A.G.)  as well as the Raymond and Beverly Sackler Foundation Fund (K.H.).
\appendix

\section{Models of lepto-axiogenesis}
\label{app:models}

The UV completion of the QCD axion model requires new particles charged under the PQ symmetry, such as heavy quarks in KSVZ models~\cite{Kim:1979if,Shifman:1979if} or two Higgs doublets in DFSZ models~\cite{Zhitnitsky:1980tq,Dine:1981rt}. In lepto-axiogenesis, the transfer of the PQ asymmetry into that of $B-L$ can be both qualitatively and quantitatively different depending on whether these new states are in thermal equilibrium. We discuss the case of heavy KSVZ quarks in \ref{model:KSVZ_heavy} in Sec.~\ref{sec:lepto-axio} and the other three possibilities below.

\begin{enumerate}[label=\textbf{KSVZ-light},ref=\underline{KSVZ-light}, wide, labelindent=0pt]
\item KSVZ model with light quarks \label{model:KSVZ_light}

When the KSVZ quarks $Q$ are in the thermal bath, the asymmetry of $P$ is first translated into the chiral asymmetry of $Q$ with a rate $\alpha_3 m_Q(f_{\rm eff})^2/T$, where $m_Q(f_{\rm eff})$ is the mass of $Q$. The chiral asymmetry of $Q$ is then translated into the quark and/or lepton chiral asymmetries via strong and electroweak sphaleron transitions.
The $B-L$ asymmetry produced by the neutrino mass interaction per Hubble time is
\begin{align}
\label{eq:nB-L_light}
\Delta n_{B-L} = \frac{\Gamma_L}{H} \times c_{B-L} \dot{\theta}T ^2 \times \min \left( 1, \frac{\Gamma_{\rm ss}}{H} \right) \times \min \left(1, \frac{\alpha_3 y_{t}^2 T}{H}\right)\times \min \left(1, \frac{\alpha_3 m_Q(f_{\rm eff})^2/T}{H}\right).
\end{align}
When the production of the chiral asymmetry of $Q$ is efficient, the last factor is unity, so that the production of $B-L$ asymmetry is also efficient, as in the case~\ref{model:KSVZ_heavy} with $Q$ heavy, so that (\ref{eq:nB-L_light}) reduces to (\ref{eq:nB-L}).

With $Q$ light, $B-L$ production does not necessarily need the strong sphaleron transition. If the quarks $Q$ have the same gauge quantum numbers as the Standard Model fermions, direct couplings between $Q$ and the SM fermions and Higgs can transfer their asymmetries. For example, if $Q=U, \bar{U}$ with $\bar{U}$ having the same gauge interactions as the right-handed up quark, we may introduce
\begin{align}
{\cal L} = y_U H^\dag q \bar{U}~~{\rm or}~~ m_{U\bar{u}} \; U\bar{u},
\end{align}
where $q$ is the quark doublet, and $\bar{u}$ is the singlet up quark. Then the chiral asymmetry of $U,\bar{U}$ is converted into the asymmetry of $q$ and $H$ or $\bar{u}$ with a rate $\alpha_3 y_U^2 T$ or $\alpha_3 m_{U\bar{u}}^2/T$. These asymmetries then create the $B-L$ asymmetry via SM interactions and the Majorana neutrino mass operator. 
The $B-L$ asymmetry produced per Hubble time via this flow is
\begin{align}
\label{eq:nB-L_light2}
\Delta n_{B-L} = \frac{\Gamma_L}{H} \times c_{B-L} \dot{\theta}T ^2 \times \min \left(1, \frac{\alpha_3 m_Q(f_{\rm eff})^2 T}{H}\right) \times
\begin{cases}
  \min \left(1, \frac{\alpha_3 y_{U}^2 T}{H}\right) \\
  \min \left(1, \frac{\alpha_3 m_{U\bar{u}}^2/ T}{H}\right)
\end{cases} .
\end{align}
The production in Eq.~(\ref{eq:nB-L_light}) is also present.
\end{enumerate}

\begin{enumerate}[label=\textbf{DFSZ-light},ref=\underline{DFSZ-light}, wide, labelindent=0pt]
\item DFSZ model with light Higgses \label{model:DFSZ_light}

\noindent

We next consider the DFSZ model with two Higgs doublets $H_u$ and $H_d$, whose vacuum expectation values give masses to up-type and down-type quarks, respectively. The PQ breaking field is coupled to these Higgs bosons by an interaction $P^m H_u H_d$.

We first assume that the masses of $H_u$ and $H_d$ are smaller than the temperature. Then the PQ charge asymmetry is transferred into those of $H_u$ and $H_d$ with a rate $\alpha_2(B\mu)^2/T^3$, where $B\mu$ is the quadratic term $B\mu H_uH_d$ given by the PQ symmetry breaking field. In supersymmetric theory, the PQ charge asymmetry is more efficiently transferred into the chiral asymmetry of Higgsinos $\tilde{H}_u$ and $\tilde{H}_d$ with a rate $\alpha_2 \mu^2 / T$. The chiral asymmetry of $H_u$ or $\tilde{H}_u$ is transferred into the $B-L$ asymmetry by $\ell\ell H_u^\dagger H_u^\dagger$.
The $B-L$ asymmetry produced per Hubble time is
\begin{align}
\label{eq:nB-L_DFSZ}
\Delta n_{B-L} = \frac{\Gamma_L}{H} \times c_{B-L} \dot{\theta}T ^2 \times
\begin{cases}
\min \left(1, \frac{\alpha_2 \left(B\mu \left(f_{\rm eff}\right) \right)^2/T^3}{H}\right) & :\text{non-SUSY} \\
\min \left(1, \frac{\alpha_2 \left(\mu \left(f_{\rm eff}\right) \right)^2/T}{H}\right) & :\text{SUSY}
\end{cases}.
\end{align}
Note that neither the strong sphaleron nor the Standard Model Yukawa interactions are necessary. If the production of the Higgs asymmetry is in thermal equilibrium, the production of $n_{B-L}$ is as efficient as the case \ref{model:KSVZ_heavy} so that (\ref{eq:nB-L_DFSZ}) reduces to (\ref{eq:nB-L}), with both strong sphaleron and Yukawa processes taken in thermal equilibrium.

\end{enumerate}

\begin{enumerate}[label=\textbf{DFSZ-heavy},ref=\underline{DFSZ-heavy}, wide, labelindent=0pt]
\item DFSZ model with heavy Higgses \label{model:DFSZ_heavy}

When the PQ symmetry breaking field $P$ gives a mass to $H_u H_d$ that is larger than the temperature, they are no longer in the thermal bath and the scattering by $\ell\ell H_u^\dagger H_u^\dagger$ is ineffective. The $B-L$ asymmetry can be still produced via the operators $\ell \ell (Q^\dag \bar{u}^\dag)^2$ and $\ell \ell (Q \bar{d})^2$ which arises after integrating out $H_u$ and $H_d$.

Depending on the couplings between $H_{u,d}$ and $P$, large field values for $H_{u,d}$ may develop in the early universe. Then $H_{u,d}$ also rotate as $P$ rotates. If the neutrino Majorana mass during this era $\vev{H_u}^2/M$ is smaller than $T$, a $B-L$ asymmetry is created directly from the rotating phase of $H_u$ via the operator $\ell\ell H_u^\dagger H_u^\dagger/M$.
\end{enumerate}

\section{Boltzmann equations}
\label{app:Boltz_eq}

In this appendix, we show the Boltzmann equations for the particle-antiparticle asymmetry $n_X - n_{\bar{X}}$, which we simply denote as $n_X$.
For an illustration, we consider only one generation in the Standard Model and the result is similar for three generations.

The Yukawa interactions and the neutrino mass operator are
\begin{align}
{\cal L} = y_u H^\dag q \bar{u} + y_d H q \bar{d} + y_e H \ell \bar{e} + \frac{1}{M}  \ell_i \ell_j H^\dag H^\dag. 
\end{align}
The Boltzmann equations for the asymmetries are given by
\begin{align}
\label{eq:BoltzEqns}
\dot{n}_q = & \, \alpha_3 y_u^2 T \left( - \frac{n_q}{6} - \frac{n_{\bar{u}}}{3} + \frac{n_H}{4} \right) + \alpha_3 y_d^2 T \left( - \frac{n_q}{6} - \frac{n_{\bar{d}}}{3} -  \frac{n_H}{4} \right) \nonumber \\
& \hspace{0.15 in} +  3 \Gamma_{\rm ws} \Big(-n_q - n_\ell -  \frac{c_W}{3} \dot{\theta} T^2  \Big) +  2 \Gamma_{\rm ss} \left( -   n_q -n_{\bar{u}}-n_{\bar{d}} - \frac{1}{2} \dot{\theta} T^2 \right), \nonumber \\
\dot{n}_{\bar{u}} = & \,\alpha_3 y_u^2 T \left( - \frac{n_q}{6} - \frac{n_{\bar{u}}}{3} + \frac{n_H}{4} \right) + \Gamma_{\rm ss} \left( -  n_q -n_{\bar{u}}-n_{\bar{d}} - \frac{1}{2} \dot{\theta} T^2 \right), \nonumber \\
\dot{n}_{\bar{d}} = & \,\alpha_3 y_d^2 T  \left( - \frac{n_q}{6} - \frac{n_{\bar{d}}}{3} -  \frac{n_H}{4} \right) + \Gamma_{\rm ss} \left( -  n_q -n_{\bar{u}}-n_{\bar{d}} - \frac{1}{2} \dot{\theta} T^2 \right), \\
\dot{n}_\ell = & \, \alpha_2 y_e^2 T \left( - \frac{n_\ell}{2} - n_{\bar{e}} -  \frac{n_H}{4} \right) +  \Gamma_{\rm ws} \Big(- n_q - n_\ell -  \frac{c_W}{3} \dot{\theta} T^2 \Big) +  2 \Gamma_L \left(-\frac{n_{\ell}}{2}+\frac{n_{H}}{4}\right) , \nonumber \\
\dot{n}_{\bar{e}} = & \,  \alpha_2 y_e^2 T \left( - \frac{n_\ell}{2} - n_{\bar{e}} -  \frac{n_H}{4} \right), \nonumber \\
\dot{n}_H = & \, - \alpha_3 y_u^2 T \left( - \frac{n_q}{6} - \frac{n_{\bar{u}}}{3} + \frac{n_H}{4} \right) + \alpha_3  y_d^2 T  \left( - \frac{n_q}{6} - \frac{n_{\bar{d}}}{3} -  \frac{n_H}{4} \right) \nonumber \\
& +  \alpha_2 y_e^2 T \left( - \frac{n_\ell}{2} - n_{\bar{e}} -  \frac{n_H}{4} \right) - 2  \Gamma_L \left(-\frac{n_{\ell}}{2}+\frac{n_{H}}{4}\right), \nonumber \\
\dot{n}_{\rm PQ} = & \,\Gamma_{\rm ss} \left( - n_q -n_{\bar{u}}-n_{\bar{d}} - \frac{1}{2} \dot{\theta} T^2 \right) + c_W \Gamma_{\rm ws} \Big(-n_q - n_\ell -  \frac{c_W}{3} \dot{\theta} T^2  \Big). \nonumber
\end{align}

If all interactions are in thermal equilibrium, by taking $\dot{n}_i = 0$ except for $n_{\rm PQ}$ and imposing the hypercharge conservation, we find the equilibrium $B-L$ number density
\begin{align}
\label{eq:nB_eq}
n_{B-L}^{\rm eq} =
\begin{cases}
\frac{7-4c_W}{26} \dot{\theta} T^2 & y_d^2 \ll y_u^2 \\
\frac{- 1 - 4 c_W}{26} \dot{\theta} T^2 & y_u^2 \ll y_d^2
\end{cases} .
\end{align}

At low temperatures, the lepton number violating interaction $\Gamma_L$ is out of thermal equilibrium. The production of $B-L$ can be obtained by considering $\Gamma_L$ as a perturbation as in Eq.~(\ref{eq:B-L_dot}). After imposing the hypercharge and the $B-L$ conservation laws, we find the equilibrium densities of $n_\ell$ and $n_H$ and obtain
\begin{align}
\label{eq:nB_pert}
\dot{n}_{B-L} = \Gamma_L \left(n_\ell - \frac{n_H}{2} \right) = \Gamma_L \times
\begin{cases}
\frac{7-4 c_W}{35} \dot{\theta} T^2 & y_d^2 \ll y_u^2  \\ 
\frac{-1-4c_W}{35} \dot{\theta} T^2 & y_u^2 \ll y_d^2
\end{cases} 
\ = \frac{26}{35} \Gamma_L \, n_{B-L}^{\rm eq}  .
\end{align} 
The results in the large $y_u$ limit of Eqs.~(\ref{eq:nB_eq}) and (\ref{eq:nB_pert}) are listed in the first two rows of Table~\ref{tab:cB_L}, where $n_{B-L} = c_{B-L} \dot\theta T^2$ is understood as opposed to Eq.~(\ref{eq:n_B-L_dot}) when $\Gamma_L > H$. The computation of $c_{B-L}$ for other sets of interactions in equilibrium can be obtained in a similar manner. Starting from the full Boltzmann equations in Eq.~(\ref{eq:BoltzEqns}), one drops the interaction terms that are out of equilibrium and includes the conservation laws that are restored due to the inactive interactions. For example, when the electroweak sphaleron process is inactive, $n_{B+L}$ becomes conserved. On the other hand, if the Yukawa interaction involving $y_e$ is inactive, then the right-handed electron number $n_{\bar{e}}$ is conserved. Generically speaking, $c_{B - L} \simeq 0.1-0.2$ along with Eq.~(\ref{eq:c_B}) motivates our choice of $c_B = 0.05-0.1$ throughout the paper.

\section{Kinetic misalignment mechanism}
\label{app:KMM}
For large values of $\dot{\theta}$ the axion can account for the observed dark matter relic abundance via the kinetic misalignment mechanism~\cite{Co:2019jts}. Let us consider the axion potential provided by non-perturbative QCD effects
\begin{align}
V = m_{a}^{2}(T)f_{a}^{2}\left(1-\cos \frac{a}{f_{a}} \right) \,,
\end{align}
where the temperature dependence of the axion mass $m_a(T)$ can be obtained from the dilute instanton gas approximation~\cite{Gross:1980br}
\begin{align}
m_a(T) \simeq m_a(0)  \left(\frac{\Lambda_{\rm QCD}}{T} \right)^{4} \,.
\end{align}
At high temperatures $T\gg \Lambda_{\rm QCD}\simeq 150 \, \rm{MeV}$ the axion is basically massless, and for small $\dot{\theta}$ the axion field is essentially frozen due to Hubble friction. However, when the mass term becomes comparable to the Hubble friction term, the axion field  begins to oscillate around the minimum of its potential. The temperature when the axion field starts oscillating, $T_{*}$, is set by $3  H(T_{*})=m_{a}(T_{*})$, leading to the conventional misalignment mechanism. On the other hand, for large values of $\dot{\theta}$ at $T_{*}$, the kinetic energy may be larger that the potential energy delaying the initiation of the axion field oscillation leading to the KMM. The boundary between kinetic and misalignment mechanism can be define as the minimum value of $\dot{\theta}$ needed such that the kinetic energy is equal to the height of the potential at $T_{*}$, i.e.,  $\dot{\theta}^{2}_{\rm{crit}}f^{2}_{a}/2 = 2 m_{a}^{2}(T_{*})f^{2}_{a}$. It is convenient to write this condition in terms of the comoving PQ charge density as
\begin{align}
 Y_{\rm PQ} = \frac{n_{\rm PQ}}{s} = \frac{\dot{\theta} f_a^2}{s}  > Y_{\rm crit} \simeq 0.07 \left( \frac{f_{a}}{10^{9} \GeV} \right)^{\scalebox{1.01}{$\frac{13}{6}$}}   \, ,
\end{align}
where we evaluated $g_{*}(T_{*})\simeq 80$. As soon as the kinetic energy drops below the potential height, the axion starts to oscillates down to the potential minimum. The axion energy density at the time of the oscillations for $Y_{\rm PQ} \gg  Y_{\rm crit}$ is 
\begin{align}
\label{eq:N_PQ_epsilon} 
\frac{\rho_{a}}{s} \simeq  2 m_{a}(0) Y_{\rm PQ} \, ,
\end{align}
where the factor of 2 is a correction from the analytical estimation~\cite{Co:2019jts}, resulting in $Y_{a}\simeq 2 Y_{\rm PQ}$. The final dark matter abundance set by KMM is
\begin{align}
\label{eq:DM_relic_KMM}    
 \Omega_{a}h^{2} \simeq 0.12 \left(\frac{Y_{\rm PQ}}{37}\right)\left(\frac{10^{9} \GeV}{f_{a}}\right) \, .
\end{align}
Notice that if we substitute $Y_{\rm crit}$ in the previous equation the kinetic misalignment mechanism cannot produce the entirety of the dark matter abundance for $f_{a}\gtrsim 2\times 10^{11} \GeV$.

\section{Initiation of rotations}
\label{app:epsilon}

For large initial field values $|P_{i}|\gg N_{\rm DW} f_{a}$, the higher-dimensional potential $V_{\cancel{\rm PQ}}(P)$ can be extremely effective in generating an initial kick for the $P$ field in the $\theta$ direction. Since $|P|$ decreases due to cosmic expansion, the influence of the higher-dimensional operator diminishes. As a result,    
the PQ charge density can be thought of as the charge associated with the global $U(1)_{\rm PQ}$ symmetry. Therefore, the corresponding Noether charge density is simply $n_{\rm PQ}= (iP\dot{P^{*}}-iP^{*}\dot{P})/N_{\rm DW} =\dot{\theta}S^{2}/ N^{2}_{\rm DW} $. 
We derive the initial asymmetry produced by $V_{\cancel{\rm PQ}}(P)$ using the equation of motion of $P$
\begin{align}
\label{eq:eom_P}
\ddot{P}+3H\dot{P}+\frac{\partial V_{\rm PQ}}{\partial P^{*}}+\frac{\partial V_{\cancel{\rm PQ}}}{\partial P^{*}}=0 \,.
\end{align}
It is useful to isolate the contribution of $V_{\cancel{\rm PQ}}(P)$ to understand the initial kick more clearly. After multiplying Eq.~(\ref{eq:eom_P}) by $P^{*}$ and subtracting the equation of motion of $P^{*}$ multiplied by $P$, we obtain
\begin{align}
S^{2}\, \ddot{\theta}+2\dot{S}S \, \dot{\theta}+3H S^{2} \, \dot{\theta} = i N_{\rm DW} \left(P^{*}\frac{\partial V_{\cancel{\rm PQ}}}{\partial P^{*}} - P\frac{ \partial V_{\cancel{\rm PQ}}}{\partial P}\right) \,.
\end{align}
Then, the first two terms on the left-hand side are just $\dot{n}_{\rm PQ}=(S^{2}\ddot{\theta}+2\dot{S}S\dot{\theta})/N^{2}_{\rm DW}$. Therefore, the evolution of the PQ charge density is described by 
\begin{align}
\dot{n}_{\rm PQ} + 3Hn_{\rm PQ} = i N^{-1}_{\rm DW} \left(P^{*}\frac{\partial V_{\cancel{\rm PQ}}}{\partial P^{*}} - P\frac{ \partial V_{\cancel{\rm PQ}}}{\partial P}\right) \,.
\end{align}
This equation is general and can be used for any $V_{\cancel{\rm PQ}}(P)$ as long as $S\gg N_{\rm DW} f_a$. Using $V_{\cancel{\rm PQ}}(P)$ of Eq.~(\ref{eq:explicit_PQV}) (or Eq.~(\ref{eq:explicit_PQV_S})) and integrating, the PQ asymmetry produced at time $t$ is
\begin{align}
\label{eq:ntheta_i}
R^{3} \, n_{\rm PQ}(t) = \frac{2n \, A}{2^{n/2} N_{\rm DW} M^{n-3}} \int \frac{R^{2} S^{n} \sin(n\theta/N_{\rm DW})}{H} \, \mathrm{d}R \,.
\end{align}
Let us examine in detail when the dominant contribution to the PQ asymmetry is generated.  Initially, the saxion is frozen due to Hubble friction and therefore the field is constant $S=S_i$. The term on the right-hand side proportional to $R^5$ in a radiation dominated universe. Hence, the dominant contribution results from later times. When $3H = m_S(S_i)$ the saxion begins to oscillate and the amplitude decreases as $S\propto R^{-k}$, with $k=1, \frac{3}{2}$ for quartic and quadratic potentials respectively. The term on the right-hand side of Eq.~(\ref{eq:ntheta_i}) goes as $R^{-nk+5}$ and as long $nk>5$ the contribution to the PQ asymmetry at later times is negligible. Therefore, $n_{\rm PQ}$ is essentially conserved as $V_{\cancel{\rm PQ}}(P)$ becomes negligible. Hence, the PQ asymmetry is dominantly produced at the onset of $S$ oscillations at $R=R_{\rm osc}$ and is approximated by evaluating the integral at $R_{\rm osc}$
\begin{align}
\label{eq:ntheta_A}
n_{\rm PQ}(t_{\rm osc}) \simeq \frac{6 n \, A S^{n}_{i} \sin(n\theta_{\rm inf}/N_{\rm DW})}{2^{n/2} \, N_{\rm DW} \, M^{n-3} \, m_S (S_{i})} \,,
\end{align}
where $\theta_{\rm inf}$ is the initial value of the phase direction of the $P$ field and it is fixed during inflation and one expects $\sin(n\theta_{\rm inf}/N_{\rm DW}) \simeq \mathcal{O}(1)$. 
For the quartic case, using Eq.~(\ref{eq:ntheta_A}) in the definition of $\epsilon$ from Eq.~(\ref{eq:epsilon_q}), gives
\begin{align}
\label{eq:epsilon_quartic}
\epsilon  \simeq \frac{6 n \, A S^{n-2}_{i}}{2^{n/2} \, M^{n-3} \, m_S^{2} (S_{i})} =  3 \left( \frac{V'_{\cancel{\rm PQ}}}{V'_{\rm PQ}} \right) \,.
\end{align}
Now for the quadratic case, the PQ asymmetry density due to the $A$-term potential is derived from Eqs.~(\ref{eq:ntheta_A}) and (\ref{eq:saxion_minimum}) and reads 
\begin{align}
n_{\rm PQ} \simeq \left (\frac{n}{N_{\rm DW} (n-1)^{1/2}} \right) A S_i^2 \,.
\end{align}
Unlike the quartic case, the definition of $\epsilon$ is slightly different, $\epsilon \equiv (N_{\rm DW} n_{\rm PQ})/2n_S$, giving
\begin{align}
\label{eq:epsilon_quadratic}
\epsilon \simeq \left( \frac{n}{(n-1)^{1/2}} \right) \frac{A}{m_S} =  3 \left( \frac{V'_{\cancel{\rm PQ}}}{V'_{\rm PQ}} \right)\,.
\end{align}

\section{Averaged angular velocity and asymmetries}
\label{app:ave_thetadot}

After initiation, the rotation of the PQ symmetry breaking field $P$ is not necessarily circular; it may have high ellipticity. The angular velocity $\dot{\theta}$ is not a constant, but oscillates in time with a period $\sim m_S^{-1}$. In this appendix, we compute the time-averaged value of $\dot{\theta}$. We also derive the conditions such that the baryon asymmetry produced per Hubble time is simply given by replacing $\dot{\theta}$ with $\vev{\dot{\theta}}$.
Here we put $N_{\rm DW}=1$. For $N_{\rm DW}>1$, the result is simply $N_{\rm DW}$ times larger.

\begin{figure}[!ht]
	\includegraphics[width= 0.6 \linewidth]{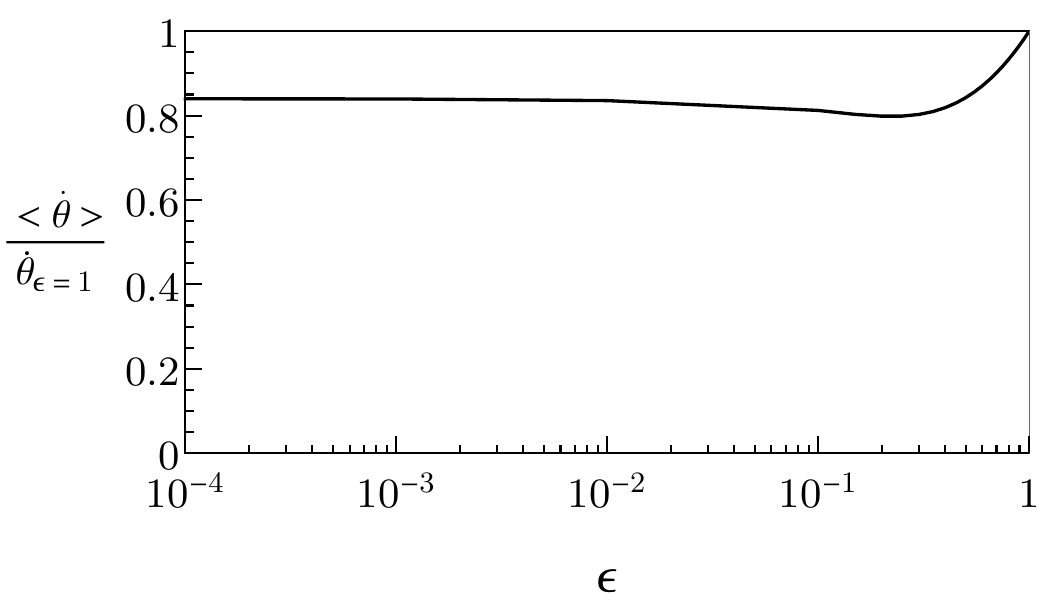}
	\caption{The time average of $\dot{\theta}$ as a function of $\epsilon$ for the quartic potential, normalized to $\dot{\theta}$ for a pure rotation ($\epsilon = 1$).}
	\label{fig:dottheta_ave}	
\end{figure}

As discussed in the main text, the order of magnitude of $\vev{\dot{\theta}}$ is independent of $\epsilon$, and the precise value becomes independent of $\epsilon$ for $\epsilon \ll 1$. In Fig.~\ref{fig:dottheta_ave}, we show $\vev{\dot{\theta}}$ as a function of $\epsilon$ for a quartic potential. 

The near independence of $\vev{\dot{\theta}}$ on $\epsilon$ can be understood as follows.
For a rotation with a maximal amplitude $S_{\rm max}$, the energy density $\rho$ and the PQ asymmetry $n_{\rm PQ}$ are
\begin{align}
\rho \simeq m_S^2 S_{\rm max}^2,~~n_{\rm PQ} \simeq \epsilon m_S S_{\rm max}^2.
\end{align}
The angular velocity $\dot\theta$ takes the maximal value when $P$ passes nearest to the origin. The maximal value can be estimated by the energy and the charge conservations,
\begin{align}
\label{eq:dtheta_max}
\dot{\theta}_{\rm max} \simeq \frac{m_S}{\epsilon}.
\end{align}
During one rotation, $\dot{\theta}$ remains of this order only for a time scale $\dot{\theta}_{\rm max}^{-1}$. The time-averaged value of $\dot{\theta}$ is then
\begin{align}
\vev{\dot{\theta}} \simeq \dot{\theta}_{\rm max}  \times \frac{m_S}{\dot{\theta}_{\rm max}} = m_S,
\end{align}
which is independent of $\epsilon$.

We use this time-averaged value for $\dot{\theta}$ in Eq.~(\ref{eq:nB-L}), but this is not correct for extremely small $\epsilon$. Eq.~(\ref{eq:nB+L}) implicitly assume $\dot{\theta} < T$ and $S> T$ which is violated by Eq.~(\ref{eq:dtheta_max}) for too small $\epsilon < m_S/T$ or $T/S_{\rm max}$. Also, when $\dot{\theta} >T$, it can no longer be treated us a background field which slowly varies in comparison with the typical time scale of thermal bath $\sim T^{-1}$. We assume $\epsilon > m_S/T$ and $T/S_{\rm max}$ in this paper.

When $\epsilon \ll1$, $\dot{\theta}$ rapidly changes around $\dot{\theta} \sim \dot{\theta}_{\rm max}\sim m_S/\epsilon$. One may then wonder that the transfer rate of $n_{\rm PQ}$ into particle charge asymmetries must be larger than $m_S/\epsilon$ rather than $H$, so that the particle charge asymmetries follow $\sim \dot{\theta} T^2$ during the rotations and our estimation of the $B-L$ asymmetry is correct. We find that such a large transfer rate is not necessary. 
To see this, let us consider the transfer of the charge $P\rightarrow \psi_1 \rightarrow \psi_2$ governed by the following Boltzmann equations,
\begin{align}
    \dot{n}_1 &= \Gamma_1(\dot{\theta} T^2 - n_1) - \Gamma_2(n_1-n_2),\nonumber \\
    \dot{n}_2& = \Gamma_2(n_1 - n_2).
\end{align}
The cycle averages of $n_{1,2}$ are
\begin{align}
    \vev{n_1} &= \frac{\omega}{2\pi}\int_0^{2\pi/\omega} {\rm d}t~n_1 = \frac{\omega}{2\pi}\int_0^{2\pi/\omega} {\rm d}t~\left( \dot{\theta}T^2 - \frac{\dot{n}_1 +\dot{n}_2 }{\Gamma_1} \right) \nonumber \\
    &= \vev{\dot{\theta}}T^2 - \frac{\omega}{2\pi \Gamma_1}\left(n_1\left(\frac{2\pi}{\omega}\right)- n_1\left(0\right)+ n_2\left(\frac{2\pi}{\omega}\right)- n_2\left(0\right) \right), \nonumber \\
    \vev{n_2} &= \vev{n_1} - \frac{\omega}{2\pi \Gamma_2}\left(n_2\left(\frac{2\pi}{\omega}\right)- n_2\left(0\right) \right) .
\end{align}
As long as $\Gamma_{1,2}$ is much larger than $H$, $n_{1,2}$ follows a quasi-equilibrium value where they change periodically, and $n_{1,2}(2\pi/\omega) -n_{1,2}(0) $  vanishes. Thus, the cycle averages of $n_{1,2}$ are given by
\begin{align}
    \vev{n_1} = \vev{\dot{\theta}} T^2,~~\vev{n_2} = \vev{n_1}.
\end{align}
One can generalize this analysis to a more generic chain of transfers and show that the cycle average of particle asymmetries follows $\sim \vev{\dot{\theta}} T^2$ as long as the transfer rates are larger than $H$. 
One can also confirm that if some transfer rate $\Gamma_i$ in the chain is below $H$, a suppression factor $\Gamma_i/H$ is present after that transfer is involved.

\begin{figure}[!ht]
	\includegraphics[width= \linewidth]{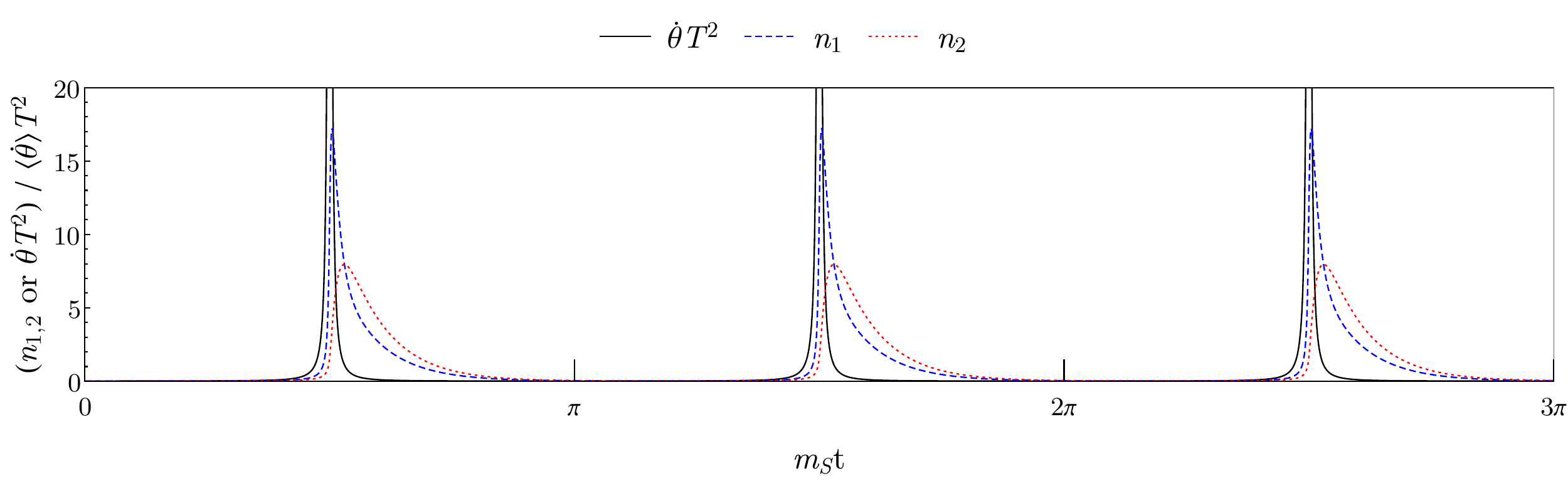}
	\includegraphics[width= \linewidth]{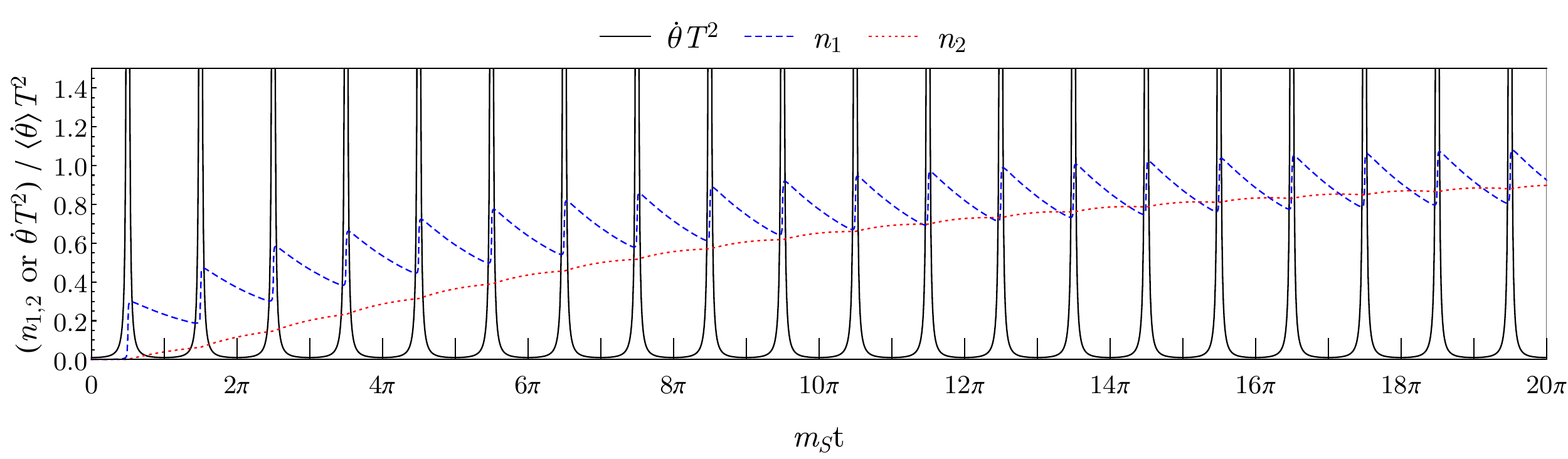}
	\caption{The evolution of asymmetries. In the upper panel, $m_S\ll \Gamma_{1,2}\ll m_S/\epsilon$, while in the lower panel $\Gamma_{1,2}\ll m_S\ll  m_S/\epsilon$.}
	\label{fig:asymmetry}	
\end{figure}

Although the cycle averages of $n_{1,2}$ do not depend on the hierarchy between $m_S$ and $\Gamma_{1,2}$, the evolution of $n_{1,2}$ during the cycle shows quite different behaviors depending on the hierarchy.
In the upper panel of Fig.~\ref{fig:asymmetry}, we show the numerical solution of the Boltzmann equation for $\epsilon = 0.01$, $\Gamma_{1,2} = 10 m_S$ with $n_{1,2}(0)=0$. Here we assume a quadratic potential of $S$, but a similar result holds for a quartic potential.
$n_1$ does not follow the equilibrium value $\dot{\theta} T^2$ since $\Gamma_1 \ll m_S/\epsilon$. In each cycle, $n_1$ is driven to a non-zero value by instantaneously large $\dot{\theta}$, and exponentially decays with a time scale $\Gamma_1^{-1}$. Since $\Gamma_1 \gg m_S$, the decay is effective and at the end of the cycle, $n_1\simeq 0$. $n_2$ shows a similar behavior driven by the rapid change of $n_1$.

In the lower panel of Fig.~\ref{fig:asymmetry}, we show the numerical solution for $\epsilon = 0.01$, $\Gamma_{1,2} = 0.1 m_S$ with $n_{1,2}(0)=0$. Initially, in each cycle, $n_1$ increases by instantaneously large $\dot{\theta}$. Since $\Gamma_1 \ll m_S$, the decay is not effective and $n_1$ is non-zero at the end of the cycle. After a time $\sim \Gamma_1^{-1}$, $n_1$ begins a periodical evolution with $\vev{n_1} = \vev{\dot{\theta}}T^2$. $n_2$ shows a similar behavior driven by $n_1$.

In the above analysis $\Gamma_1$ is assumed to be constant during the cycle. This is the case, for example, if $\Gamma_1$ is given by the sphaleron transition. In the \ref{model:KSVZ_light} or \ref{model:DFSZ_light}, $\Gamma_1$ depends on $S$ and rapidly changes during the cycle. Let us consider the following Boltzmann equation,
\begin{align}
\label{eq:Boltzmann_light}
    \dot{n}_1 &= \Gamma_{0} \frac{S^2}{S_{\rm max}^2}(\dot{\theta} T^2 - n_1),
\end{align}
which is applicable to the case where $P$ linearly couples to light particles. 
By numerically solving the Boltzmann equation, we find that the cycle averaged $n_1$ is well-fitted by 
\begin{align}
\label{eq:nA_light}
   \vev{n_1} \simeq  \vev{\dot{\theta}}T^2 \times
    \begin{cases}
      2\epsilon & : \Gamma_0 \ll m_S \\
     \epsilon \left(\frac{\Gamma_0}{m_S}\right)^{1/3} & m_S \ll \Gamma_0 \ll m_S/\epsilon^3 \\
     1 & m_S/\epsilon^3 \ll \Gamma_0
    \end{cases}.
\end{align}
The result can be understood in the following way. For $\Gamma_0\ll m_S$, the evolution of $n_1$ is slow in comparison with that of $S$, and we may take the time average of Eq.~(\ref{eq:Boltzmann_light}) over a time scale longer than $m_S^{-1}$. Taking $\vev{\dot{n}_1}=0$, we obtain the first line in Eq.~(\ref{eq:nA_light}).
For $\Gamma_0 \gg m_S$, $n_1$ follows an equilibrium value $\dot{\theta} T^2$ when $S$ is close to $S_{\rm max}$. As $S$ becomes smaller and $\dot{\theta}$ becomes larger, $n_1$ fails to follow $\dot{\theta}T^2$, which occurs when
\begin{align}
    \frac{\ddot{\theta}}{\dot{\theta}}\simeq \Gamma_0 \frac{S^2}{S_{\rm max}^2}.
\end{align}
Using the charge conservation $\dot{\theta}S^2 = \epsilon m_S S_{\rm max}^2$ and the energy conservation $|\dot{S}| \simeq m_S S_{\rm max}$, we find that this occurs when
\begin{align}
    \dot{\theta} \simeq \epsilon\Gamma_0^{1/3} m_S^{1/3} \equiv \dot{\theta}_d.
\end{align}
After this occurs, for a time scale $\Delta t_d \simeq (\dot{\theta}/\ddot{\theta})|_{\dot{\theta} \simeq \dot{\theta}_d} \simeq \Gamma_0^{-1/3} m_S^{-2/3}$, $n_1$ is as large as $\dot{\theta}_d T^2$. The averaged $n_1$ is then $\dot{\theta}_d T^2 \Delta t_d m_S$, which is the second line of Eq.~(\ref{eq:nA_light}). When $\Gamma_0 \gg m_S/\epsilon^3$, $n_1$ follows $\dot{\theta} T^2$ whole time during the cycle, and we obtain the third line of Eq.~(\ref{eq:nA_light}).

\section{Scaling laws in various cosmological eras for quadratic potentials}
\label{app:scaling}

\begin{figure}[!ht]
	\includegraphics[width= \linewidth]{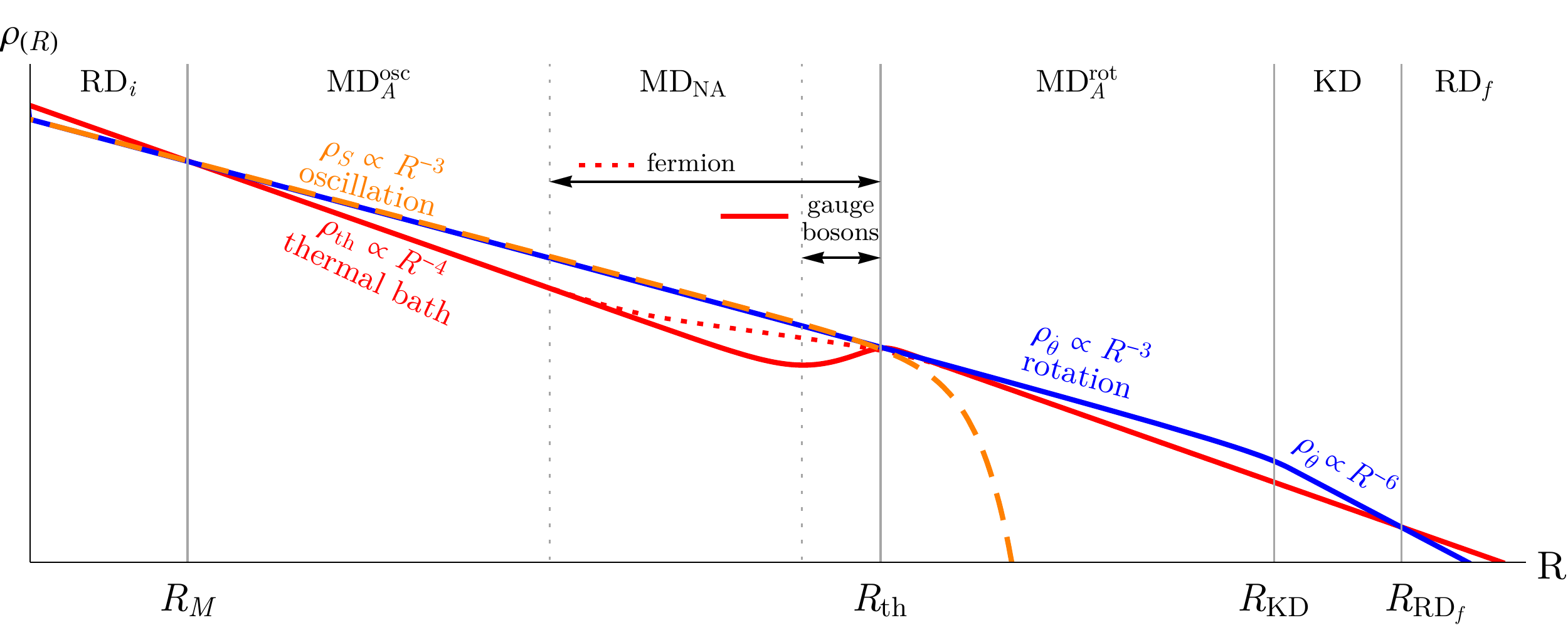}
	\caption{Evolution of various energy densities as functions of the scale factor $R$ for the quadratic potential. The red, orange dashed, and blue curves are respectively the energy densities of the thermal bath, the oscillation (saxion), and the rotation (axion). The red solid and dotted curves distinguish the different thermalization channels-- scattering with the fermion and the gluon. The vertical lines separate different cosmological eras: $R_M$ the beginning of matter domination MD, $R_{\rm th}$ the completion of thermalization, $R_{\rm KD}$ the beginning of kination domination KD, and $R_{{\rm RD}_f}$ the beginning of final radiation domination RD$_f$.}
	\label{fig:evolution_quadratic}	
\end{figure}

If the universe is radiation dominated when the PQ symmetry breaking field begins rotation, as shown in Fig.~\ref{fig:evolution_quadratic}, the universe experiences the following six eras: 1) the first radiation domination, RD$_{i}$, 2) the first matter domination with adiabatic evolution of the thermal bath, MD$_{\rm A}^{\rm osc}$, 3) matter domination with non-adiabatic evolution of the thermal bath, MD$_{\rm NA}^{\rm osc}$, 4) the second matter domination with adiabatic evolution of the thermal bath, MD$_{\rm A}^{\rm rot}$, 5) kination domination, KD, and 6) the second radiation domination, RD$_{f}$. The second matter-dominated era MD$_{\rm A}^{\rm rot}$ is possible because thermalization processes are PQ-conserving and cannot deplete the energy density associated with the rotation that carries the PQ charge. Importantly, this MD$_{\rm A}^{\rm rot}$ era does not lead to subsequent entropy production due to the rapid redshifting during kination, when the saxion field value is relaxed to $f_a$.

We derive in this section the scaling of the baryon asymmetry produced per Hubble time during these various eras. For pedagogical purposes, we show explicit discussions only for the case where $T_{\rm EW} < T < T_{\rm ss}$ so the strong sphaleron processes are in thermal equilibrium and the rotation begins in a radiation-dominated universe. A summary of all other relevant cases is given below in Table~\ref{tab:scalings}.

Since entropy is produced at thermalization of the rotating PQ symmetry breaking field $P$, before the completion of the thermalization it is convenient to normalize the baryon number by the energy density of $P$, which scales as $\rho_P \propto R^{-3}$.

1) During the first radiation domination, $T \propto R^{-1}$, $\Gamma_{L} \propto T^{3} \propto R^{-3}$, and $H \propto R^{-2}$. Since $\dot{\theta} = $ const for $\bar{S}\gg f_{a}$, using (\ref{eq:nBquad}) we find
\begin{equation}
\frac{\Delta n_{B}}{\rho_P}\propto R^{0}\,.
\end{equation}
Baryon production proceeds with equal efficiency at all temperatures.

2) During the first matter dominated adiabatic era, we have $T \propto R^{-1}$, $\Gamma_{L} \propto R^{-3}$ and $H \propto R^{-3/2}$, so
\begin{equation}
\frac{\Delta n_{B}}{\rho_P} \propto R^{-1/2}.
\end{equation}
Baryon production is UV dominated and most efficient when the universe becomes matter dominated at $T = T_{\rm M}$.

3) During matter domination with non-adiabatic evolution,
$T \propto R^{3/2}$ and $R^{-1/2}$ for the thermalization by gluon scattering and fermion scattering, respectively.
Then
\begin{equation}
\frac{\Delta n_{B}}{\rho_P} \propto R^{12}~{\rm and}~R^{2},
\end{equation}
respectively. The production is IR dominated and hence the baryon asymmetry is dominantly produced at $T_{\rm th}$ when thermalization is completed.

4) After the completion of thermalization, the universe is in the second matter dominated adiabatic era. Since no more entropy is created, it is now convenient to normalize the asymmetry by the entropy density. Similar to the first matter dominated adiabatic era,
\begin{equation}
\frac{\Delta n_{B}}{s} \propto R^{-1/2}
\end{equation}
and hence the production of baryon asymmetry is dominated at $R_{\rm th}$.

5) The universe enters kination domination when $S \simeq f_{a}$. In kination domination $\dot\theta \propto R^{-3}$ and $H \propto R^{-3}$. This leads to
\begin{equation}
\frac{\Delta n_{B}}{s} \propto R^{-2},
\end{equation}
which is UV dominated.

6) During the second radiation domination, $H \propto R^{-2}$, leading to
\begin{equation}
\frac{\Delta n_{B}}{s} \propto R^{-3},
\end{equation}
which is again UV dominated.

\begin{table}
\begin{center}
\makebox[\textwidth][c]{
\begin{tabular}{|c|c|c|c|c|c|c|c|c|c|c|}
 \hline
\multirow{2}{*}{Epoch} & \multirow{2}{*}{$H$} & \multirow{2}{*}{$T$} &  \multirow{2}{*}{$\Gamma_L$} &\multirow{2}{*}{$\Gamma_{\rm ss}$} & \multirow{2}{*}{$\rho_P$} & \multirow{2}{*}{$\dot\theta$} &   \multicolumn{2}{c|}{$T<T_{\rm ss}$} &  \multicolumn{2}{c|}{$T_{\rm ss} < T < T_L$}     \\   \cline{8-11} 
 & & & & & & & $\frac{\Delta n_B}{s}$ & $\frac{\Delta n_B}{\rho_P}$ & \ $\frac{\Delta n_B}{s}$ \  & $\frac{\Delta n_B}{\rho_P}$ \\
\hline
\rowcolor{gray!20} MD$_{\rm NA}^{\rm inf}$ & $R^{-\frac{3}{2}}$ & $R^{-\frac{3}{8}}$ & $R^{-\frac{9}{8}}$ & $R^{-\frac{3}{8}}$ & $R^{-3}$ &  $R^{0}$  & -- & $R^{\frac{21}{8}}$ & -- & $R^{\frac{15}{4}}$ \\ \hline 
\rowcolor{red!30} RD$_{i}$ & $R^{-2}$ & $R^{-1}$ & $R^{-3}$ & $R^{-1}$ & $R^{-3}$ &  $R^{0}$  & -- & $R^{0}$ & -- & $R^{1}$  \\ \hline 
MD$_{\rm A}^{\rm osc}$ & $R^{-\frac{3}{2}}$ & $R^{-1}$ & $R^{-3}$ & $R^{-1}$ & $R^{-3}$ &  $R^{0}$  & -- & $R^{-\frac{1}{2}}$ & -- & $R^{0}$  \\ \hline 
MD$_{\rm NA}^{\rm osc}$ $\begin{dcases} {\scalebox{1.3}{$^{\rm gauge}_{\rm bosons}$}} \\ {\rm fermion} \end{dcases}$ & \makecell[c]{$R^{-\frac{3}{2}}$ \\ $R^{-\frac{3}{2}}$}  & \makecell[c]{$R^{\frac{3}{2}}$ \\ $R^{-\frac{1}{2}}$}  & \makecell[c]{$R^{\frac{9}{2}}$ \\ $R^{-\frac{3}{2}}$}  & \makecell[c]{$R^{\frac{3}{2}}$ \\ $R^{-\frac{1}{2}}$} &  \makecell[c]{$R^{-3}$ \\ $R^{-3}$} &   \makecell[c]{$R^{0}$ \\ $R^{0}$} &  \makecell[c]{-- \\ --} & \makecell[c]{$R^{12}$ \\ $R^{2}$}& \makecell[c]{-- \\ --} & \makecell[c]{$R^{15}$ \\ $R^{3}$}  \\ \hline
\rowcolor{blue!20} MD$_{\rm A}^{\rm rot}$ & $R^{-\frac{3}{2}}$ & $R^{-1}$ & $R^{-3}$ & $R^{-1}$ & $R^{-3}$ &  $R^{0}$ & $R^{-\frac{1}{2}}$ & -- & $R^{0}$  & --  \\ \hline
\rowcolor{blue!20} KD & $R^{-3}$ & $R^{-1}$ & $R^{-3}$ & $R^{-1}$ & $R^{-6}$ &  $R^{-3}$ & $R^{-2}$ & --  & $R^{0}$ & --  \\ \hline
\rowcolor{red!30} RD$_{f}$ & $R^{-2}$ & $R^{-1}$ & $R^{-3}$ & $R^{-1}$ & $R^{-6}$ &  $R^{-3}$ & $R^{-3}$ & -- & $R^{-2}$ & -- \\ \hline
\end{tabular}
}
\end{center}
\caption{Scalings of physical quantities relevant for the estimation of the baryon asymmetry.}
\label{tab:scalings}
\end{table}

We summarize the results in Table~\ref{tab:scalings}. If the saxion oscillates during the inflation matter-dominated era, the universe starts with the gray row MD$_{\rm NA}^{\rm inf}$ and moves to the first red row RD$_{i}$ after inflationary reheating completes. Otherwise, the universe starts from RD$_{i}$ when the saxion oscillates. If the saxion energy comes to dominate, then the universe evolves through the white rows MD$_{\rm A, NA}^{\rm osc}$. In this case, the MD$_{\rm A}^{\rm rot}$ and KD eras in the blue rows are expected to occur before the final red row RD$_{f}$ arrives, unless $\epsilon$ is sufficiently small. On the other hand, if the saxion thermalizes before dominating, then the universe does not evolve through MD$_{\rm A, NA}^{\rm osc}$, MD$_{\rm A}^{\rm rot}$, and KD and instead goes directly to RD$_{f}$ from RD$_{i}$. In each row, the scaling laws of various relevant quantities are listed; a derived quantity involving $\Delta n_B$, which is otherwise conserved in the absence of a source, is presented for each case of $T$ in relation to $T_{\rm ws}$ and $T_L$. From this derived quantity, one can determine  by the scaling with $R$ whether the baryon asymmetry production per unit Hubble time is UV- or IR-dominated or neither.

Finally, we note that this table still does not exhaust all the possibilities because we have assumed that $S$ settles to $f_a$ at a temperature lower than any other relevant temperatures. A sufficiently large $S_i$ will validate this assumption.

\section{Parametric resonance for the nearly quadratic potential}
\label{app:PR_quadratic}

In this appendix, we discuss parametric resonance in theories with nearly quadratic potentials.
We first consider the theory with the potential in Eq.~(\ref{eq:dim_trans}),
and show the critical value of $\epsilon$ below which parametric resonance occurs. 

We decompose the field $P$ as
\begin{align}
 P = \frac{S_0}{\sqrt{2}} \left( X(t) + \chi (x,t) \right) e^{ i (\theta(t) + \alpha(x,t))},
\end{align}
and scale the space-time variables as
\begin{align}
 (x,t) \rightarrow \frac{1}{\omega_0}(x,t),~~ \omega_0 \equiv m_S \left( 2\ln \frac{S_0}{f_a} \right)^{ \scalebox{1.01}{$\frac{1}{2}$} }.
\end{align}
We expect that the growth rate of fluctuations is typically at most $\mathcal{O}(m_S)$, and parametric resonance becomes effective only after tens of oscillations. At that stage, since $m_S / H \gg 1$, we may neglect the effect of cosmic expansion.
The equations of motion of the zero modes $X$ and $\theta$ are
\begin{align}
    \ddot{X} - \dot{\theta}^2 X + \frac{\ln \left(r^2 X^2\right)}{\ln r^2  } X = & 0,~~ r \equiv \frac{S_0}{f_a} \,, \\
    \ddot{\theta} + \frac{2}{X} \dot{X} \dot{\theta} = &  0 \,.
\end{align}
We take the initial time at a point with $\dot{X}(0)=0$. By adjusting $S_0$, we can take $X(0) = 1$.
The solution for $\theta$ is
\begin{align}
    \dot{\theta} = \frac{X(0)^2 \dot{\theta}(0)}{X^2} = \frac{\epsilon}{X^2},~\epsilon \equiv \dot{\theta}(0) \,,
\end{align}
which is nothing but conservation of the PQ charge. The equation of motion of $X$ is
\begin{align}
\ddot{X} - \frac{\epsilon^2}{X^4} X + \frac{\ln \left(r^2 X^2\right)}{\ln r^2  } X =  0 \,.
\end{align}

\begin{figure}[!t]
	\includegraphics[width=0.5 \linewidth]{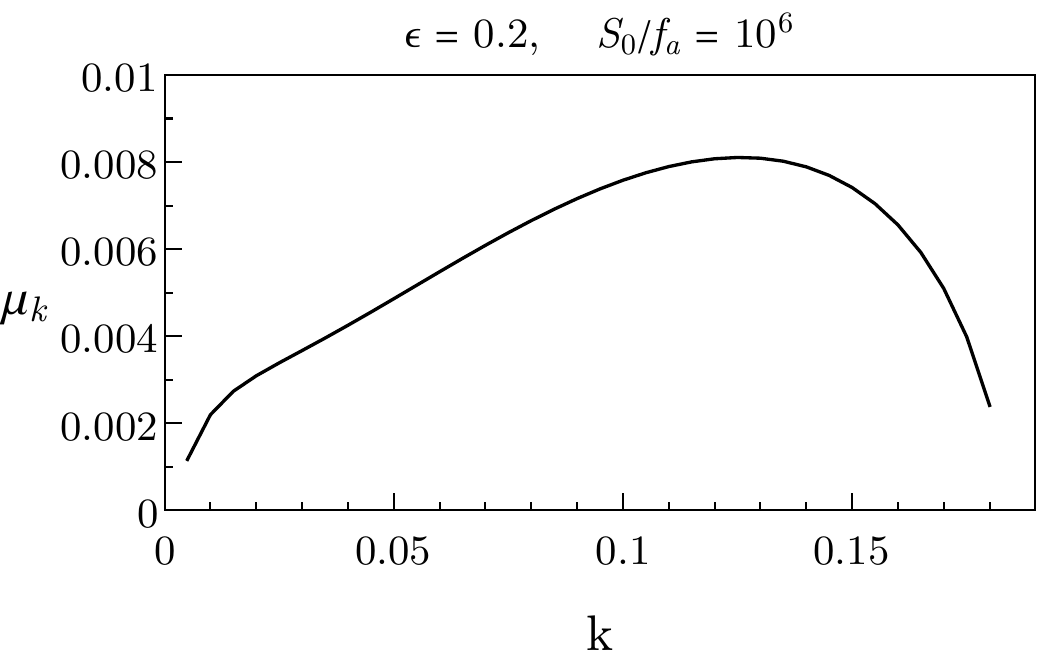}
	\caption{The growth rate the axion fluctuations for $\epsilon = 0.2$ and $S_0/f_a=10^6$.}
	\label{fig:mu_k}	
\end{figure}

The equations of motion for the fluctuations $\chi$ and $\alpha$ in momentum space are 
\begin{align}
 \ddot{\chi}_k - 2 X \dot{\theta} \dot{\alpha}_k + \left( k^2 + \frac{2 + \ln\left(r^2 X^2\right) }{\ln r^2} - \dot{\theta}^2 \right) \chi_k = 0 \,, \\
 \ddot{\alpha}_k + \frac{2}{X}\left(\dot{\theta}\dot{\chi}_k+ \dot{X}\dot{\alpha}_k\right) + k^2 \alpha_k + \frac{\ddot{\theta}}{X} \chi_k = 0 \,.
\end{align}
We solve these equations of motion numerically, and compute the growth rate $\mu_k$ of the amplitudes of the fluctuations,
\begin{align}
    \chi_k, \alpha_k \sim e^{\mu_k t} \,.
\end{align}
We find that the axion fluctuations $\alpha$ generically grow faster than the fluctuations of $\chi$. In Fig.~\ref{fig:mu_k}, we show $\mu_k$ of the axion fluctuations for $\epsilon=0.2$ and $S_0/f_a = 10^6$.

For a given $\epsilon$, we obtain the value of the wavenumber $k_{\rm max}$ with the maximal growth rate $\mu_{\rm max}$, both of which are shown in Table~\ref{tab:k_mu}. Here we take $S_0 / f_a = 10^6$ and $10^4$. $(k_{\rm max},\mu_{\rm max})$ stays nearly constant for $\epsilon \ll 1$, and becomes smaller as $\epsilon$ approaches $\mathcal{O}(1)$. We find no resonant bands for $\epsilon > 0.5 (0.6)$ for $S_0/f_a = 10^6 (10^4)$. The resonance enhancement becomes stronger for smaller $S_0/f_a$ as the potential deviates from a quadratic one for smaller $S_0$.

\begin{table}[]
    \centering
    \begin{tabular}{c| c c c c c c c c c}
    \multicolumn{10}{c}{$S_0/f_a = 10^6$}\\ \hline
    $\epsilon$ & 0.001 & 0.01 & 0.1 & 0.2 & 0.3 & 0.4 & 0.5 & 0.6 & 0.7   \\ \hline
   $k_{\rm max}$ & 0.19 & 0.19 & 0.15 & 0.13 & 0.10 & 0.08 & 0.06 & N/A & N/A   \\
   $\mu_{\rm max}$ & 0.018 & 0.017 & 0.012 & 0.008 & 0.006 & 0.005 & 0.002 & N/A & N/A \\
       \multicolumn{10}{c}{$S_0/f_a = 10^4$}\\ \hline
           $\epsilon$ & 0.001 & 0.01 & 0.1 & 0.2 & 0.3 & 0.4 & 0.5 & 0.6 & 0.7 \\ \hline
   $k_{\rm max}$ & 0.23 & 0.23 & 0.19 & 0.15 & 0.12 & 0.1 & 0.08 & 0.06 & N/A \\
   $\mu_{\rm max}$ & 0.027 & 0.026  & 0.018 & 0.012 & 0.008 & 0.006 & 0.006 & 0.002 & N/A
    \end{tabular}
    \caption{The wavenumber $k_{\rm max}$ having the maximal growth rate $\mu_{\rm max}$, for the theory of Eq.~(\ref{eq:dim_trans}) with a logarithmically evolving soft mass.}
    \label{tab:k_mu}
\end{table}

\begin{table}[]
    \centering
    \begin{tabular}{c | c c c c c c c c}
    \multicolumn{9}{c}{$\epsilon=0.2$}\\ \hline
    $S_0 / \sqrt{2} v_{\rm PQ}$ & 10 & 20 & 30 & 40 & 50 & 60 & 70 & 80    \\ \hline
   $k_{\rm max}$ & 0.72 & 0.20 & 0.09 & 0.05 & 0.03 & 0.02 & 0.02 & N/A  \\
   $\mu_{\rm max}$ & 0.29 & 0.028 & 0.016 & 0.0018 & 0.0015 & 0.0006 & 0.0003 & N/A \\
   \multicolumn{9}{c}{$\epsilon = 0.4$}\\ \hline
          $S_0 / \sqrt{2} v_{\rm PQ}$ & 10 & 20 & 30 & 40 & 50 & 60 & 70 & 80    \\ \hline
   $k_{\rm max}$ & 0.14 & 0.03 & 0.02 & N/A \\
   $\mu_{\rm max}$ & 0.010 & 0.0024 & 0.00025 & N/A \\
       \multicolumn{9}{c}{$\epsilon = 0.5$}\\ \hline
          $S_0 / \sqrt{2} v_{\rm PQ}$ & 10 & 20 & 30 & 40 & 50 & 60 & 70 & 80    \\ \hline
   $k_{\rm max}$ & 0.08 & N/A  \\
   $\mu_{\rm max}$ & 0.0060 & N/A
    \end{tabular}
    \caption{The wave number $k_{\rm max}$ having maximal growth rate $\mu_{\rm max}$, for the theory with the superpotential $W = X(P\bar{P}- v_{\rm PQ}^2)$.}
    \label{tab:mu_moduli}
\end{table}

We next consider the theory with the superpotential in Eq.~(\ref{eq:two_field}).
Without loss of generality we assume that $P\gg v_{\rm PQ}$. We may integrate out the heavy mode with a mass $\sim P$ by the constraint $P\bar{P} =v_{\rm PQ}^2$,
\begin{align}
    {\cal L}_{\rm eff} =\left( 1 + \frac{v_{\rm PQ}^4}{|P|^4} \right) \partial P \partial P^\dag  - m_S^2|P|^2 \left( 1 + \frac{v_{\rm PQ}^4}{|P|^4} \right),
\end{align}
where we assume $m_P = m_{\bar{P}} \equiv m_S$, for simplicity. We decompose $P$ as
\begin{align}
    P = \frac{S_0}{\sqrt{2}}Y \left( 1+ \frac{1}{6 r^4 Y^4} \right)e^{ i (\theta(t) + \alpha(x,t))},\nonumber\\
  r \equiv \frac{S_0}{\sqrt{2} v_{\rm PQ}},~Y=X(t) + \chi(x,t).
\end{align}
Then, to the next-leading order in $v_{\rm PQ}/ |P|$, the equations of motion of the zero modes $X$ and $\theta$ are
\begin{align}
    \ddot{X}  + (1- \dot{\theta}^2)\left( 1- \frac{4 }{3 r^4 X^4}  \right) X=0, \nonumber \\
    \ddot{\theta} + \frac{2}{X} \left( 1- \frac{4 }{3 r^4 X^4}  \right) \dot{\theta}\dot{X}=0.
\end{align}
We take the initial time at a point with $\dot{X}(0)=0$. By adjusting $S_0$, we can take $X(0)=1$. The solution for $\theta$ is
\begin{align}
 \dot{\theta} = \frac{\epsilon}{X^2} \left( 1 - \frac{2}{3 r^4} \left( \frac{1}{X^4}-1 \right) \right),~~\epsilon = \dot{\theta}(0).
\end{align}
The equations of motion of the $\chi$ and $\alpha$ fluctuations, in momentum space, are
\begin{align}
    \ddot{\chi}_k - 2 X \left( 1- \frac{4 }{3 r^4 X^4}  \right) \dot{\theta} \dot{\alpha}_k+ \left( k^2 + \left(1- \dot{\theta}^2\right)\left( 1+  \frac{4}{r^4 X^4}\right) \right)\chi_k =0 \nonumber \\
    \ddot{\alpha}_k + \frac{2}{X}\left( 1- \frac{4 }{3 r^4 X^4}  \right)\left(\dot{\theta}\dot{\chi}_k+ \dot{X}\dot{\alpha}_k\right) + k^2 \alpha_k - \frac{2}{X^2}\left( 1- \frac{20}{3 r^4 X^4}  \right) \dot{\theta}\dot{X} \chi_k=0.
\end{align}

Values of $k_{\rm max}$ and $\mu_{\rm max}$ are shown in Table~\ref{tab:mu_moduli}. PR becomes weak as $\epsilon$ and/or $r$ becomes larger. For $\epsilon =0.2$, resonant bands exist only for $r < 80$. As $\epsilon$ becomes larger, the upper bound on $r$ for a resonance band to exist becomes stronger.

\bibliography{leptoaxio} 

\end{document}